\documentclass[a4paper,aps,pra,twocolumn,floatfix,superscriptaddress,notitlepage,usenames,dvipsnames,svgnames,table,nofootinbib,longbibliography]{revtex4-1}
\usepackage[utf8]{inputenc}
\usepackage[T1]{fontenc}
\usepackage{graphicx}
\usepackage{grffile}
\usepackage{wrapfig}
\usepackage{rotating}
\usepackage[normalem]{ulem}
\usepackage{amsmath}
\usepackage{textcomp}
\usepackage{amssymb}
\usepackage{capt-of}

\usepackage{parskip}
\usepackage{mathrsfs}
\usepackage[margin= 0.75in]{geometry}
\usepackage[braket, qm]{qcircuit}

\usepackage[english]{babel}
\usepackage{letltxmacro}
\LetLtxMacro{\ORIGselectlanguage}{\selectlanguage}
\makeatletter
\DeclareRobustCommand{\selectlanguage}[1]{%
  \@ifundefined{alias@\string#1}
    {\ORIGselectlanguage{#1}}
    {\begingroup\edef\x{\endgroup
      \noexpand\ORIGselectlanguage{\@nameuse{alias@#1}}}\x}%
}
\newcommand{\definelanguagealias}[2]{%
  \@namedef{alias@#1}{#2}%
}
\makeatother
\definelanguagealias{en}{english}
\definelanguagealias{eng}{english}

\usepackage{bbm}
\usepackage{etoolbox}
\usepackage{tcolorbox} 
\usepackage{tikz}
\usepackage{amsthm}
\usepackage{amssymb}
\usetikzlibrary{arrows.meta}
\usepackage{mathtools}

\newcommand{\prlsection}[1]{{\em {#1}.---~}}

\usepackage{bm}
\usepackage{dsfont}
\usepackage[font=small,labelfont=bf,justification=RaggedRight,format=plain]{caption}
\usepackage{subcaption}

\usepackage{thmtools}
\usepackage{thm-restate}

\newtheorem{definition}{Definition}
\newtheorem{proposition}{Proposition}

\newtheorem{cor}{Corollary}

\usepackage{microtype}
\usepackage{dsfont}
\usepackage{tensor}

\usepackage{physics}
\usepackage{mathrsfs}
\usepackage{mathtools}
\usepackage{relsize}

\DeclarePairedDelimiterX\phys[2]{\langle}{\rangle}{#1 \delimsize\vert\mathopen{} #2}

\theoremstyle{remark}
\newtheorem{exmp}{Example}
\newcommand{\naturalto}{%
  \mathrel{\vbox{\offinterlineskip
    \mathsurround=0pt
    \ialign{\hfil##\hfil\cr
      \normalfont\scalebox{1.2}{.}\cr
      $\longrightarrow$\cr}
  }}%
}

\definecolor{blue-violet}{rgb}{0.54, 0.17, 0.89}
\usepackage{hyperref}
\hypersetup{
 pdfauthor={Faidon Andreadakis, Namit Anand, and Paolo Zanardi},
 pdftitle={Scrambling of Algebras in Open Quantum Systems},
 pdflang={English},colorlinks=true,linkcolor=RubineRed,citecolor=blue-violet,urlcolor=Cerulean} 
\usepackage[capitalise]{cleveref}

\begin{document}

\title{Scrambling of Algebras in Open Quantum Systems}

\author{Faidon Andreadakis}
\email [e-mail: ]{fandread@usc.edu}

\author{Namit Anand}
\email [e-mail: ]{namitana@usc.edu}

\author{Paolo Zanardi}
\email [e-mail: ]{zanardi@usc.edu}

\affiliation{Department of Physics and Astronomy, and Center for Quantum Information Science and Technology, University of Southern California, Los Angeles, California 90089-0484, USA}

\date{\today}

\begin{abstract}
Many quantitative approaches to the dynamical scrambling of information in quantum systems involve the study of out-of-time-ordered correlators (OTOCs). In this paper, we introduce an algebraic OTOC ($\mathcal{A}$-OTOC) that allows us to study information scrambling of \textit{generalized quantum subsystems} under quantum channels. For closed quantum systems, this algebraic framework was recently employed to unify quantum information-theoretic notions of operator entanglement, coherence-generating power, and Loschmidt echo. The main focus of this work is to provide a natural generalization of these techniques to open quantum systems. We first show that, for unitary dynamics, the $\mathcal{A}$-OTOC quantifies a generalized notion of information scrambling, namely between a subalgebra of observables and its commutant. For open quantum systems, on the other hand, we find a competition between the \textit{global} environmental decoherence and the \textit{local} scrambling of information. We illustrate this interplay by analytically studying various examples of algebras and quantum channels. To complement our analytical results, we perform numerical simulations of two paradigmatic systems: the PXP model and the Heisenberg XXX model, under dephasing. Our numerical results reveal connections with many-body scars and the stability of decoherence-free subspaces.
\end{abstract}
\maketitle
\section{Introduction} \label{sec1}
\par \emph{Quantum information scrambling}, in its purest form, refers to the ability of quantum systems to generate entanglement and correlations under time evolution \cite{ovchinnikov_quasiclassical_nodate,kitaev_simple_2015,maldacena_bound_2016,lashkari_towards_2013,roberts_diagnosing_2015,polchinski_spectrum_2016,mezei_entanglement_2017,roberts_chaos_2017,hayden_black_2007,shenker_black_2014,swingle_unscrambling_2018,xu_swingle_tutorial_2022}. In the Schr{\"o}dinger picture, this is typically characterized by starting from \textit{distinguishable}, low-entanglement states (e.g., orthogonal product states), which, under unitary dynamics become more and more \textit{indistinguishable} to local measurements. In a similar spirit, in the Heisenberg picture, scrambling manifests as the growth of the support of initially local operators under time evolution and their subsequent noncommutativity with operators supported on distinct subsystems. This spreading of initially localized information (\emph{delocalization}) allows for the emergence of nonlocal quantum correlations, which are linked to many-body phenomena such as thermalization \cite{Gogolin_2016_review} and quantum chaos \cite{rigol2016quantum, maldacena_bound_2016,xu_does_2020}, among others. A central quantitative approach to information scrambling has been the study of out-of-time-order-correlators (OTOCs), which possess a principal position in theoretical insights into scrambling dynamics for a variety of phenomena, ranging from, e.g. many-body chaos to black hole physics \cite{ovchinnikov_quasiclassical_nodate,kitaev_simple_2015,maldacena_bound_2016,lashkari_towards_2013,roberts_diagnosing_2015,polchinski_spectrum_2016,mezei_entanglement_2017,roberts_chaos_2017,hayden_black_2007,shenker_black_2014,swingle_unscrambling_2018,xu_swingle_tutorial_2022}. This theoretical investigation of OTOCs has been accompanied by a number of state-of-the-art experimental implementations \cite{mi_information_2021,braumuller_probing_2022,wei_exploring_2018,li_measuring_2017,nie_detecting_2019,nie_experimental_2020,garttner_measuring_2017,joshi_quantum_2020,meier_exploring_2019,chen_detecting_2020,landsman_verified_2019}. 
\par Recent works have revealed connections between OTOCs and prominent quantum information-theoretic concepts, such as operator entanglement and entropy production \cite{styliaris_information_2021,zanardi_information_2021,anand_brotocs_2021}, quantum coherence \cite{anand_quantum_2021}, Loschmidt echo \cite{yan_information_2020}, quasiprobabilities \cite{PhysRevA.97.042105}, multiple-quantum coherences \cite{PhysRevLett.120.040402}, among others \cite{PhysRevE.99.052143,yoshida_disentangling_2019,touil_information_2021}. In several of these studies, the OTOCs were averaged over an appropriate \textit{class} of randomly distributed operators, thereby extracting features of the OTOC that are \textit{independent} of the specific choice of operators involved, manifesting instead the typical features of the \textit{class} of operators. These results suggest that the averaged OTOC is a promising tool for investigating scrambling properties of dynamical systems, revealing connections to many-body phenomena such as integrability, localization, and quantum chaos.
\par Ref. \cite{zanardi_quantum_nodate} considers unitary dynamics and provides a generalized formalism in which the notion of locality is with respect to a generalized subsystem structure, described by a $*$-closed unital algebra of observables. This gives rise to a natural geometrical picture which connects information scrambling to a distance between algebras, while also conceptually unifying many of the aforementioned results. In this paper, we provide a quantitative framework for analyzing scrambling at the algebra level when the evolution is allowed to be a unital quantum channel (completely positive trace-preserving map) in the Heisenberg picture, thus incorporating open quantum system effects, e.g., decoherence. 
\par Disentangling the contribution of \textit{environmental} decoherence from \textit{unitary} scrambling has been studied in previous works using a host of ideas and techniques. To this end, Ref. \cite{zanardi_information_2021}, by a subset of the current authors introduced Haar averaged OTOCs for open quantum systems, Ref. \cite{yoshida_disentangling_2019} introduced a quantum teleportation based decoding protocol, Ref. \cite{touil_information_2021} used the quantum mutual information between the system and environment, and Ref. \cite{swingle_resilience_2018} introduced an interferometric and weak-measurement based scheme, to list a few. These works have focused on specific forms of environmental decoherence or techniques to disentangle it from scrambling. Our algebraic approach, we believe, may provide a much broader framework for this task.
\par This paper is structured as follows. In \cref{sec2}, we present general results (in the form of Propositions) that combine and extend ideas of Refs. \cite{styliaris_information_2021,anand_quantum_2021,zanardi_information_2021}. In \cref{sec3}, we treat analytically a few illustrative cases of algebras and channels, that complement the general results and reveal the ``competition'' between decoherence and information scrambling. In \cref{sec4}, we study numerically the application of our tools in representative open quantum spin chain models with selected algebras that probe their respective physical properties. In \cref{sec5}, we conclude with a brief discussion of the results. The detailed proofs of the technical results are included in the Supplemental Material \ref{append}.
\section{Theoretical Results} \label{sec2}
Let $\mathcal{H} \cong \mathbb{C}^d$ be a finite $d$-dimensional Hilbert space representing a quantum system and $\mathcal{L}(\mathcal{H})$ be the space of linear operators on $\mathcal{H}$. The space $\mathcal{L}(\mathcal{H})$ endowed with the Hilbert-Schmidt inner product $\left\langle X,Y\right\rangle:=\Tr\left[ X^\dagger Y \right]$ is a Hilbert space and the associated Hilbert-Schmidt norm is the 2-norm $\left\lVert X \right\rVert_2 := \sqrt{\left\langle X,X \right\rangle}$. Quantum states are identified as $\rho \in \mathcal{L}(\mathcal{H})$ with $\rho \geq 0$ and $\Tr [\rho]=1$.
\subsection{Preliminaries} \label{subsec2.1}
The Schr{\"o}dinger picture evolution of quantum states is described by quantum channels, i.e., completely positive trace-preserving (CPTP) superoperators $\mathcal{E}^\dagger$ on $\mathcal{L}(\mathcal{H})$. The Heisenberg picture evolution of observables is described by the adjoint channel $\mathcal{E}$ identified by $\langle X, \mathcal{E}^\dagger (Y) \rangle = \langle \mathcal{E} (X) , Y \rangle$. Since $\mathcal{E}^\dagger$ is completely positive (CP) \& trace-preserving, it follows that $\mathcal{E}$ is CP \& \emph{unital} (namely the identity is a fixed point). Although the concepts presented in the paper largely do not depend on this, it is convenient to assume that $\mathcal{E}^\dagger$ is also unital, which means that $\mathcal{E}$ is also trace-preserving.
\par Given a quantum channel $\mathcal{E}$, the object we will use to quantify scrambling dynamics is the norm of the commutator \cite{zanardi_information_2021}
\begin{equation} \label{eq1}
C_{V,W}(\mathcal{E}) := \frac{1}{2d} \lVert \left[\mathcal{E} (V), W \right] \rVert_2^2.
\end{equation}
To illustrate the intuition behind this quantity and the connection with the OTOC, assume that $V,W$ are local operators that initially commute and the time evolution is unitary ($\mathcal{E}=\mathcal{U}_t$, where $\mathcal{U}_t(V) \equiv V_t= U_t^\dagger V U_t$, $U_t \in \mathcal{L}(\mathcal{H})$ are unitary operators depending on time $t$). Then, under time evolution the support of $V_t$ grows, leading, after sufficient amount of time, to potential non-commutativity with $W$, which is understood as scrambling of information initially localized in the support of $V$. If in addition we assume that $V,W$ are unitaries, then 
\begin{equation*}
C_{V,W}(\mathcal{U}_t)=1-\frac{1}{d}\Re F_{V,W}(\mathcal{U}_t)
\end{equation*}
where
\begin{align}
F_{V,W}(\mathcal{U}_t):=\frac{1}{d} \Tr\left[V_t^\dagger W^\dagger V_t W \right] \label{eq2}
\end{align}
is the four point correlation function referred to as the OTOC  \footnote{We focus on the infinite temperature case where the correlation functions are over the Gibbs state $\rho_{\beta =0} = \frac{\mathds{1}}{d}$, hence the factor of $\frac{1}{d}$.}. Notice that, as the norm of the commutator grows, the OTOC decays. If we allow for open system dynamics, then $C_{V,W}(\mathcal{E})$ will also incorporate effects of decoherence \cite{zanardi_information_2021}.
\par The main mathematical structures of interest are $*$-closed unital algebras of observables $\mathcal{A}$ and their commutants,
\begin{align*}
\mathcal{A}^\prime = \left\{ Y \in \mathcal{L}(\mathcal{H}) \; | \; [X,Y]=0 \; \; \forall \; X \in \mathcal{A} \right\}.
\end{align*}
We denote the center of $\mathcal{A}$ as $\mathcal{Z}(\mathcal{A}) := \mathcal{A} \cap \mathcal{A}^\prime$. Note that by virtue of the double commutant theorem $\left( \mathcal{A}^\prime \right)^\prime = \mathcal{A}$ \cite{davidson_c-algebras_1996}.
\par A fundamental structure theorem for $C^*$-algebras states that there is an algebra-induced decomposition of $\mathcal{H}$ into $d_{\mathcal{Z}}=\dim\mathcal{Z}(\mathcal{A})$ blocks of the form \cite{davidson_c-algebras_1996}
\begin{equation} \label{eq3}
\begin{split}
&\mathcal{H} \cong \oplus_{J=1}^{d_{\mathcal{Z}}} \,  \mathbb{C}^{n_J} \otimes \mathbb{C}^{d_J}, \\
&\mathcal{A} \cong \oplus_{J=1}^{d_{\mathcal{Z}}} \,  \mathds{1}_{n_J} \otimes \mathcal{L}(\mathbb{C}^{d_J}), \\
&\mathcal{A}^\prime \cong \oplus_{J=1}^{d_Z} \, \mathcal{L}(\mathbb{C}^{n_J}) \otimes \mathds{1}_{d_J}.
\end{split}
\end{equation}
On account of the above decomposition,
\begin{equation*}
\begin{split}
&\dim\mathcal{H}\equiv d = \sum_{J=1}^{d_{\mathcal{Z}}} n_J d_J, \\
&\dim\mathcal{A} = \sum_{J=1}^{d_{\mathcal{Z}}} d_J^2 =: d(\mathcal{A}), \\
&\dim\mathcal{A}^\prime = \sum_{J=1}^{d_{\mathcal{Z}}} n_J^2 =: d(\mathcal{A}^\prime).
\end{split}
\end{equation*}
\par For any algebra $\mathcal{A}$ there exists a projection CP map $\mathbb{P}_{\mathcal{A}}$, such that $\mathbb{P}_{\mathcal{A}}^\dagger = \mathbb{P}_{\mathcal{A}}$, $\mathbb{P}_{\mathcal{A}}^2=\mathbb{P}_{\mathcal{A}}$, $\Im \mathbb{P}_{\mathcal{A}}= \mathcal{A}$. Such a map can be written in a Kraus operator sum representation (OSR) form as 
\begin{equation*}
\mathbb{P}_\mathcal{A} [\bullet ] = \sum_{\gamma =1}^{d(\mathcal{A}^\prime)} f_\gamma \bullet f_\gamma^\dagger
\end{equation*}where $\{ f_\gamma \}_{\gamma =1}^{d(\mathcal{A}^\prime)}$ is a \emph{suitable orthogonal} basis of $\mathcal{A}^\prime$ \cite{zanardi_quantum_nodate}. Similarly,
\begin{equation*}
\mathbb{P}_{\mathcal{A}^\prime} [\bullet ] = \sum_{\alpha =1}^{d(\mathcal{A})} e_\alpha \bullet e_\alpha^\dagger,
\end{equation*}
where $\{ e_\alpha \}_{\alpha =1}^{d(\mathcal{A})}$ is a \emph{suitable orthogonal} basis of $\mathcal{A}$.

By virtue of the Cauchy-Schwarz inequality $d^2 \leq d(\mathcal{A}) \, d(\mathcal{A}^\prime)$. The equality is satisfied when $d_J = \lambda\, n_J \; \forall J$ (for some $\lambda \in \mathbb{Z}$), in which case we say that the pair  $(\mathcal{A}, \mathcal{A}^\prime)$ is \emph{collinear}. We note that in the above decomposition, the Hilbert space is broken into orthogonal blocks with a virtual (algebra induced) ``local'' structure. These observations are exemplified by two physically relevant choices of collinear algebras \cite{zanardi_quantum_nodate}: (i) For $d_\mathcal{Z}=1$, the algebra $\mathcal{A}$ induces a bipartition into virtual subsystems, $\mathcal{H} \cong \mathbb{C}^{n_1} \otimes \mathbb{C}^{d_1}$ \cite{zanardi_virtual_2001,zanardi_quantum_2004}. \\ (ii) For $n_J=1 \; \forall J$, the algebra $\mathcal{A}$ induces a decomposition in super-selection sectors, $\mathcal{H} \cong \oplus_{J=1}^{d_\mathcal{Z}} \mathbb{C}^{d_J}$ \cite{giulini_decoherence_2000}. The case of a maximal abelian algebra $\mathcal{A}$ ($n_J = d_J =1 \; \forall J$) is, in fact, intimately related to the study of the dynamical generation of quantum coherence \cite{zanardi_coherence-generating_2017,zanardi_quantum_2018}.

\subsection{$\mathcal{A}$-OTOC} \label{subsec2.2}
We are now ready to define the main object of this study, which we refer to as the $\mathcal{A}$-OTOC.

\begin{definition} \label{def1}
Let $\mathcal{E}:\mathcal{L}(\mathcal{H}) \rightarrow \mathcal{L}(\mathcal{H})$ be a unital CPTP map. We define the open (averaged) $\mathcal{A}$-OTOC as:
\begin{equation} \label{eq4}
{G}_{\mathcal{A}}(\mathcal{E}) := \frac{1}{2d} \; {\mathlarger{\mathbb{E}}}_{X_\mathcal{A},Y_{\mathcal{A}^\prime}} \left[ \left\Vert \left[ X_\mathcal{A}, \mathcal{E}(Y_{\mathcal{A}^\prime}) \right] \right\Vert_2^2 \right].
\end{equation}
\end{definition}
where ${\mathlarger{\mathbb{E}}}_{X_\mathcal{A},Y_{\mathcal{A}^\prime}}[\bullet ]:= \int_{\mathrm{Haar}} [\bullet ] \; dX_{\mathcal{A}} dY_{\mathcal{A}^\prime}$ denotes averaging over the Haar measures on the unitary subgroups of operators in $\mathcal{A}$ and $\mathcal{A}^\prime$.
Note that the above definition is closely related to but distinct from the geometric algebra anti-correlator (GAAC) introduced in Ref. \cite{zanardi_quantum_nodate} for the case of unitary dynamics. The key difficulty in generalizing the GAAC to open systems is that the algebra structure is, in general, \textit{not} preserved under the mapping $\mathcal{E}(\cdot)$. Hence, the geometric interpretation of scrambling as a distance between algebras ceases to be straightforward. However, as we will see throughout this paper, the $\mathcal{A}$-OTOC can help mitigate this issue. First, it provides a natural generalization to open quantum systems, capturing both scrambling and decoherence and hence generalizing the results of Ref. \cite{zanardi_information_2021}, which was focused on the bipartite algebra case. Second, when restricted to unitary dynamics and collinear algebras, it turns out to be exactly equal to the GAAC, thereby retaining the intuitive geometric notion of distance between algebras.

\par In order to perform the averaging in \cref{eq4}, we consider the replica space, $\mathcal{H}^{\otimes 2}=\mathcal{H}\otimes \mathcal{H}$ and let $S$ denote the swap operator between the two copies.

\begin{proposition} \label{prop1}
\begin{align}
{G}_{\mathcal{A}}(\mathcal{E})=\frac{1}{d} \Tr\left[ S(\mathds{1}_{d^2}-\Omega_{\mathcal{A}})\; \mathcal{E}^{\otimes 2}(\Omega_{\mathcal{A}^\prime})\right], \label{eq5}
\end{align}
where 
\begin{align*}
&\Omega_{\mathcal{A}}:= \sum_{\alpha =1}^{d(\mathcal{A})} e_\alpha \otimes e_\alpha^\dagger, \\
&\Omega_{\mathcal{A}^\prime} := \sum_{\gamma =1}^{d(\mathcal{A}^\prime)} f_\gamma \otimes f_\gamma^\dagger,
\end{align*} and $\{ e_\alpha \}_{\alpha =1}^{d(\mathcal{A})}$, $\{ f_\gamma \}_{\gamma =1}^{d(\mathcal{A}^\prime)}$ are suitable orthogonal bases of $\mathcal{A}$, $\mathcal{A}^\prime$ respectively.
\end{proposition}

Note that the orthogonal bases $\{ e_\alpha \}_{\alpha =1}^{d(\mathcal{A})}$, $\{ f_\gamma \}_{\gamma =1}^{d(\mathcal{A}^\prime)}$ are defined up to unitary transformations. The doubled Hilbert space in \cref{eq5} is the usual cost one has to pay when linearizing \cref{eq4}. In the special case of a bipartite system $\mathcal{H}\cong \mathcal{H}_A \otimes \mathcal{H}_B$ ($\mathcal{H}_A \cong \mathbb{C}^{d_A}$, $\mathcal{H}_B \cong \mathbb{C}^{d_B}$) with $\mathcal{A} \cong \mathds{1}_A \otimes \mathcal{L} (\mathcal{H}_B )$, $\mathcal{A}^\prime \cong \mathcal{L} (\mathcal{H}_A ) \otimes \mathds{1}_{B}$, the $\mathcal{A}$-OTOC reduces to the open (averaged) bipartite OTOC \cite{zanardi_information_2021} and if we further restrict to unitary dynamics generated by $U_t$ one recovers the bipartite OTOC \cite{styliaris_information_2021}, which coincides with the operator entanglement of $U_t$ \cite{zanardi_entanglement_2001,wang_quantum_2002}. The quantities $\Omega_\mathcal{A}$, $\Omega_{\mathcal{A}^\prime}$, while abstract at first sight, provide the input of the algebra. For example, for the bipartite case they reduce just to swaps between the subsystem copies, $\Omega_\mathcal{A}= S_{BB^\prime}/d_B$, $\Omega_{\mathcal{A}^\prime}=S_{AA^\prime}/d_A$. 

\par A corollary of \autoref{prop1} is that the $\mathcal{A}$-OTOC can be expressed in terms of 2-point correlation functions as stated below.
\begin{cor}\label{cor1}
\begin{align}
{G}_{\mathcal{A}} (\mathcal{E} ) =\frac{1}{d}\sum_{\gamma^\prime =1}^{d(\mathcal{A}^\prime )} \lVert \mathcal{E} (f_{\gamma^\prime} ) \rVert_2^2 - \frac{1}{d} \sum_{\gamma ,\gamma^\prime =1}^{d(\mathcal{A}^\prime )}  
\left\lvert \langle \tilde{f}_\gamma^\dagger , \mathcal{E} (f_{\gamma^\prime })\rangle\right\rvert^2 \label{eq6}
\end{align}
where $\tilde{f}_\gamma := \frac{f_\gamma}{\lVert f_\gamma \rVert_2}$ is the normalized basis of $\mathcal{A}^\prime$.
\end{cor}
This formula is practically useful as it allows the direct computation of the $\mathcal{A}$-OTOC for specific examples of algebras and channels (\cref{sec3}) and also suggests how one could potentially measure the $\mathcal{A}$-OTOC by a ``process tomography'' of the channel $\mathcal{E}$.

\prlsection{Decoherence and scrambling} From Definition \ref{def1} we can deduce that a sufficient condition for the $\mathcal{A}$-OTOC to vanish is that the channel $\mathcal{E}$ does not map elements of $\mathcal{A}^\prime$ outside of $\mathcal{A}^\prime$, i.e., commutativity with $\mathcal{A}$ is preserved under time evolution. The following result shows that the $\mathcal{A}^\prime$ invariance is also a necessary condition for the vanishing of the $\mathcal{A}$-OTOC.

\begin{proposition} \label{prop2}
\begin{equation} \label{eq7}
{G}_{\mathcal{A}}(\mathcal{E}) = 0 \Leftrightarrow \mathcal{E}(\mathcal{A}^\prime ) \subseteq \mathcal{A}^\prime
\end{equation}
\end{proposition}

This suggests that the $\mathcal{A}$-OTOC quantifies the deviation of $\mathcal{A}^\prime$ from itself under the map $\mathcal{E}$. Intuitively, if we denote by $\mathcal{A}^\prime$ the relevant degrees of freedom of our quantum system, then, as long as they are mapped within the set, there is no scrambling of information within the system. Indeed, this becomes apparent by manipulating \cref{eq5} into the following form.

\begin{proposition} \label{prop3}
\begin{align}
\begin{split}
{G}_{\mathcal{A}}(\mathcal{E}) &= \frac{1}{d} \sum_{\gamma =1}^{d(\mathcal{A}^\prime )} \left( \lVert \mathcal{E}(f_\gamma ) \rVert_2^2 -\lVert \mathbb{P}_{\mathcal{A}^\prime}\, \mathcal{E} (f_\gamma )\rVert_2^2 \right) \label{eq8}\\
&= \frac{1}{d} \sum_{\gamma =1}^{d(\mathcal{A}^\prime )} \lVert (\mathcal{I}-\mathbb{P}_{\mathcal{A}^\prime})\,\mathcal{E} (f_\gamma ) \rVert_2^2 
\end{split}
\end{align}
\end{proposition}
The second formula in \cref{eq8} shows that the quantification we alluded to is obtained exactly by the norm of the components of $\mathcal{E}(\mathcal{A}^\prime)$ that are in the orthogonal complement of $\mathcal{A}^\prime$. In addition, the first formula in \cref{eq8} breaks the $\mathcal{A}$-OTOC into two terms, ${G}_{\mathcal{A}}^{(1)}(\mathcal{E}) := \frac{1}{d} \sum_{\gamma =1}^{d(\mathcal{A}^\prime )} \lVert \mathcal{E}(f_\gamma ) \rVert_2^2$ and ${G}_{\mathcal{A}}^{(2)}(\mathcal{E}) := \frac{1}{d} \sum_{\gamma =1}^{d(\mathcal{A}^\prime )} \lVert \mathbb{P}_{\mathcal{A}^\prime}\, \mathcal{E} (f_\gamma )\rVert_2^2$. Similar to Ref. \cite{zanardi_information_2021}, we associate the first term with decoherence effects and the second term with information scrambling inside the system. Notice that indeed the decoherence term is in general upper bounded by one and is exactly equal to one if we restrict to unitary dynamics, where there is no decoherence. Thus, decoherence and scrambling have competing roles in the $\mathcal{A}$-OTOC, which roughly corresponds to information (as seen by $\mathcal{A}^\prime$) becoming inaccessible, even if the system is otherwise maximally scrambling. This observation will become explicit in the physical example of stabilizer algebras and dephasing channels in Sec. \ref{sec3}. This type of competition between information scrambling and decoherence in OTOCs has been previously studied in terms of the mutual information for the case of a Hilbert space that is bipartitioned in subsystems \cite{yoshida_disentangling_2019,touil_information_2021}. Other approaches for isolating only the information scrambling in the presence of environmental noise involve attempts to "normalize" the information scrambling term \cite{zhang_information_2019} and time-reversal experimental protocols that account for imperfections \cite{swingle_resilience_2018,dominguez_dynamics_2021}.

\par As a corollary of \autoref{prop3}, we can express the scrambling term of the $\mathcal{A}$-OTOC in terms of out-of-time-ordered four point correlation functions involving structural elements of the pair $(\mathcal{A}, \mathcal{A}^\prime )$.

\begin{cor} \label{cor2}
\begin{align}
{G}_{\mathcal{A}}^{(2)}(\mathcal{E})=\frac{1}{d} \sum_{\gamma =1}^{d(\mathcal{A}^\prime )}  \sum_{\alpha =1}^{d(\mathcal{A} )} \Tr \left[ \mathcal{E} (f_\gamma )^\dagger \, e_\alpha^\dagger \, \mathcal{E} (f_\gamma ) \, e_\alpha \right] \label{eq9}
\end{align}
\end{cor}
This recovers an analog expression of \cref{eq2} in the framework of the $\mathcal{A}$-OTOC, where the out-of-time-order correlation function now contains uniform contributions from all choices of operators in some bases of $\mathcal{A}$ and $\mathcal{A}^\prime$. Since the OTOC $F_{V,W}$  is related to information scrambling in terms of the local structure of $V$ and $W$, the term $G_\mathcal{A}^{(2)}$ is understood to relate to the average information scrambling in terms of the structure induced by the pair $(\mathcal{A},\mathcal{A}^\prime)$.

\prlsection{Upper bound, GAAC, \& typical value} Given an algebra $\mathcal{A}$, a natural question concerns the upper bound of scrambling as quantified by the $\mathcal{A}$-OTOC, which we address in the following proposition.

\begin{proposition} \label{prop4}
\begin{align}
{G}_{\mathcal{A}}(\mathcal{E}) \leq \min \left\{ 1-\frac{1}{d(\mathcal{A})}, \; 1- \frac{1}{d(\mathcal{A}^\prime)} \right\} \label{eq10}
\end{align}
\end{proposition}

Situations where the bound \cref{eq10} is saturated, e.g. the ones described in the collinear case below and the examples in Section \ref{sec3}, are identified with maximal scrambling of the algebra degrees of freedom.

In order to understand the scrambling ability of open quantum systems as characterized by the $\mathcal{A}$-OTOC, it is useful to first focus on the case of unitary channels. In this case, we find that the double commutant theorem and the unitary invariance of the 2-norm imply that there is a simple relation when exchanging the role of $\mathcal{A}$ and $\mathcal{A}^\prime$ as emphasized by the following result.

\begin{proposition} \label{prop5}
For a unitary channel $\; \mathcal{U}[\bullet ] = U \bullet  U^\dagger$ ($U$ is a unitary operator on $\mathcal{L}(\mathcal{H})$) it follows that
\begin{align} \label{eq11}
{G}_{\mathcal{A}} (\mathcal{U})={G}_{\mathcal{A}^\prime} (\mathcal{U^\dagger})
\end{align}
\end{proposition}
Namely, for unitary dynamics, exchanging the role of \(\mathcal{A} \leftrightarrow \mathcal{A}^\prime\) is akin to \(U \leftrightarrow U^{\dagger}\). Furthermore, we find that if $\mathcal{A}$ is collinear, then the $\mathcal{A}$-OTOC coincides with the GAAC.

\begin{proposition} \label{prop6}
For the collinear case ($d_J = \lambda \; n_J \; \forall J$) and a unitary channel $\mathcal{U}$ it follows that
\begin{align}
{G}_{\mathcal{A}}(\mathcal{U} ) = \widetilde{G}_{\mathcal{A}}(\mathcal{U}) \label{eq12}
\end{align}
where \footnote{The superoperator space $\mathcal{L}(\mathcal{L}(\mathcal{H}))$ is endowed with the inner product $\langle \mathcal{X} , \mathcal{Y} \rangle_{HS} := \Tr_{HS} \left[ \mathcal{X}^\dagger \mathcal{Y} \right] := \sum_{k=1}^{d^2} \langle \mathcal{X} (B_k) , \mathcal{Y}(B_k) \rangle$, where $\{ B_k \}_{k=1}^{d^2}$ is an orthonormal basis of $\mathcal{L}(\mathcal{H})$. The corresponding inner product norm is $\lVert \mathcal{X} \rVert_{HS} := \sqrt{\langle \mathcal{X} , \mathcal{X} \rangle_{HS}}$. }
\begin{align*}
\widetilde{G}_{\mathcal{A}}(\mathcal{U}) = 1-\frac{\langle \mathbb{P}_{\mathcal{A}^\prime} , \mathbb{P}_{\mathcal{U}(\mathcal{A}^\prime)}\rangle_{HS}}{\lVert \mathbb{P}_{\mathcal{A}^\prime}\rVert_{HS}^2}
\end{align*}
is the GAAC \cite{zanardi_quantum_nodate}.
\end{proposition}

It follows that even though the definition of the $\mathcal{A}$-OTOC in \cref{eq4} was an algebraic construction, by restricting to unitary channels (and collinear algebras) one recovers the geometrical intuition associated with the GAAC, i.e., the distance between the commutant and its dynamically evolved image. Note that the GAAC was found to be upper-bounded by exactly the same quantity as in \cref{eq10} \cite{zanardi_quantum_nodate}. Moreover, for the collinear case, the bound is achievable if and only if $\mathbb{P}_{\mathcal{A}^\prime} \, \mathcal{U} \, \mathbb{P}_{\mathcal{A}^\prime} = \mathcal{T}$ or $\mathbb{P}_{\mathcal{A}} \, \mathcal{U} \, \mathbb{P}_{\mathcal{A}} = \mathcal{T}$, where $\mathcal{T} [\bullet ] = \Tr [ \bullet ] \, \frac{\mathds{1}}{d}$ is the completely depolarizing channel, whence from the point of view of $\mathcal{A}$ (or $\mathcal{A}^\prime$) the degrees of freedom are maximally scrambled.\\

To gain insights into the scrambling ability of unitary dynamics, we consider the dynamics generated by the ensemble of Haar random unitaries, also known as the circular unitary ensemble (CUE) in the theory of random matrices \cite{mehta_random_2004,guhr_random-matrix_1998}. For finite-dimensional systems, they provide a natural proxy for ``maximally scrambling'' evolutions \cite{xu_swingle_tutorial_2022}. In particular, while local quantum many-body systems cannot scramble information as quickly, random unitaries provide an analytically tractable case to quantitatively estimate how close a quantum system is to maximally scrambling information. It is important to note, however, that for the bipartite case, namely, when $\mathcal{A} \cong \mathds{1}_A \otimes \mathcal{L} (\mathcal{H}_B )$, locally interacting, chaotic many-body systems can quickly equilibrate near to the random matrix theory predicted value, see, e.g., the numerical results in Refs. \cite{styliaris_information_2021,anand_brotocs_2021}. To this end, we compute the typical value of the $\mathcal{A}$-OTOC for Haar unitary channels.

\begin{proposition} \label{prop7}
The average of the $\mathcal{A}$-OTOC over Haar distributed unitary channels $\mathcal{U}$ is:
\begin{align}
\overline{{G}_{\mathcal{A}} (\mathcal{U} )}^{\, \mathcal{U}}=\frac{\left( d^2 - d( \mathcal{A} ) \right) \left( d^2 - d( \mathcal{A}^\prime ) \right)}{d^2 (d^2 -1)} \label{eq13}
\end{align}
\end{proposition}

The symmetry of this average value in $\mathcal{A}$, $\mathcal{A}^\prime$ is a direct consequence of Proposition \ref{prop5}. As anticipated $\overline{{G}_{\mathcal{A}} (\mathcal{U} )}^{\, \mathcal{U}}=0$ if and only if $d(\mathcal{A})=d^2$ or $d(\mathcal{A}^\prime )=d^2$, in which cases $\mathcal{A}^\prime = \mathbf{C}\mathds{1}$ and $\mathcal{A}^\prime=\mathcal{L}(\mathcal{H})$ respectively, implying that $\mathcal{A}^\prime$ is (trivially) unitarily invariant. Notions of $\mathcal{A}$-chaoticity can emerge by comparing typical values as in \cref{eq13} with infinite-time averages \cite{zanardi_quantum_nodate}.

\section{Special algebras and channels} \label{sec3}
In order to concretely illustrate our formalism, we will now apply it to a few analytically tractable physical choices of algebras and channels. To that end, given an algebra $\mathcal{A}$, we will use \cref{eq6} to calculate the $\mathcal{A}$-OTOC for some channel $\mathcal{E}$.
\subsection{Maximal abelian subalgebras \& coherence-generating power}
We consider the algebra $\mathcal{A}_B$ of operators diagonal with respect to an orthonormal basis $B:= \{ \ket{\mu} \}_{\mu=1}^d$, i.e., $\mathcal{A}_B = \{ P_\mu = \ket{\mu}\bra{\mu}\}_{\mu =1}^d$. This is a $d$-dimensional maximal Abelian subalgebra of $\mathcal{L}(\mathcal{H} )$, so as $\mathcal{A}_B =\mathcal{A}_B^\prime$ (which on account of \cref{eq3} corresponds to $d_J=n_J=1 \; \forall J$). Then, one has $\{ f_\gamma \}_{\gamma =1}^{d(\mathcal{A}^\prime)} = \{P_\mu \}_{\mu =1}^{d}$, and thus we find \cite{styliaris_coherence-generating_2018}
\begin{equation}
\begin{split}
{G}_{\mathcal{A}_B} (\mathcal{E} ) &= \frac{1}{d} \left( \sum_{\mu =1}^d \lVert \mathcal{E} (P_\mu ) \rVert_2^2 -  \sum_{\mu , \mu^\prime =1}^d \lvert \Tr \left[ P_{\mu^\prime} \mathcal{E} (P_\mu )\right] \rvert^2\right) \label{eq14}\\
&= \frac{1}{d} \sum_{\mu =1}^d \lVert \mathbb{Q}_B \, \mathcal{E} (P_\mu ) \rVert_2^2
\end{split}
\end{equation}
where $\mathbb{Q}_B := \mathds{1} - \mathbb{P}_{\mathcal{A}_B}$ is the projector on the orthogonal complement of $\mathcal{A}_B$. The quantity in \cref{eq14} is a \emph{coherence generating power} (CGP) \cite{zanardi_coherence-generating_2017,styliaris_quantum_2019,styliaris_coherence-generating_2018} measure for CP unital maps \cite{zanardi_measures_2017}. Operationally, the CGP expresses the average coherence generated by the map $\mathcal{E}$ on initially incoherent states (identified as states that are $B$-diagonal). In \cref{eq14} the averaging is taken over the basis states $\ket{\mu}$, i.e., the extremal points of the simplex $I_B$ formed by the set of $B$-diagonal states \cite{zanardi_measures_2017}. The CGP has been used as a signature of localization transitions in many-body systems \cite{styliaris_quantum_2019} and as a diagnostic tool for quantum chaos \cite{anand_quantum_2021}.
\par Note that the bound $1-d^{-1}$ of \cref{eq10} can be achieved by a unitary channel \cite{zanardi_coherence-generating_2017} $\mathcal{E} [\bullet ] = U \bullet  U^\dagger$  with $\left\lvert\mel{\mu^\prime}{U}{\mu}\right\rvert=d^{\, -1/2}$, e.g., a unitary that is \textit{mutually unbiased} with respect to the basis $B$ \cite{durt2010mutually}.
\par Let us specify $\mathcal{E}$ in two physically relevant examples of Lindbladian dynamics $\mathcal{E}_t = e^{\mathcal{L}t}$ for the case of $n$-qubit systems $\mathcal{H} \cong {\mathbb{C}^2}^{\otimes n}$ ($d=2^n$). These examples illustrate effects of open dynamics, similarly observed in the bipartite case \cite{zanardi_information_2021}.
\begin{exmp} \label{exmp1}
Consider the Lindbladian 
\begin{align*}
\mathcal{L}_1 = \text{Ad}M - \mathcal{I}
\end{align*}
where $\text{Ad}M [\bullet ]:= M \bullet  M^\dagger$, $M\equiv H_0^{\otimes n}$ and $H_0$ is the Hadamard gate. Then, $M=M^\dagger = M^{-1}$ and, letting $B$ be the computational basis, $\left\lvert \mel{\mu^\prime}{M}{\mu} \right\rvert=2^{-n/2}$. By direct exponentiation, one then finds that the evolution is the convex combination 
\begin{align*}
{\mathcal{E}_1}_t = \alpha (t) \, \mathcal{I} + \beta (t) \, \text{Ad}M
\end{align*}
with $\alpha (t) = \left( 1+ e^{-2t} \right)/2$, $\beta (t) = \left( 1-e^{-2t} \right) /2$. Then, the $\mathcal{A}$-OTOC becomes
\begin{equation*}
{G}_{\mathcal{A}_B} ({\mathcal{E}_1}_t ) = \beta^2 (t) \left(1-\frac{1}{2^n} \right)
\end{equation*}
The evolution is, by construction, a convex combination of the identity and a unitary evolution generated by $M$ with time dependent probabilities. The identity evolution is non-scrambling, while the evolution by $\text{Ad}M$ is maximally scrambling. For $t\rightarrow 0$ only the identity evolution is present, whereas for $t\rightarrow \infty$ both evolutions become equiprobable. The resulting $\mathcal{A}$-OTOC depends only on the $\text{Ad}M$ evolution, starting from zero and tending asymptotically to
\begin{align*}
    {G}_{\mathcal{A}_B} (\mathcal{E}_{\infty}) = \frac{1}{4} \left( 1- \frac{1}{2^n} \right).
\end{align*}
\end{exmp}
\begin{exmp} \label{exmp2}
Consider the Lindbladian 
\begin{align*}
\mathcal{L}_2=i\; \text{ad}H + \lambda (\mathcal{D}_H - \mathcal{I})
\end{align*}
where $\text{ad}H [\bullet ] := [H, \bullet  ]$ corresponds to a Hamiltonian evolution and $\mathcal{D}_H [\bullet ]=\sum_{i=1}^{2^n} \Pi_i \bullet  \Pi_i$ is dephasing generated by one-dimensional eigenprojectors $\Pi_i$ of $H$. By exponentiation, one finds that the evolution is a convex combination of a unitary channel and dephasing 
\begin{align*}
{\mathcal{E}_2}_t=a(t)\, e^{it \; \text{ad}H} + (1-a(t)) \, \mathcal{D}_H
\end{align*}
with $a(t) = e^{-\lambda t}$. Letting $B$ be the computational basis, we assume that $H=\sigma_x^{\otimes n}$, where $\sigma_x$ is the x Pauli operator, and thus $\Tr \left[ P_\mu \Pi_i \right] = 2^{-n}$. Then, the $\mathcal{A}$-OTOC becomes
\begin{equation*}
{G}_{\mathcal{A}_B} ({\mathcal{E}_2}_t)= a^2 (t) \, {G}_{\mathcal{A}_B} (e^{it \; \text{ad}H})
\end{equation*}
Furthermore, we have $H^2=\mathds{1}$ and $\mel{\mu^\prime}{H}{\mu}= \delta_{\mu^\prime \bar{\mu}}$, where $\ket{\bar{\mu}} \equiv H \ket{\mu}$. Then,
\begin{equation*}
{G}_{\mathcal{A}_B} ({\mathcal{E}_2}_t)=a^2(t) \, \frac{\sin^2(2t)}{2} 
\end{equation*}which corresponds to \emph{damped} oscillations.
\end{exmp}
\subsection{Projector algebra \& Loschmidt echo}
Let $\ket{\psi}\in \mathcal{H}$ be a quantum state. We consider the algebra $\mathcal{A}_{LE}$ of operators that leave the subspace $\mathbf{C}\ket{\psi}$ and its orthogonal complement invariant. Then, $\mathcal{A}_{LE}^\prime=\mathbf{C}\{\mathds{1},\Pi = \ket{\psi}\bra{\psi}\}$\footnote{$\mathbf{C}G$ denotes the group algebra of $G$} is the unital $*$-closed algebra generated by $\Pi$. One then has $\{f_\gamma \}_{\gamma =1}^{d(\mathcal{A}^\prime)}=\{\Pi , \mathds{1}-\Pi \}$, and thus we find:
\begin{align}
{G}_{\mathcal{A}_{LE}}(\mathcal{E})=\frac{2}{d} \left( \lVert \mathcal{E} (\Pi ) \rVert_2^2 - \frac{ \mathcal{L}_2 (d\mathcal{L}_2 -2 )+1}{d-1} \right) \label{eq15}
\end{align}
where $\mathcal{L}_2 := \Tr\left[ \Pi \, \mathcal{E} (\Pi ) \right]$ and reduces to a Loschmidt echo for unitary dynamics. In \cref{sec4}, we analyze \cref{eq15} in the context of quantum many-body scars \cite{turner_weak_2018,serbyn_quantum_2021}. The maximum value of \cref{eq15} as a function of $\mathcal{L}_2$ is $2/d \, (\lVert \mathcal{E}(\Pi ) \rVert_2^2 - 1/d)$ and is achieved for $\mathcal{L}_2 = 1/d$, which corresponds to the intuition of $\ket{\psi}$ being scrambled into an equal weight superposition in some basis of $\mathcal{L}(\mathcal{H} )$. The bound $1/2$ of \cref{eq10} is achievable only for $d=2$ and is realized by $\mathcal{E} (\Pi ) = 1/d \, \sum_{i,j=1}^d \ket{i}\bra{j}$, where $\{ \ket{i} \}_{i=1}^d$ is an orthonormal basis of $\mathcal{L}(\mathcal{H})$ containing $\ket{\psi}$.
\subsection{Stabilizer algebra \& Dephasing}
\begin{table}
\centering
\begin{tabular}{c|ccccc}
\hline
$J$                & $S^g_1$  &   $S^g_2$  &    \dots    &  $S^g_{n-k-1}$ & $S^g_{n-k}$   \\ \hline
$1$                &  $+$     &    $+$     &    \dots    &      $+$       &   $+$       \\
$2$                &  $+$     &    $+$     &    \dots    &      $+$       &   $-$       \\
\hspace{2pt}\vdots & \vdots   &  \vdots    &    \vdots   &     \vdots     &  \vdots     \\
$2^{n-k}$          &  $-$     &    $-$     &    \dots    &      $-$       &   $-$       \\ \hline
\end{tabular}
\caption{Irreducible representations of stabilizer group. Each irrep $J$ is characterized by the 1-dimensional representation of the $n-k$ stabilizers, that generate the stabilizer group ($ \{ S_\mu \}_{\mu=1}^{2^{n-k}}=\langle S^g_\alpha\rangle_{\alpha=1}^{n-k}$).}
\label{T1}
\end{table}
Let $\{ S_\mu \}_{\mu=1}^{2^{n-k}}$ be a stabilizer group identified as an Abelian subgroup of the n-qubit $(d=2^n)$ Pauli group such that $S_\mu^2=\mathds{1} \; \forall \mu$ \cite{lidar_quantum_2013}. We consider the algebra $\mathcal{A}_{st}$, such that $\mathcal{A}^\prime_{st}= \mathbf{C} \{ S_\mu \}_{\mu=1}^{2^{n-k}}$. Then, the Hilbert space decomposes into $2^k$-dimensional sectors $\mathcal{H} \cong \oplus_{J=1}^{2^{n-k}}\mathbb{C}^{2^k}$ and $\mathcal{A}_{st}$ contains all the stabilizers and additionally all logical-error operators of the corresponding stabilizer code. The stabilizers act as scalars on each sector (see \cref{T1}). We consider a dephasing channel $\mathcal{D} [\bullet ]= \sum_{i=1}^d \Pi_i \bullet \Pi_i$ generated by rank-1 orthogonal projectors . Denoting as $\mathds{1}_J$ the restriction of the identity onto the irrep $J$, we have $\{f_\gamma \}_{\gamma =1}^{d(\mathcal{A}^\prime)}= \{\mathds{1}_J \}_{J=1}^{2^{n-k}}$. This is unitarily equivalent to $\{\hat{f}_\delta \}_{\delta =1}^{d(\mathcal{A}^\prime)}=\{S_\mu/2^{(n-k)/2} \}_{\mu=1}^{2^{n-k}}$, which contains an element proportional to the identity (see \cref{eqa26}). Using the latter orthogonal basis of $\mathcal{A}^\prime_{st}$, the expression for the $\mathcal{A}$-OTOC is
\begin{align}
\begin{split}
{G}_{\mathcal{A}_{st}}(\mathcal{D})= &\frac{2^k}{2^{2n}} \sum_{\mu=1}^{2^{n-k}} \langle S_\mu , \mathcal{D}(S_\mu ) \rangle \label{eq16}\\
&- \frac{2^{k}}{2^{3n}} \sum_{\mu ,\mu^\prime =1}^{2^{n-k}} \lvert\langle S_{\mu^\prime} , \mathcal{D}(S_\mu) \rangle\rvert^2 
\end{split}
\end{align}
For simplicity, let us make a specific choice for the dephasing operators. For each irrep $J$ of the stabilizer group we choose $\chi$ of the $\Pi_i$'s to project onto a state in $J$, while the rest $2^n-2^{n-k} \, \chi$ of the $\Pi_i$'s to project onto a uniform superposition of states from each irrep. Then,
\begin{align}
{G}_{\mathcal{A}_{st}}(\mathcal{D_\chi})= \left(1-\frac{2^k}{2^n}\right)\frac{\chi}{2^k}\left(1-\frac{\chi}{2^k}\right) \label{eq17}
\end{align}
The $\mathcal{A}$-OTOC depends on the ratio $\frac{\chi}{2^k}$ and is directly related to the average information obtained for $\mathcal{A}_{st}$ by measuring the stabilizers $\mathcal{A}_{st}^\prime$ and knowing the form of the channel $D_{\chi}$. Notice that the result does not depend on the choice of states on each irrep $J$, as $\mathcal{A}_{st}^\prime$ is insensitive to transformations inside an irrep (logical errors). Moreover, the $\mathcal{A}$-OTOC is zero if $\chi=2^k$ or $\chi=0$. The former case is a model that contains only logical errors and from the perspective of $\mathcal{A}_{st}^\prime$ this is equivalent to no scrambling; in this case both terms in \cref{eq16} are equal to one. The latter case is a model of ``white noise'', where all states are projected to equivalent (from the perspective of $\mathcal{A}_{st}^\prime$) superpositions; in this case both terms in \cref{eq16} equal to $2^k/2^n$. Despite the scrambling being intuitively maximal, decoherence (from the perspective of $\mathcal{A}_{st}^\prime$) is also maximal. As a result, the $\mathcal{A}$-OTOC vanishes, thereby showing a \emph{competition} between information scrambling and decoherence in accordance with observations in the case of a bipartite algebra \cite{zanardi_information_2021,yoshida_disentangling_2019,touil_information_2021}.
\section{Quantum spin-chain models} \label{sec4}
\begin{figure*}[th]
\centering
\begin{subfigure}{0.4\textwidth}
  \centering
  \includegraphics[width=1\linewidth]{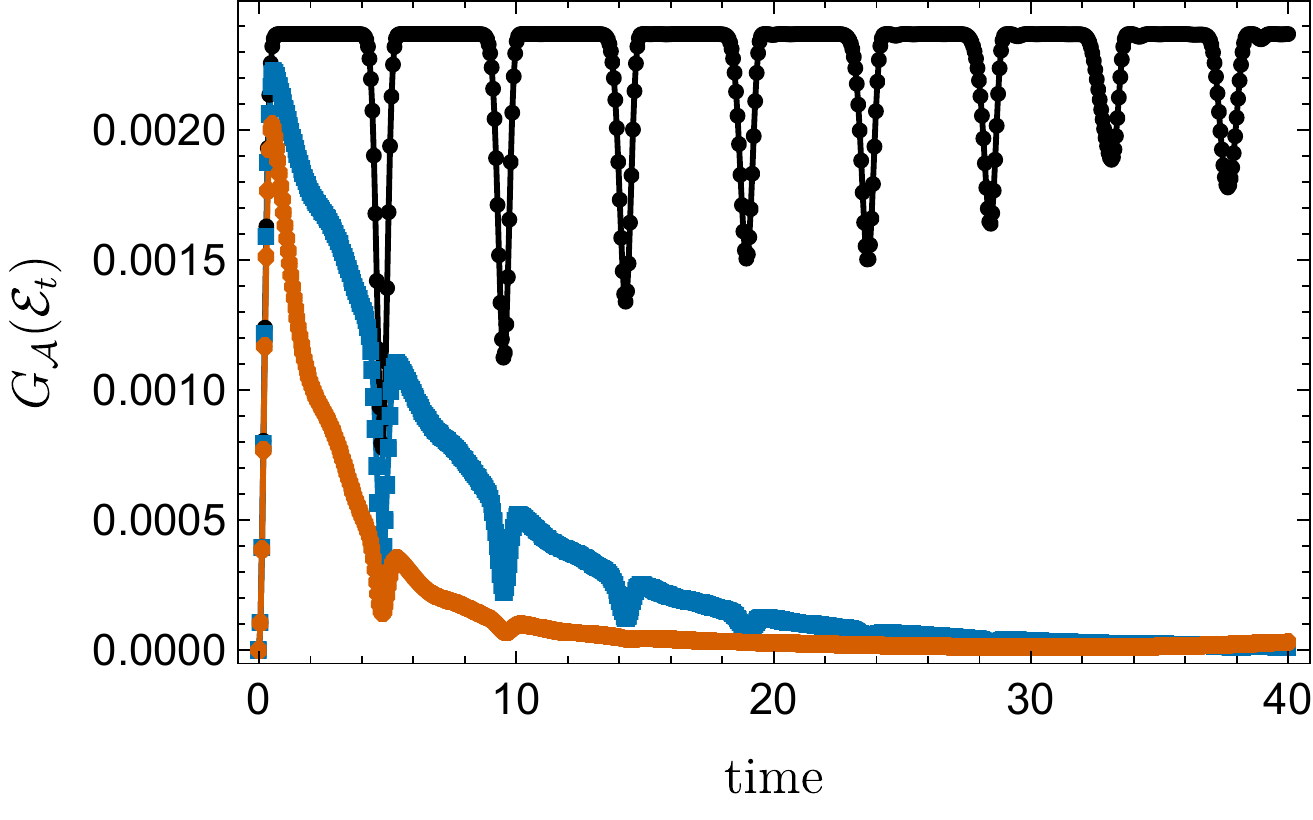}
  \caption{}
  \label{fig1s1}
\end{subfigure}%
\begin{subfigure}{0.18\textwidth}
  \centering
  \includegraphics[width=0.95\linewidth]{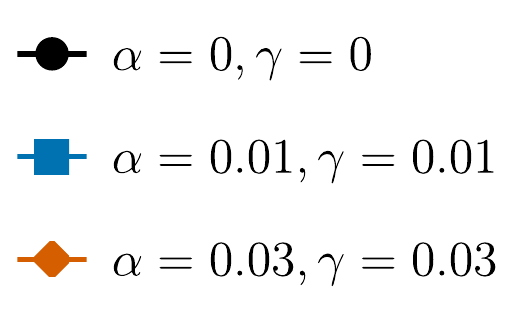}
\end{subfigure}%
\begin{subfigure}{0.4\textwidth}
  \centering
  \includegraphics[width=1\linewidth]{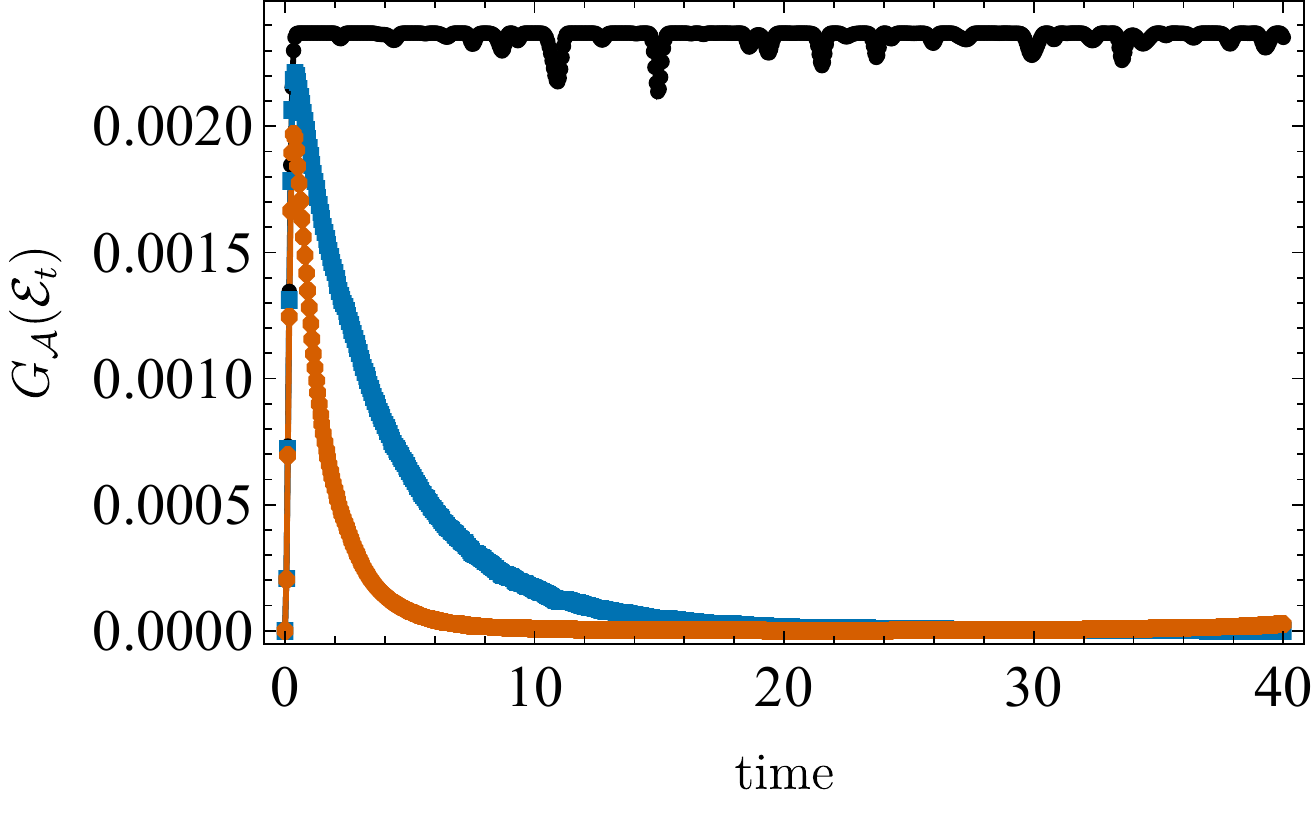}
  \caption{}
  \label{fig1s2}
\end{subfigure}
\caption{Temporal variation of the $\mathcal{A}$-OTOC ${G}_{\mathcal{A}}(\mathcal{E} )$ for the PXP model with $N=14$ and $\mathcal{A}^\prime= \mathbf{C} \{\mathds{1}, \Pi \}$ for (a) $\Pi = \ket{\mathbb{Z}_2}\bra{\mathbb{Z}_2}$, (b) $\Pi = \ket{0}\bra{0}$ and varying dissipation strengths $\alpha$, $\gamma$. The characteristic periodic recurrences of scar dynamics are present in \cref{fig1s1}, but are increasingly suppressed as  we scale up $\alpha$, $\gamma$, obfuscating the distinction with the thermal behavior in \cref{fig1s2}. }
\label{fig1}
\end{figure*}
\begin{figure*}
\centering
\begin{subfigure}{.38\textwidth}
  \centering
  \includegraphics[width=1\linewidth]{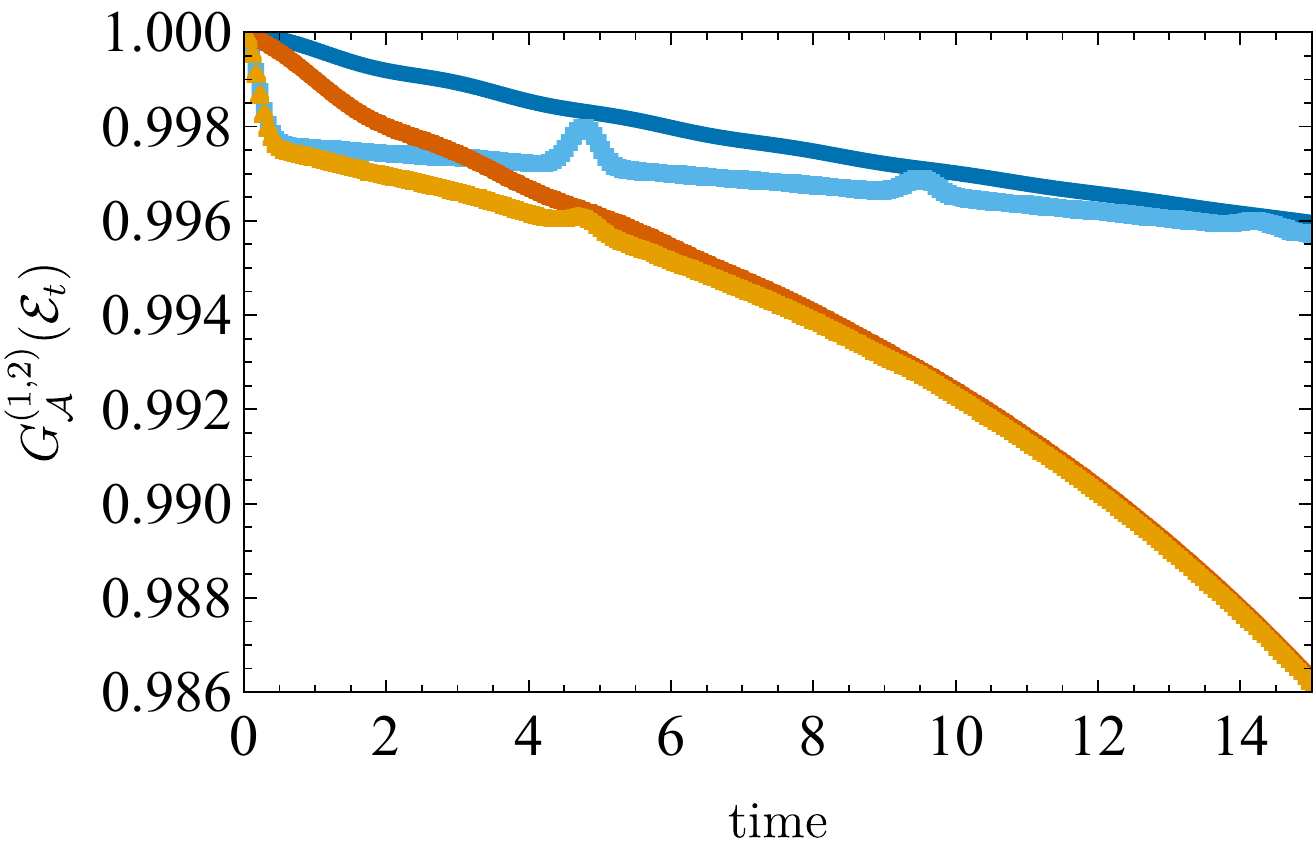}
  \caption{}
  \label{fig2s1}
\end{subfigure}%
\begin{subfigure}{.23\textwidth}
  \centering
  \includegraphics[width=.9\linewidth]{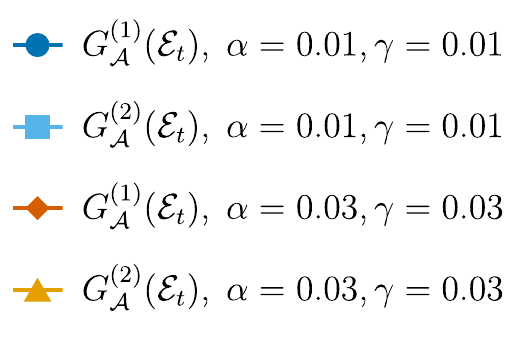}
\end{subfigure}%
\begin{subfigure}{.38\textwidth}
  \centering
  \includegraphics[width=1\linewidth]{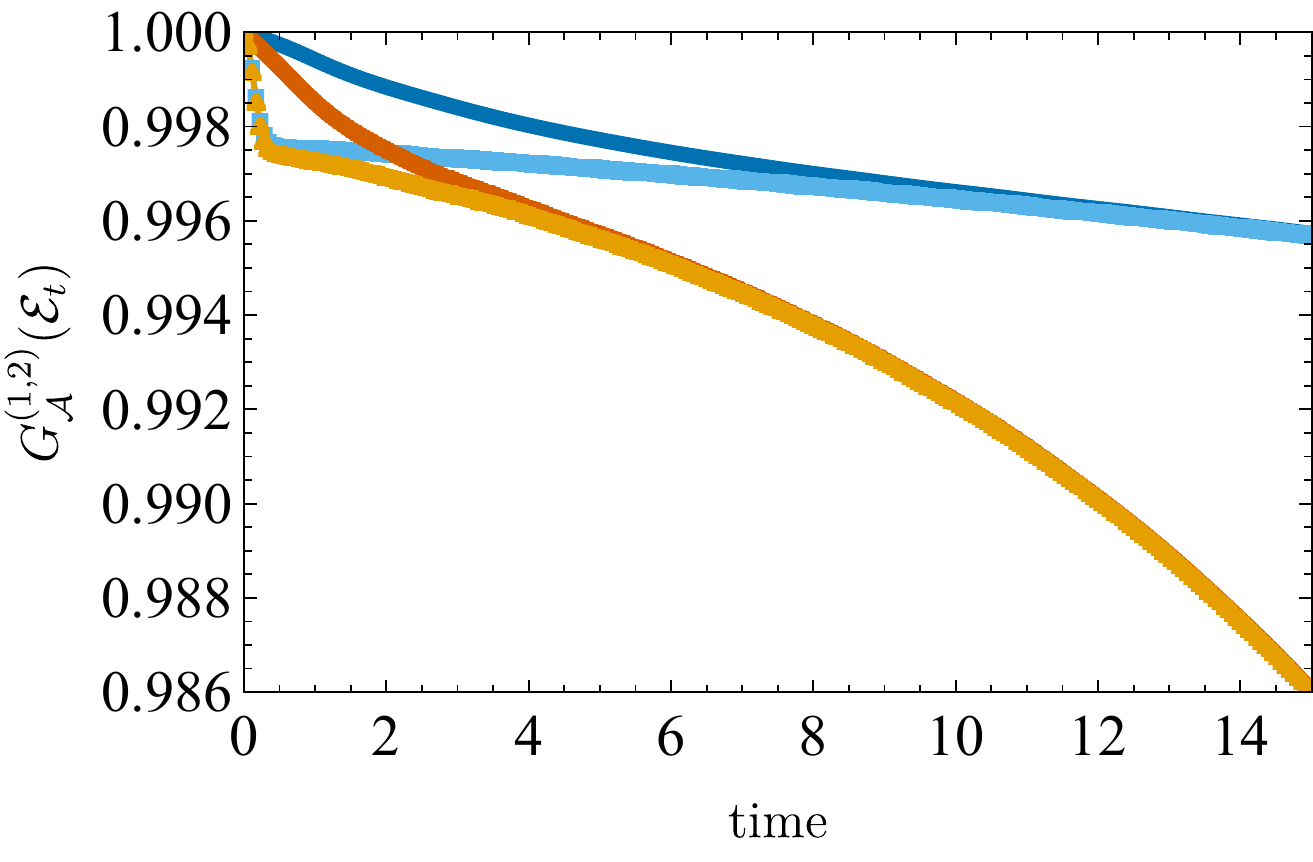}
  \caption{}
  \label{fig2s2}
\end{subfigure}
\caption{Temporal variation of the separated terms ${G}_{\mathcal{A}}^{{(1)}}(\mathcal{E} )=\frac{1}{d} \sum_{\gamma =1}^{d(\mathcal{A}^\prime )} \lVert \mathcal{E}(f_\gamma ) \rVert_2^2$, ${G}_\mathcal{A}^{(2)}(\mathcal{E} )=\frac{1}{d} \sum_{\gamma =1}^{d(\mathcal{A}^\prime )} \lVert \mathbb{P}_{\mathcal{A}^\prime}\, \mathcal{E} (f_\gamma )\rVert_2^2$ for the PXP model with $N=14$ and $\mathcal{A}^\prime= \mathbf{C} \{\mathds{1}, \Pi \}$ for (a) $\Pi = \ket{\mathbb{Z}_2}\bra{\mathbb{Z}_2}$, (b) $\Pi = \ket{0}\bra{0}$ and varying dissipation strengths $\alpha$, $\gamma$. After a time-scale that depends on the system-environment coupling the open system effects dominate and the scrambling is ``saturated''.}
\label{fig2}
\end{figure*}
As a physical application of the $\mathcal{A}$-OTOC formalism, we consider representative spin-chain models with open system dynamics, which are a result of system-bath interactions. For systems where the bath is Markovian, the system evolution is described by a continuous, one-parameter family of dynamical maps \footnote{Note that this is the Heisenberg picture evolution map.} $\mathcal{E}_t=e^{t\mathcal{L}}, \, t\geq 0$ generated by the Lindbladian \cite{breuer_theory_2007}
\begin{align}
\mathcal{L} [\bullet ]= i [H^\dagger, \bullet ] + \sum_j \left( L_j^\dagger \bullet  L_j - \frac{1}{2} \{ L_j^\dagger L_j, \bullet  \} \right) \label{eq18}
\end{align}
where $H$ is the Hamiltonian and $\{ L_j \}_j$ are the Lindblad operators that describe the system-bath interactions.
\par To numerically simulate the evolution, we vectorize the Hilbert-Schmidt space \cite{am-shallem_three_2015} and the Linbladian $\mathcal{L}$  is represented in matrix form as
\begin{align}
\mathcal{L} \naturalto \, &i \left( \mathds{1} \otimes H^\dagger - H^* \otimes \mathds{1} \right) + \sum_j \left( L_j^\mathrm{T} \otimes L_j^\dagger \right. \nonumber\\
&\left. - \frac{1}{2} \mathds{1} \otimes L_j^\dagger L_j - \frac{1}{2} L_j^\mathrm{T} L_j^* \otimes \mathds{1} \right) \label{eq19}
\end{align}
where $X^\mathrm{T}$ and $X^*$ denote the matrix transpose and matrix conjugate of $X$,  respectively.
\begin{figure*}[t]
\centering
\begin{subfigure}{.5\textwidth}
  \centering
  \includegraphics[width=1\linewidth]{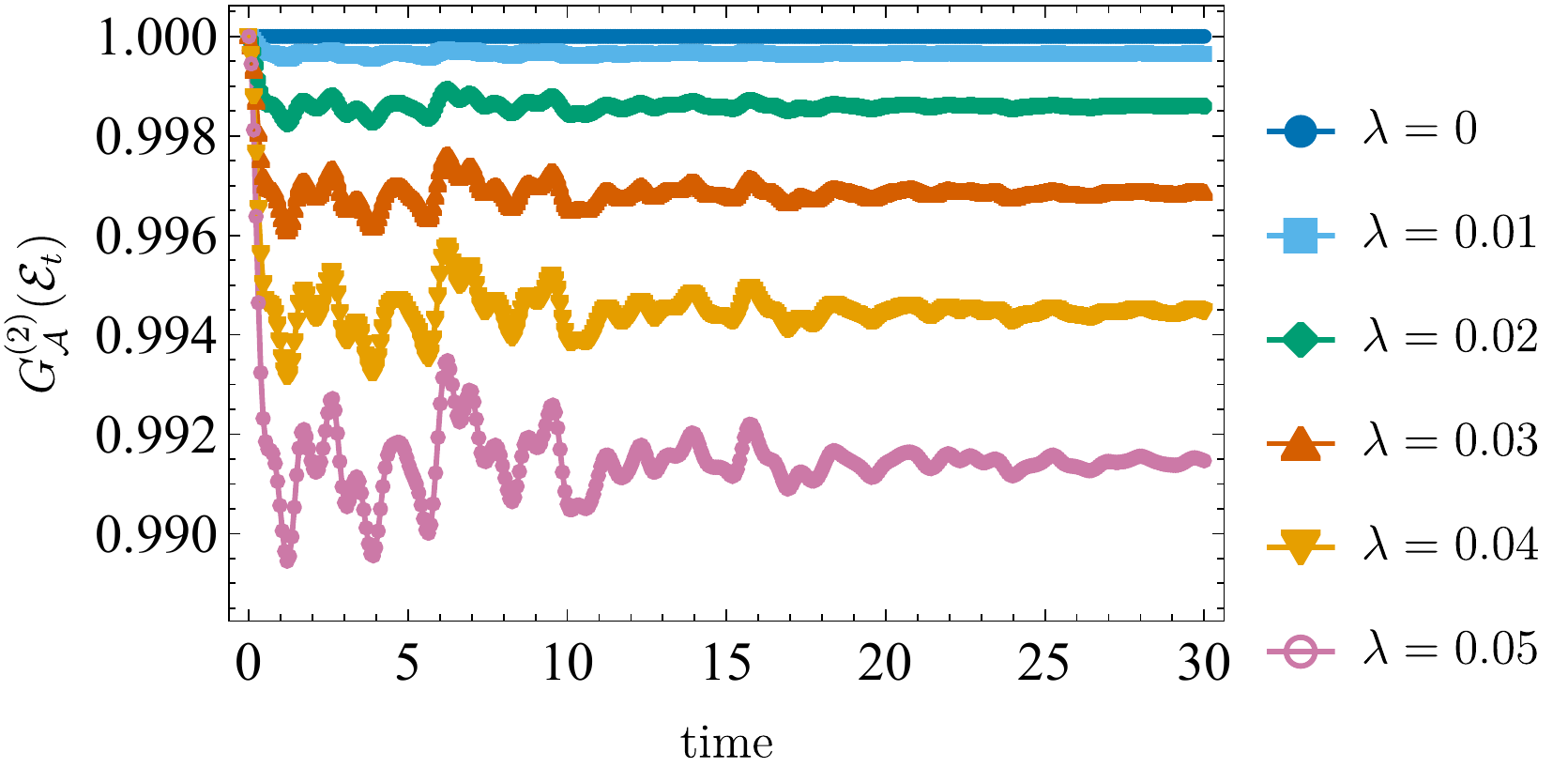}
  \caption{}
  \label{fig3s1}
\end{subfigure}%
\begin{subfigure}{.5\textwidth}
  \centering
  \includegraphics[width=.85\linewidth]{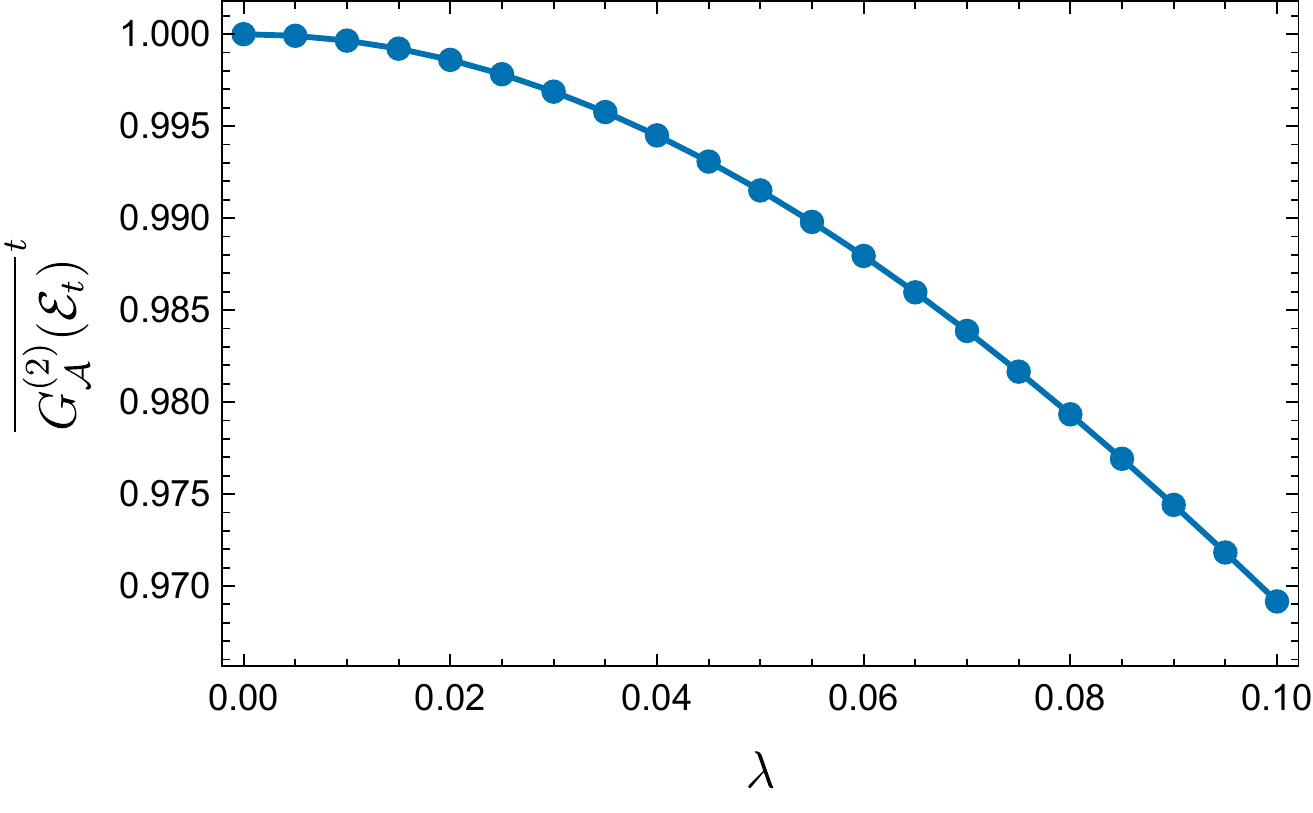}
  \caption{}
  \label{fig3s2}
\end{subfigure}
\caption{(a) Temporal variation of the scrambling term $G_{\mathcal{A}}^{(2)}(\mathcal{E}_t)=\frac{1}{d} \sum_{\gamma =1}^{d(\mathcal{A}^\prime )} \lVert \mathbb{P}_{\mathcal{A}^\prime}\, \mathcal{E} (f_\gamma )\rVert_2^2$ of $\lambda$-perturbed DFS subspace algebras for the Heisenberg XXX model with collective decoherence and $N=6$, $\alpha = \gamma = 0.05$. $G_{\mathcal{A}}^{(2)}(\mathcal{E}_t)$ exhibits a decaying oscillatory behavior and as we scale up the rotation strength $\lambda$, the long-time limit decreases (which corresponds to increased scrambling). (b) The long-time average of the $G_{\mathcal{A}}^{(2)}(\mathcal{E}_t)$ depends quadratically on the rotation strength $\lambda$, showing a first-order stability of the DFS in terms of information scrambling.}
\label{fig3}
\end{figure*}
\subsection{PXP model}
We consider a 1D spin-1/2 chain model with $N$ sites, periodic boundary conditions and Hamiltonian dynamics given as:
\begin{align}
H_{PXP} = \mathcal{J} \sum_{j=1}^{N} P_{j-1} \sigma_{j}^x P_{j+1} \label{eq20}
\end{align}
where $P_j := \left( 1 - \sigma_{j}^z \right)/2$ and $\{ \sigma_j^\alpha \}_{\alpha= x,y,z}$ are the Pauli operators. In our numerical simulations, we set $\mathcal{J}=1$, which sets the energy scale of the Hamiltonian, and in turn the timescale of the dynamics. This model is relevant in Rydberg atom experiments \cite{bernien_probing_2017} in the limit of Rydberg blockade \cite{lesanovsky_interacting_2012,saffman_quantum_2010}. The projectors $P_j$ effectively truncate the Hilbert space so as to exclude states with neighboring excitations (here corresponding to $\ket{\uparrow}$). Scrambling properties of the PXP model were recently studied using OTOCs for specific choices of local observables \cite{yuan_quantum_2022}.
\par  In terms of level-statistics the PXP Hamiltonian was shown to exhibit level repulsion \cite{turner_weak_2018}, a characteristic of non-integrable systems. However, the system exhibits weak ergodicity breaking that has been associated with a small set of special many-body eigenstates (\emph{scars}) \cite{turner_weak_2018,turner_quantum_2018}. Specifically, quenching the system from states inside the scar subspace leads to \emph{revivals} of the wavefunction and local observable correlations. A prototypical example is the Ne{\' e}l state $\ket{\mathbb{Z}_2}:=\ket{\uparrow\downarrow\uparrow\downarrow\cdots}$, which was shown to have an unusually large overlap with the scar eigenstates compared to eigenstates with similar energy $E$, violating the strong eigenstate thermalization hypothesis (ETH) \cite{deutsch_quantum_1991,srednicki_chaos_1994} conjectured for quantum ergodic systems. On the contrary, other initial sates, like the ferromagnetic state $\ket{0}:=\ket{\downarrow \downarrow \cdots}$, quickly thermalize without revivals \cite{turner_quantum_2018}. Generally, this scar behavior is sensitive to perturbations, which can make the model integrable \cite{fendley_competing_2004} or thermalizing \cite{turner_quantum_2018}, although some robustness is exhibited with respect to disorder \cite{mondragon-shem_fate_2021}.
\par We consider Lindbladian dynamics of \cref{eq18}, with Lindblad operators corresponding to bulk-dephasing $L^{(z)}_j = \sqrt{\alpha} \, \sigma_j^z$ and bulk-driving $L^{\pm}_j=\sqrt{\gamma} \, \sigma_j^{\pm}$, where $\sigma_j^{\pm} = 1/2 \left( \sigma_j^x \pm i \, \sigma_j^y \right)$. Simulating exact dynamics for $N=14$, we compute the corresponding $\mathcal{A}$-OTOCs as a function of time for the algebra $\mathcal{A}_{LE}$ with $\Pi = \ket{\mathbb{Z}_2}\bra{\mathbb{Z}_2}$ and $\Pi = \ket{0}\bra{0}$. As we increase the system-bath couplings $\mathcal{\alpha}$,$\mathcal{\gamma}$, we observe that the $\mathcal{A}$-OTOC starts decaying from its closed system value ($\alpha = \gamma = 0$) due to open system effects (Fig. \ref{fig1}). At the same time, the scar dynamics (revivals) that clearly distinguish the scar from the thermal dynamics in the closed system case are still present but become less apparent as we scale $\alpha$ and $\gamma$.
\par Given the intuition following \autoref{prop3}, we compute the terms ${G}_\mathcal{A}^{(1)}(\mathcal{E}_t)=\frac{1}{d} \sum_{\gamma =1}^{d(\mathcal{A}^\prime )} \lVert \mathcal{E}(f_\gamma ) \rVert_2^2$ and ${G}_\mathcal{A}^{(2)}(\mathcal{E}_t)=\frac{1}{d} \sum_{\gamma =1}^{d(\mathcal{A}^\prime )} \lVert \mathbb{P}_{\mathcal{A}^\prime}\, \mathcal{E} (f_\gamma )\rVert_2^2$ separately in \cref{fig2}. We observe that the decoherence term ${G}_\mathcal{A}^{(1)}(\mathcal{E}_t)$  provides an enveloping function to the information scrambling term ${G}_\mathcal{A}^{(2)}(\mathcal{E}_t)$. After a certain timescale, the distance between the functions diminishes and the system (in terms of scrambling of the algebra degrees of freedom) becomes ``saturated'', in the sense that open system effects have dominated and the interesting information scrambling behavior is suppressed.
\subsection{Heisenberg model \& DFS}
Consider the Hamiltonian of a 1D spin-1/2 Heisenberg XXX model with $N$ sites and periodic boundary conditions
\begin{align}
H_{XXX}=\mathcal{J} \sum_{j=1}^{N} \left( \sigma_j^x \sigma_{j+1}^x + \sigma_j^y \sigma_{j+1}^y + \sigma_j^z \sigma_{j+1}^z \right) \label{eq21}
\end{align}
In our numerical simulations, we set $\mathcal{J}=1$, which sets the energy scale of the Hamiltonian, and in turn the timescale of the dynamics.
\par Let us assume that the evolution is described by Lindbladian dynamics as in \cref{eq18} with Lindblad operators corresponding to collective decoherence $L^{(z)}=\sqrt{\alpha} \sum_{j=1}^N \sigma_j^z, \, L^{(\pm)}=\sqrt{\gamma} \sum_{j=1}^N \sigma_j^{\pm}$. Then, for even $N$ there exists a decoherence-free subspace (DFS) \cite{zanardi_noiseless_1997,zanardi_error_1997,lidar_decoherence-free_1998} spanned by the spin=0 eigenstates of $S^2:=\sum_{\alpha = x,y,z} \left( \sum_{j=1}^N \sigma_j^\alpha \right)^2$. In fact, the underlying structure is exactly as in \cref{eq3}, where $J$ now labels the irreducible representations of $\mathfrak{sl}(2)$ on $\mathbb{C}^{\otimes N}$ and the DFS simply corresponds to the singlets $J=0$.
\par We consider the unital algebra of observables $\mathcal{A}_{DFS}$ that act non-trivially only on the orthogonal complement of the DFS. Then, for the commutant $\mathcal{A}_{DFS}^\prime$ we have the orthogonal basis
\begin{align*}
\{ f_\gamma \}_{\gamma =1}^{d_D^2+1} = \left\{ \mathds{1}_\perp, \frac{\ket{p }\bra{q}}{\sqrt{d_{D}}} \right\}_{p ,q =1}^{d_D}
\end{align*}
where $\{  \ket{p} \}_{p =1}^{d_D}$ is an orthonormal basis of the DFS and $\mathds{1}_\perp := \mathds{1} - \sum_{p =1}^{d_D} \ket{p} \bra{p}$. We also consider $\lambda$-perturbed algebras defined by unitary rotations $U(\lambda ) := exp \{i \sum_{j=1}^N \lambda \, \vec{\eta}_j \cdot \vec{\sigma_j}\}$ via $\ket{p_\lambda} := U(\lambda ) \ket{p}$, where $\vec{\eta}_j$ is a uniformly distributed vector on the unit sphere in $\mathbb{R}^3$ and $\vec{\sigma_j} \equiv (\sigma_j^x, \sigma_j^y, \sigma_j^z )$. The free parameter $\lambda$ provides a simple representation of departure from exact DFS dynamics due to model inaccuracies. Simulating exact dynamics for $N=6$, we compute the corresponding scrambling terms $G_{\mathcal{A}}^{(2)}(\mathcal{E}_t)$ of the $\mathcal{A}$-OTOCs as functions of time for various values of $\lambda$. Naturally, for $\lambda=0$ there is no scrambling, as the DFS is invariant under both the Heisenberg XXX Hamiltonian and collective decoherence and thus  $G_{\mathcal{A}}^{(2)}(\mathcal{E}_t)$ is constant in time. As $\lambda$ is scaled up, effects of decoherence and information scrambling strengthen and $G_{\mathcal{A}}^{(2)}(\mathcal{E}_t)$ exhibits a decaying oscillatory behavior (\cref{fig3s1}). In the long-time limit the system generally transitions to fixed points that are not entirely in the $\lambda$-perturbed subspace. As an example, for $N=2$, the singlet is invariant, the diagonal elements of the triplet subspace transition to $\mathds{1}_\perp$, while all non-diagonal elements vanish. As we increase $\lambda$, the $\lambda$-perturbed algebra of observables moves further from the fixed points, leading to increased scrambling under evolution (which corresponds to the decreased long-time limit of $G_{\mathcal{A}}^{(2)}(\mathcal{E} )$).
\par In order to gauge the stability of the DFS in the $\lambda$-perturbation, we time average $G_{\mathcal{A}}^{(2)}(\mathcal{E}_t)$ for each $\lambda$ in the time interval $[0,30]$ using a time-step $\Delta t = 0.075$. We find that $\overline{{G}_{\mathcal{A}}^{(2)} (\mathcal{E}_t )}^{\, t}$ depends quadratically on the parameter $\lambda$ (\cref{fig3s2}), which shows a first-order stability of the DFS in terms of scrambling. This result is in accordance with previous stability considerations under addition of symmetry-breaking Hamiltonian terms \cite{lidar_decoherence-free_1998}.
\section{Conclusion} \label{sec5}
In this paper, we have established a formalism for quantitatively describing scrambling at the level of algebras of observables in open quantum systems. In doing so, we have defined an algebraic (averaged) out-of-time-order correlator, termed the $\mathcal{A}$-OTOC, generalizing the open bipartite OTOC to arbitrary algebras of observables that correspond to the relevant physical quantities of interest of the system. Explicit analytic calculations showed that the $\mathcal{A}$-OTOC quantifies the degree of deviation of $\mathcal{A}^\prime$ from its dynamically evolved image $\mathcal{E}(\mathcal{A}^\prime)$ (\autoref{prop3}) and allowed for the identification of a competing role of the effects of decoherence and information scrambling in the $\mathcal{A}$-OTOC. For unitary dynamics and a collinear algebra, we have shown that the $\mathcal{A}$-OTOC is exactly equal to the geometric algebra anti-correlator \cite{zanardi_quantum_nodate}. We also computed its typical value for Haar random unitaries, thereby providing a quantitative estimate for the $\mathcal{A}$-OTOC in chaotic quantum systems, which, after an initial transient, are expected to equilibrate to this random matrix theory value.
\par Additionally, we have studied concrete, physically motivated examples of algebras and channels, showing that the $\mathcal{A}$-OTOC recovers, as special cases, the open-system extensions of information-theoretic notions like the coherence-generating power (CGP) and Loschmidt echo. Analytic calculations for a stabilizer algebra, as well as numerical simulations for the Loschmidt echo algebra, demonstrate how decoherence, after a certain timescale, suppresses the signatures of information scrambling. The competing effects are described by separated terms, with ``saturation'' occurring when open system effects dominate. A concrete manifestation of this phenomenon was observed in the PXP model, where the characteristic revivals related to the quantum scars are suppressed as open dynamics become predominant. In addition, we have analyzed subspace algebras for the Heisenberg XXX model with collective decoherence, where the subspaces are obtained by unitary rotations of the decoherence-free subspace (DFS) and determined that the DFS is stable to first-order in terms of the time averaged scrambling term of the $\mathcal{A}$-OTOC.
\par A worthwhile direction for future investigation is the detailed characterization of the separate contributions of decoherence and information scrambling in the generalized framework introduced in this paper, so that one can \textit{disentangle} their contributions, both in principle and in experimental setups. Additionally, it is of interest to consider the role of the $\mathcal{A}$-OTOC framework in general classifications of ergodicity-breaking in physical models, e.g., with regards to the spectrum-generating algebra of scar systems \cite{serbyn_quantum_2021} or Hilbert space fragmentation \cite{moudgalya_hilbert_2021}.
\section{Acknowledgments}
The authors acknowledge the Center for Advanced Research Computing (CARC) at the University of Southern California for providing computing resources that have contributed to the research results reported within this publication. URL: \url{https://carc.usc.edu}. P.Z. acknowledges partial support from the NSF award PHY-1819189. This research was (partially) sponsored by the Army Research Office and was accomplished under Grant Number W911NF-20-1-0075. The views and conclusions contained in this document are those of the authors and should not be interpreted as representing the official policies, either expressed or implied, of the Army Research Office or the U.S. Government. The U.S. Government is authorized to reproduce and distribute reprints for Government purposes notwithstanding any copyright notation herein.
\bibliographystyle{apsrev4-1}
\bibliography{main}

\begin{thebibliography}{75}%
\makeatletter
\providecommand \@ifxundefined [1]{%
 \@ifx{#1\undefined}
}%
\providecommand \@ifnum [1]{%
 \ifnum #1\expandafter \@firstoftwo
 \else \expandafter \@secondoftwo
 \fi
}%
\providecommand \@ifx [1]{%
 \ifx #1\expandafter \@firstoftwo
 \else \expandafter \@secondoftwo
 \fi
}%
\providecommand \natexlab [1]{#1}%
\providecommand \enquote  [1]{``#1''}%
\providecommand \bibnamefont  [1]{#1}%
\providecommand \bibfnamefont [1]{#1}%
\providecommand \citenamefont [1]{#1}%
\providecommand \href@noop [0]{\@secondoftwo}%
\providecommand \href [0]{\begingroup \@sanitize@url \@href}%
\providecommand \@href[1]{\@@startlink{#1}\@@href}%
\providecommand \@@href[1]{\endgroup#1\@@endlink}%
\providecommand \@sanitize@url [0]{\catcode `\\12\catcode `\$12\catcode
  `\&12\catcode `\#12\catcode `\^12\catcode `\_12\catcode `\%12\relax}%
\providecommand \@@startlink[1]{}%
\providecommand \@@endlink[0]{}%
\providecommand \url  [0]{\begingroup\@sanitize@url \@url }%
\providecommand \@url [1]{\endgroup\@href {#1}{\urlprefix }}%
\providecommand \urlprefix  [0]{URL }%
\providecommand \Eprint [0]{\href }%
\providecommand \doibase [0]{http://dx.doi.org/}%
\providecommand \selectlanguage [0]{\@gobble}%
\providecommand \bibinfo  [0]{\@secondoftwo}%
\providecommand \bibfield  [0]{\@secondoftwo}%
\providecommand \translation [1]{[#1]}%
\providecommand \BibitemOpen [0]{}%
\providecommand \bibitemStop [0]{}%
\providecommand \bibitemNoStop [0]{.\EOS\space}%
\providecommand \EOS [0]{\spacefactor3000\relax}%
\providecommand \BibitemShut  [1]{\csname bibitem#1\endcsname}%
\let\auto@bib@innerbib\@empty
\bibitem [{\citenamefont {Larkin}\ and\ \citenamefont
  {Ovchinnikov}(1969)}]{ovchinnikov_quasiclassical_nodate}%
  \BibitemOpen
  \bibfield  {author} {\bibinfo {author} {\bibfnamefont {I.~A.}\ \bibnamefont
  {Larkin}}\ and\ \bibinfo {author} {\bibfnamefont {Y.~N.}\ \bibnamefont
  {Ovchinnikov}},\ }\href@noop {} {\bibfield  {journal} {\bibinfo  {journal}
  {Journal of Experimental and Theoretical Physics}\ }\textbf {\bibinfo
  {volume} {28}},\ \bibinfo {pages} {2262} (\bibinfo {year}
  {1969})}\BibitemShut {NoStop}%
\bibitem [{\citenamefont {Kitaev}(2015)}]{kitaev_simple_2015}%
  \BibitemOpen
  \bibfield  {author} {\bibinfo {author} {\bibfnamefont {A.}~\bibnamefont
  {Kitaev}},\ }\href {https://online.kitp.ucsb.edu/online/entangled15/kitaev/}
  {\enquote {\bibinfo {title} {A simple model of quantum holography},}\ }
  (\bibinfo {year} {2015})\BibitemShut {NoStop}%
\bibitem [{\citenamefont {Maldacena}\ \emph {et~al.}(2016)\citenamefont
  {Maldacena}, \citenamefont {Shenker},\ and\ \citenamefont
  {Stanford}}]{maldacena_bound_2016}%
  \BibitemOpen
  \bibfield  {author} {\bibinfo {author} {\bibfnamefont {J.}~\bibnamefont
  {Maldacena}}, \bibinfo {author} {\bibfnamefont {S.~H.}\ \bibnamefont
  {Shenker}}, \ and\ \bibinfo {author} {\bibfnamefont {D.}~\bibnamefont
  {Stanford}},\ }\href {\doibase 10.1007/JHEP08(2016)106} {\bibfield  {journal}
  {\bibinfo  {journal} {Journal of High Energy Physics}\ }\textbf {\bibinfo
  {volume} {2016}},\ \bibinfo {pages} {106} (\bibinfo {year}
  {2016})}\BibitemShut {NoStop}%
\bibitem [{\citenamefont {Lashkari}\ \emph {et~al.}(2013)\citenamefont
  {Lashkari}, \citenamefont {Stanford}, \citenamefont {Hastings}, \citenamefont
  {Osborne},\ and\ \citenamefont {Hayden}}]{lashkari_towards_2013}%
  \BibitemOpen
  \bibfield  {author} {\bibinfo {author} {\bibfnamefont {N.}~\bibnamefont
  {Lashkari}}, \bibinfo {author} {\bibfnamefont {D.}~\bibnamefont {Stanford}},
  \bibinfo {author} {\bibfnamefont {M.}~\bibnamefont {Hastings}}, \bibinfo
  {author} {\bibfnamefont {T.}~\bibnamefont {Osborne}}, \ and\ \bibinfo
  {author} {\bibfnamefont {P.}~\bibnamefont {Hayden}},\ }\href {\doibase
  10.1007/JHEP04(2013)022} {\bibfield  {journal} {\bibinfo  {journal} {Journal
  of High Energy Physics}\ }\textbf {\bibinfo {volume} {2013}} (\bibinfo {year}
  {2013}),\ 10.1007/JHEP04(2013)022}\BibitemShut {NoStop}%
\bibitem [{\citenamefont {Roberts}\ and\ \citenamefont
  {Stanford}(2015)}]{roberts_diagnosing_2015}%
  \BibitemOpen
  \bibfield  {author} {\bibinfo {author} {\bibfnamefont {D.~A.}\ \bibnamefont
  {Roberts}}\ and\ \bibinfo {author} {\bibfnamefont {D.}~\bibnamefont
  {Stanford}},\ }\href {\doibase 10.1103/PhysRevLett.115.131603} {\bibfield
  {journal} {\bibinfo  {journal} {Physical Review Letters}\ }\textbf {\bibinfo
  {volume} {115}},\ \bibinfo {pages} {131603} (\bibinfo {year}
  {2015})}\BibitemShut {NoStop}%
\bibitem [{\citenamefont {Polchinski}\ and\ \citenamefont
  {Rosenhaus}(2016)}]{polchinski_spectrum_2016}%
  \BibitemOpen
  \bibfield  {author} {\bibinfo {author} {\bibfnamefont {J.}~\bibnamefont
  {Polchinski}}\ and\ \bibinfo {author} {\bibfnamefont {V.}~\bibnamefont
  {Rosenhaus}},\ }\href {\doibase 10.1007/JHEP04(2016)001} {\bibfield
  {journal} {\bibinfo  {journal} {Journal of High Energy Physics}\ }\textbf
  {\bibinfo {volume} {2016}},\ \bibinfo {pages} {1} (\bibinfo {year}
  {2016})}\BibitemShut {NoStop}%
\bibitem [{\citenamefont {Mezei}\ and\ \citenamefont
  {Stanford}(2017)}]{mezei_entanglement_2017}%
  \BibitemOpen
  \bibfield  {author} {\bibinfo {author} {\bibfnamefont {M.}~\bibnamefont
  {Mezei}}\ and\ \bibinfo {author} {\bibfnamefont {D.}~\bibnamefont
  {Stanford}},\ }\href {\doibase 10.1007/JHEP05(2017)065} {\bibfield  {journal}
  {\bibinfo  {journal} {Journal of High Energy Physics}\ }\textbf {\bibinfo
  {volume} {2017}},\ \bibinfo {pages} {65} (\bibinfo {year}
  {2017})}\BibitemShut {NoStop}%
\bibitem [{\citenamefont {Roberts}\ and\ \citenamefont
  {Yoshida}(2017)}]{roberts_chaos_2017}%
  \BibitemOpen
  \bibfield  {author} {\bibinfo {author} {\bibfnamefont {D.~A.}\ \bibnamefont
  {Roberts}}\ and\ \bibinfo {author} {\bibfnamefont {B.}~\bibnamefont
  {Yoshida}},\ }\href {\doibase 10.1007/JHEP04(2017)121} {\bibfield  {journal}
  {\bibinfo  {journal} {Journal of High Energy Physics}\ }\textbf {\bibinfo
  {volume} {2017}},\ \bibinfo {pages} {121} (\bibinfo {year}
  {2017})}\BibitemShut {NoStop}%
\bibitem [{\citenamefont {Hayden}\ and\ \citenamefont
  {Preskill}(2007)}]{hayden_black_2007}%
  \BibitemOpen
  \bibfield  {author} {\bibinfo {author} {\bibfnamefont {P.}~\bibnamefont
  {Hayden}}\ and\ \bibinfo {author} {\bibfnamefont {J.}~\bibnamefont
  {Preskill}},\ }\href {\doibase 10.1088/1126-6708/2007/09/120} {\bibfield
  {journal} {\bibinfo  {journal} {Journal of High Energy Physics}\ }\textbf
  {\bibinfo {volume} {2007}},\ \bibinfo {pages} {120} (\bibinfo {year}
  {2007})}\BibitemShut {NoStop}%
\bibitem [{\citenamefont {Shenker}\ and\ \citenamefont
  {Stanford}(2014)}]{shenker_black_2014}%
  \BibitemOpen
  \bibfield  {author} {\bibinfo {author} {\bibfnamefont {S.~H.}\ \bibnamefont
  {Shenker}}\ and\ \bibinfo {author} {\bibfnamefont {D.}~\bibnamefont
  {Stanford}},\ }\href {\doibase 10.1007/JHEP03(2014)067} {\bibfield  {journal}
  {\bibinfo  {journal} {Journal of High Energy Physics}\ }\textbf {\bibinfo
  {volume} {2014}},\ \bibinfo {pages} {67} (\bibinfo {year}
  {2014})}\BibitemShut {NoStop}%
\bibitem [{\citenamefont {Swingle}(2018)}]{swingle_unscrambling_2018}%
  \BibitemOpen
  \bibfield  {author} {\bibinfo {author} {\bibfnamefont {B.}~\bibnamefont
  {Swingle}},\ }\href {\doibase 10.1038/s41567-018-0295-5} {\bibfield
  {journal} {\bibinfo  {journal} {Nature Physics}\ }\textbf {\bibinfo {volume}
  {14}},\ \bibinfo {pages} {988} (\bibinfo {year} {2018})}\BibitemShut
  {NoStop}%
\bibitem [{\citenamefont {Xu}\ and\ \citenamefont
  {Swingle}(2022)}]{xu_swingle_tutorial_2022}%
  \BibitemOpen
  \bibfield  {author} {\bibinfo {author} {\bibfnamefont {S.}~\bibnamefont
  {Xu}}\ and\ \bibinfo {author} {\bibfnamefont {B.}~\bibnamefont {Swingle}},\
  }\href {\doibase 10.48550/ARXIV.2202.07060} {\enquote {\bibinfo {title}
  {Scrambling dynamics and out-of-time ordered correlators in quantum many-body
  systems: a tutorial},}\ } (\bibinfo {year} {2022})\BibitemShut {NoStop}%
\bibitem [{\citenamefont {Gogolin}\ and\ \citenamefont
  {Eisert}(2016)}]{Gogolin_2016_review}%
  \BibitemOpen
  \bibfield  {author} {\bibinfo {author} {\bibfnamefont {C.}~\bibnamefont
  {Gogolin}}\ and\ \bibinfo {author} {\bibfnamefont {J.}~\bibnamefont
  {Eisert}},\ }\href {\doibase 10.1088/0034-4885/79/5/056001} {\bibfield
  {journal} {\bibinfo  {journal} {Reports on Progress in Physics}\ }\textbf
  {\bibinfo {volume} {79}},\ \bibinfo {pages} {056001} (\bibinfo {year}
  {2016})}\BibitemShut {NoStop}%
\bibitem [{\citenamefont {D'Alessio}\ \emph {et~al.}(2016)\citenamefont
  {D'Alessio}, \citenamefont {Kafri}, \citenamefont {Polkovnikov},\ and\
  \citenamefont {Rigol}}]{rigol2016quantum}%
  \BibitemOpen
  \bibfield  {author} {\bibinfo {author} {\bibfnamefont {L.}~\bibnamefont
  {D'Alessio}}, \bibinfo {author} {\bibfnamefont {Y.}~\bibnamefont {Kafri}},
  \bibinfo {author} {\bibfnamefont {A.}~\bibnamefont {Polkovnikov}}, \ and\
  \bibinfo {author} {\bibfnamefont {M.}~\bibnamefont {Rigol}},\ }\href@noop {}
  {\bibfield  {journal} {\bibinfo  {journal} {Advances in Physics}\ }\textbf
  {\bibinfo {volume} {65}},\ \bibinfo {pages} {239} (\bibinfo {year}
  {2016})}\BibitemShut {NoStop}%
\bibitem [{\citenamefont {Xu}\ \emph {et~al.}(2020)\citenamefont {Xu},
  \citenamefont {Scaffidi},\ and\ \citenamefont {Cao}}]{xu_does_2020}%
  \BibitemOpen
  \bibfield  {author} {\bibinfo {author} {\bibfnamefont {T.}~\bibnamefont
  {Xu}}, \bibinfo {author} {\bibfnamefont {T.}~\bibnamefont {Scaffidi}}, \ and\
  \bibinfo {author} {\bibfnamefont {X.}~\bibnamefont {Cao}},\ }\href {\doibase
  10.1103/PhysRevLett.124.140602} {\bibfield  {journal} {\bibinfo  {journal}
  {Physical Review Letters}\ }\textbf {\bibinfo {volume} {124}},\ \bibinfo
  {pages} {140602} (\bibinfo {year} {2020})}\BibitemShut {NoStop}%
\bibitem [{\citenamefont {Mi}\ \emph {et~al.}(2021)\citenamefont {Mi},
  \citenamefont {Roushan}, \citenamefont {Quintana}, \citenamefont {Mandrà},
  \citenamefont {Marshall}, \citenamefont {Neill}, \citenamefont {Arute},
  \citenamefont {Arya}, \citenamefont {Atalaya}, \citenamefont {Babbush},
  \citenamefont {Bardin}, \citenamefont {Barends}, \citenamefont {Basso},
  \citenamefont {Bengtsson}, \citenamefont {Boixo}, \citenamefont {Bourassa},
  \citenamefont {Broughton}, \citenamefont {Buckley}, \citenamefont {Buell},
  \citenamefont {Burkett}, \citenamefont {Bushnell}, \citenamefont {Chen},
  \citenamefont {Chiaro}, \citenamefont {Collins}, \citenamefont {Courtney},
  \citenamefont {Demura}, \citenamefont {Derk}, \citenamefont {Dunsworth},
  \citenamefont {Eppens}, \citenamefont {Erickson}, \citenamefont {Farhi},
  \citenamefont {Fowler}, \citenamefont {Foxen}, \citenamefont {Gidney},
  \citenamefont {Giustina}, \citenamefont {Gross}, \citenamefont {Harrigan},
  \citenamefont {Harrington}, \citenamefont {Hilton}, \citenamefont {Ho},
  \citenamefont {Hong}, \citenamefont {Huang}, \citenamefont {Huggins},
  \citenamefont {Ioffe}, \citenamefont {Isakov}, \citenamefont {Jeffrey},
  \citenamefont {Jiang}, \citenamefont {Jones}, \citenamefont {Kafri},
  \citenamefont {Kelly}, \citenamefont {Kim}, \citenamefont {Kitaev},
  \citenamefont {Klimov}, \citenamefont {Korotkov}, \citenamefont {Kostritsa},
  \citenamefont {Landhuis}, \citenamefont {Laptev}, \citenamefont {Lucero},
  \citenamefont {Martin}, \citenamefont {McClean}, \citenamefont {McCourt},
  \citenamefont {McEwen}, \citenamefont {Megrant}, \citenamefont {Miao},
  \citenamefont {Mohseni}, \citenamefont {Montazeri}, \citenamefont
  {Mruczkiewicz}, \citenamefont {Mutus}, \citenamefont {Naaman}, \citenamefont
  {Neeley}, \citenamefont {Newman}, \citenamefont {Niu}, \citenamefont
  {O’Brien}, \citenamefont {Opremcak}, \citenamefont {Ostby}, \citenamefont
  {Pato}, \citenamefont {Petukhov}, \citenamefont {Redd}, \citenamefont
  {Rubin}, \citenamefont {Sank}, \citenamefont {Satzinger}, \citenamefont
  {Shvarts}, \citenamefont {Strain}, \citenamefont {Szalay}, \citenamefont
  {Trevithick}, \citenamefont {Villalonga}, \citenamefont {White},
  \citenamefont {Yao}, \citenamefont {Yeh}, \citenamefont {Zalcman},
  \citenamefont {Neven}, \citenamefont {Aleiner}, \citenamefont {Kechedzhi},
  \citenamefont {Smelyanskiy},\ and\ \citenamefont
  {Chen}}]{mi_information_2021}%
  \BibitemOpen
  \bibfield  {author} {\bibinfo {author} {\bibfnamefont {X.}~\bibnamefont
  {Mi}}, \bibinfo {author} {\bibfnamefont {P.}~\bibnamefont {Roushan}},
  \bibinfo {author} {\bibfnamefont {C.}~\bibnamefont {Quintana}}, \bibinfo
  {author} {\bibfnamefont {S.}~\bibnamefont {Mandrà}}, \bibinfo {author}
  {\bibfnamefont {J.}~\bibnamefont {Marshall}}, \bibinfo {author}
  {\bibfnamefont {C.}~\bibnamefont {Neill}}, \bibinfo {author} {\bibfnamefont
  {F.}~\bibnamefont {Arute}}, \bibinfo {author} {\bibfnamefont
  {K.}~\bibnamefont {Arya}}, \bibinfo {author} {\bibfnamefont {J.}~\bibnamefont
  {Atalaya}}, \bibinfo {author} {\bibfnamefont {R.}~\bibnamefont {Babbush}},
  \bibinfo {author} {\bibfnamefont {J.~C.}\ \bibnamefont {Bardin}}, \bibinfo
  {author} {\bibfnamefont {R.}~\bibnamefont {Barends}}, \bibinfo {author}
  {\bibfnamefont {J.}~\bibnamefont {Basso}}, \bibinfo {author} {\bibfnamefont
  {A.}~\bibnamefont {Bengtsson}}, \bibinfo {author} {\bibfnamefont
  {S.}~\bibnamefont {Boixo}}, \bibinfo {author} {\bibfnamefont
  {A.}~\bibnamefont {Bourassa}}, \bibinfo {author} {\bibfnamefont
  {M.}~\bibnamefont {Broughton}}, \bibinfo {author} {\bibfnamefont {B.~B.}\
  \bibnamefont {Buckley}}, \bibinfo {author} {\bibfnamefont {D.~A.}\
  \bibnamefont {Buell}}, \bibinfo {author} {\bibfnamefont {B.}~\bibnamefont
  {Burkett}}, \bibinfo {author} {\bibfnamefont {N.}~\bibnamefont {Bushnell}},
  \bibinfo {author} {\bibfnamefont {Z.}~\bibnamefont {Chen}}, \bibinfo {author}
  {\bibfnamefont {B.}~\bibnamefont {Chiaro}}, \bibinfo {author} {\bibfnamefont
  {R.}~\bibnamefont {Collins}}, \bibinfo {author} {\bibfnamefont
  {W.}~\bibnamefont {Courtney}}, \bibinfo {author} {\bibfnamefont
  {S.}~\bibnamefont {Demura}}, \bibinfo {author} {\bibfnamefont {A.~R.}\
  \bibnamefont {Derk}}, \bibinfo {author} {\bibfnamefont {A.}~\bibnamefont
  {Dunsworth}}, \bibinfo {author} {\bibfnamefont {D.}~\bibnamefont {Eppens}},
  \bibinfo {author} {\bibfnamefont {C.}~\bibnamefont {Erickson}}, \bibinfo
  {author} {\bibfnamefont {E.}~\bibnamefont {Farhi}}, \bibinfo {author}
  {\bibfnamefont {A.~G.}\ \bibnamefont {Fowler}}, \bibinfo {author}
  {\bibfnamefont {B.}~\bibnamefont {Foxen}}, \bibinfo {author} {\bibfnamefont
  {C.}~\bibnamefont {Gidney}}, \bibinfo {author} {\bibfnamefont
  {M.}~\bibnamefont {Giustina}}, \bibinfo {author} {\bibfnamefont {J.~A.}\
  \bibnamefont {Gross}}, \bibinfo {author} {\bibfnamefont {M.~P.}\ \bibnamefont
  {Harrigan}}, \bibinfo {author} {\bibfnamefont {S.~D.}\ \bibnamefont
  {Harrington}}, \bibinfo {author} {\bibfnamefont {J.}~\bibnamefont {Hilton}},
  \bibinfo {author} {\bibfnamefont {A.}~\bibnamefont {Ho}}, \bibinfo {author}
  {\bibfnamefont {S.}~\bibnamefont {Hong}}, \bibinfo {author} {\bibfnamefont
  {T.}~\bibnamefont {Huang}}, \bibinfo {author} {\bibfnamefont {W.~J.}\
  \bibnamefont {Huggins}}, \bibinfo {author} {\bibfnamefont {L.~B.}\
  \bibnamefont {Ioffe}}, \bibinfo {author} {\bibfnamefont {S.~V.}\ \bibnamefont
  {Isakov}}, \bibinfo {author} {\bibfnamefont {E.}~\bibnamefont {Jeffrey}},
  \bibinfo {author} {\bibfnamefont {Z.}~\bibnamefont {Jiang}}, \bibinfo
  {author} {\bibfnamefont {C.}~\bibnamefont {Jones}}, \bibinfo {author}
  {\bibfnamefont {D.}~\bibnamefont {Kafri}}, \bibinfo {author} {\bibfnamefont
  {J.}~\bibnamefont {Kelly}}, \bibinfo {author} {\bibfnamefont
  {S.}~\bibnamefont {Kim}}, \bibinfo {author} {\bibfnamefont {A.}~\bibnamefont
  {Kitaev}}, \bibinfo {author} {\bibfnamefont {P.~V.}\ \bibnamefont {Klimov}},
  \bibinfo {author} {\bibfnamefont {A.~N.}\ \bibnamefont {Korotkov}}, \bibinfo
  {author} {\bibfnamefont {F.}~\bibnamefont {Kostritsa}}, \bibinfo {author}
  {\bibfnamefont {D.}~\bibnamefont {Landhuis}}, \bibinfo {author}
  {\bibfnamefont {P.}~\bibnamefont {Laptev}}, \bibinfo {author} {\bibfnamefont
  {E.}~\bibnamefont {Lucero}}, \bibinfo {author} {\bibfnamefont
  {O.}~\bibnamefont {Martin}}, \bibinfo {author} {\bibfnamefont {J.~R.}\
  \bibnamefont {McClean}}, \bibinfo {author} {\bibfnamefont {T.}~\bibnamefont
  {McCourt}}, \bibinfo {author} {\bibfnamefont {M.}~\bibnamefont {McEwen}},
  \bibinfo {author} {\bibfnamefont {A.}~\bibnamefont {Megrant}}, \bibinfo
  {author} {\bibfnamefont {K.~C.}\ \bibnamefont {Miao}}, \bibinfo {author}
  {\bibfnamefont {M.}~\bibnamefont {Mohseni}}, \bibinfo {author} {\bibfnamefont
  {S.}~\bibnamefont {Montazeri}}, \bibinfo {author} {\bibfnamefont
  {W.}~\bibnamefont {Mruczkiewicz}}, \bibinfo {author} {\bibfnamefont
  {J.}~\bibnamefont {Mutus}}, \bibinfo {author} {\bibfnamefont
  {O.}~\bibnamefont {Naaman}}, \bibinfo {author} {\bibfnamefont
  {M.}~\bibnamefont {Neeley}}, \bibinfo {author} {\bibfnamefont
  {M.}~\bibnamefont {Newman}}, \bibinfo {author} {\bibfnamefont {M.~Y.}\
  \bibnamefont {Niu}}, \bibinfo {author} {\bibfnamefont {T.~E.}\ \bibnamefont
  {O’Brien}}, \bibinfo {author} {\bibfnamefont {A.}~\bibnamefont {Opremcak}},
  \bibinfo {author} {\bibfnamefont {E.}~\bibnamefont {Ostby}}, \bibinfo
  {author} {\bibfnamefont {B.}~\bibnamefont {Pato}}, \bibinfo {author}
  {\bibfnamefont {A.}~\bibnamefont {Petukhov}}, \bibinfo {author}
  {\bibfnamefont {N.}~\bibnamefont {Redd}}, \bibinfo {author} {\bibfnamefont
  {N.~C.}\ \bibnamefont {Rubin}}, \bibinfo {author} {\bibfnamefont
  {D.}~\bibnamefont {Sank}}, \bibinfo {author} {\bibfnamefont {K.~J.}\
  \bibnamefont {Satzinger}}, \bibinfo {author} {\bibfnamefont {V.}~\bibnamefont
  {Shvarts}}, \bibinfo {author} {\bibfnamefont {D.}~\bibnamefont {Strain}},
  \bibinfo {author} {\bibfnamefont {M.}~\bibnamefont {Szalay}}, \bibinfo
  {author} {\bibfnamefont {M.~D.}\ \bibnamefont {Trevithick}}, \bibinfo
  {author} {\bibfnamefont {B.}~\bibnamefont {Villalonga}}, \bibinfo {author}
  {\bibfnamefont {T.}~\bibnamefont {White}}, \bibinfo {author} {\bibfnamefont
  {Z.~J.}\ \bibnamefont {Yao}}, \bibinfo {author} {\bibfnamefont
  {P.}~\bibnamefont {Yeh}}, \bibinfo {author} {\bibfnamefont {A.}~\bibnamefont
  {Zalcman}}, \bibinfo {author} {\bibfnamefont {H.}~\bibnamefont {Neven}},
  \bibinfo {author} {\bibfnamefont {I.}~\bibnamefont {Aleiner}}, \bibinfo
  {author} {\bibfnamefont {K.}~\bibnamefont {Kechedzhi}}, \bibinfo {author}
  {\bibfnamefont {V.}~\bibnamefont {Smelyanskiy}}, \ and\ \bibinfo {author}
  {\bibfnamefont {Y.}~\bibnamefont {Chen}},\ }\href {\doibase
  10.1126/science.abg5029} {\bibfield  {journal} {\bibinfo  {journal}
  {Science}\ }\textbf {\bibinfo {volume} {374}},\ \bibinfo {pages} {1479}
  (\bibinfo {year} {2021})},\ \bibinfo {note} {publisher: American Association
  for the Advancement of Science}\BibitemShut {NoStop}%
\bibitem [{\citenamefont {Braumüller}\ \emph {et~al.}(2022)\citenamefont
  {Braumüller}, \citenamefont {Karamlou}, \citenamefont {Yanay}, \citenamefont
  {Kannan}, \citenamefont {Kim}, \citenamefont {Kjaergaard}, \citenamefont
  {Melville}, \citenamefont {Niedzielski}, \citenamefont {Sung}, \citenamefont
  {Vepsäläinen}, \citenamefont {Winik}, \citenamefont {Yoder}, \citenamefont
  {Orlando}, \citenamefont {Gustavsson}, \citenamefont {Tahan},\ and\
  \citenamefont {Oliver}}]{braumuller_probing_2022}%
  \BibitemOpen
  \bibfield  {author} {\bibinfo {author} {\bibfnamefont {J.}~\bibnamefont
  {Braumüller}}, \bibinfo {author} {\bibfnamefont {A.~H.}\ \bibnamefont
  {Karamlou}}, \bibinfo {author} {\bibfnamefont {Y.}~\bibnamefont {Yanay}},
  \bibinfo {author} {\bibfnamefont {B.}~\bibnamefont {Kannan}}, \bibinfo
  {author} {\bibfnamefont {D.}~\bibnamefont {Kim}}, \bibinfo {author}
  {\bibfnamefont {M.}~\bibnamefont {Kjaergaard}}, \bibinfo {author}
  {\bibfnamefont {A.}~\bibnamefont {Melville}}, \bibinfo {author}
  {\bibfnamefont {B.~M.}\ \bibnamefont {Niedzielski}}, \bibinfo {author}
  {\bibfnamefont {Y.}~\bibnamefont {Sung}}, \bibinfo {author} {\bibfnamefont
  {A.}~\bibnamefont {Vepsäläinen}}, \bibinfo {author} {\bibfnamefont
  {R.}~\bibnamefont {Winik}}, \bibinfo {author} {\bibfnamefont {J.~L.}\
  \bibnamefont {Yoder}}, \bibinfo {author} {\bibfnamefont {T.~P.}\ \bibnamefont
  {Orlando}}, \bibinfo {author} {\bibfnamefont {S.}~\bibnamefont {Gustavsson}},
  \bibinfo {author} {\bibfnamefont {C.}~\bibnamefont {Tahan}}, \ and\ \bibinfo
  {author} {\bibfnamefont {W.~D.}\ \bibnamefont {Oliver}},\ }\href {\doibase
  10.1038/s41567-021-01430-w} {\bibfield  {journal} {\bibinfo  {journal}
  {Nature Physics}\ }\textbf {\bibinfo {volume} {18}},\ \bibinfo {pages} {172}
  (\bibinfo {year} {2022})}\BibitemShut {NoStop}%
\bibitem [{\citenamefont {Wei}\ \emph {et~al.}(2018)\citenamefont {Wei},
  \citenamefont {Ramanathan},\ and\ \citenamefont
  {Cappellaro}}]{wei_exploring_2018}%
  \BibitemOpen
  \bibfield  {author} {\bibinfo {author} {\bibfnamefont {K.~X.}\ \bibnamefont
  {Wei}}, \bibinfo {author} {\bibfnamefont {C.}~\bibnamefont {Ramanathan}}, \
  and\ \bibinfo {author} {\bibfnamefont {P.}~\bibnamefont {Cappellaro}},\
  }\href {\doibase 10.1103/PhysRevLett.120.070501} {\bibfield  {journal}
  {\bibinfo  {journal} {Physical Review Letters}\ }\textbf {\bibinfo {volume}
  {120}},\ \bibinfo {pages} {070501} (\bibinfo {year} {2018})}\BibitemShut
  {NoStop}%
\bibitem [{\citenamefont {Li}\ \emph {et~al.}(2017)\citenamefont {Li},
  \citenamefont {Fan}, \citenamefont {Wang}, \citenamefont {Ye}, \citenamefont
  {Zeng}, \citenamefont {Zhai}, \citenamefont {Peng},\ and\ \citenamefont
  {Du}}]{li_measuring_2017}%
  \BibitemOpen
  \bibfield  {author} {\bibinfo {author} {\bibfnamefont {J.}~\bibnamefont
  {Li}}, \bibinfo {author} {\bibfnamefont {R.}~\bibnamefont {Fan}}, \bibinfo
  {author} {\bibfnamefont {H.}~\bibnamefont {Wang}}, \bibinfo {author}
  {\bibfnamefont {B.}~\bibnamefont {Ye}}, \bibinfo {author} {\bibfnamefont
  {B.}~\bibnamefont {Zeng}}, \bibinfo {author} {\bibfnamefont {H.}~\bibnamefont
  {Zhai}}, \bibinfo {author} {\bibfnamefont {X.}~\bibnamefont {Peng}}, \ and\
  \bibinfo {author} {\bibfnamefont {J.}~\bibnamefont {Du}},\ }\href {\doibase
  10.1103/PhysRevX.7.031011} {\bibfield  {journal} {\bibinfo  {journal}
  {Physical Review X}\ }\textbf {\bibinfo {volume} {7}},\ \bibinfo {pages}
  {031011} (\bibinfo {year} {2017})}\BibitemShut {NoStop}%
\bibitem [{\citenamefont {Nie}\ \emph {et~al.}(2019)\citenamefont {Nie},
  \citenamefont {Zhang}, \citenamefont {Zhao}, \citenamefont {Xin},
  \citenamefont {Lu},\ and\ \citenamefont {Li}}]{nie_detecting_2019}%
  \BibitemOpen
  \bibfield  {author} {\bibinfo {author} {\bibfnamefont {X.}~\bibnamefont
  {Nie}}, \bibinfo {author} {\bibfnamefont {Z.}~\bibnamefont {Zhang}}, \bibinfo
  {author} {\bibfnamefont {X.}~\bibnamefont {Zhao}}, \bibinfo {author}
  {\bibfnamefont {T.}~\bibnamefont {Xin}}, \bibinfo {author} {\bibfnamefont
  {D.}~\bibnamefont {Lu}}, \ and\ \bibinfo {author} {\bibfnamefont
  {J.}~\bibnamefont {Li}},\ }\href {http://arxiv.org/abs/1903.12237} {\bibfield
   {journal} {\bibinfo  {journal} {arXiv:1903.12237 [quant-ph]}\ } (\bibinfo
  {year} {2019})},\ \bibinfo {note} {arXiv: 1903.12237}\BibitemShut {NoStop}%
\bibitem [{\citenamefont {Nie}\ \emph {et~al.}(2020)\citenamefont {Nie},
  \citenamefont {Wei}, \citenamefont {Chen}, \citenamefont {Zhang},
  \citenamefont {Zhao}, \citenamefont {Qiu}, \citenamefont {Tian},
  \citenamefont {Ji}, \citenamefont {Xin}, \citenamefont {Lu},\ and\
  \citenamefont {Li}}]{nie_experimental_2020}%
  \BibitemOpen
  \bibfield  {author} {\bibinfo {author} {\bibfnamefont {X.}~\bibnamefont
  {Nie}}, \bibinfo {author} {\bibfnamefont {B.-B.}\ \bibnamefont {Wei}},
  \bibinfo {author} {\bibfnamefont {X.}~\bibnamefont {Chen}}, \bibinfo {author}
  {\bibfnamefont {Z.}~\bibnamefont {Zhang}}, \bibinfo {author} {\bibfnamefont
  {X.}~\bibnamefont {Zhao}}, \bibinfo {author} {\bibfnamefont {C.}~\bibnamefont
  {Qiu}}, \bibinfo {author} {\bibfnamefont {Y.}~\bibnamefont {Tian}}, \bibinfo
  {author} {\bibfnamefont {Y.}~\bibnamefont {Ji}}, \bibinfo {author}
  {\bibfnamefont {T.}~\bibnamefont {Xin}}, \bibinfo {author} {\bibfnamefont
  {D.}~\bibnamefont {Lu}}, \ and\ \bibinfo {author} {\bibfnamefont
  {J.}~\bibnamefont {Li}},\ }\href {\doibase 10.1103/PhysRevLett.124.250601}
  {\bibfield  {journal} {\bibinfo  {journal} {Physical Review Letters}\
  }\textbf {\bibinfo {volume} {124}},\ \bibinfo {pages} {250601} (\bibinfo
  {year} {2020})}\BibitemShut {NoStop}%
\bibitem [{\citenamefont {Gärttner}\ \emph {et~al.}(2017)\citenamefont
  {Gärttner}, \citenamefont {Bohnet}, \citenamefont {Safavi-Naini},
  \citenamefont {Wall}, \citenamefont {Bollinger},\ and\ \citenamefont
  {Rey}}]{garttner_measuring_2017}%
  \BibitemOpen
  \bibfield  {author} {\bibinfo {author} {\bibfnamefont {M.}~\bibnamefont
  {Gärttner}}, \bibinfo {author} {\bibfnamefont {J.~G.}\ \bibnamefont
  {Bohnet}}, \bibinfo {author} {\bibfnamefont {A.}~\bibnamefont
  {Safavi-Naini}}, \bibinfo {author} {\bibfnamefont {M.~L.}\ \bibnamefont
  {Wall}}, \bibinfo {author} {\bibfnamefont {J.~J.}\ \bibnamefont {Bollinger}},
  \ and\ \bibinfo {author} {\bibfnamefont {A.~M.}\ \bibnamefont {Rey}},\ }\href
  {\doibase 10.1038/nphys4119} {\bibfield  {journal} {\bibinfo  {journal}
  {Nature Physics}\ }\textbf {\bibinfo {volume} {13}},\ \bibinfo {pages} {781}
  (\bibinfo {year} {2017})}\BibitemShut {NoStop}%
\bibitem [{\citenamefont {Joshi}\ \emph {et~al.}(2020)\citenamefont {Joshi},
  \citenamefont {Elben}, \citenamefont {Vermersch}, \citenamefont {Brydges},
  \citenamefont {Maier}, \citenamefont {Zoller}, \citenamefont {Blatt},\ and\
  \citenamefont {Roos}}]{joshi_quantum_2020}%
  \BibitemOpen
  \bibfield  {author} {\bibinfo {author} {\bibfnamefont {M.~K.}\ \bibnamefont
  {Joshi}}, \bibinfo {author} {\bibfnamefont {A.}~\bibnamefont {Elben}},
  \bibinfo {author} {\bibfnamefont {B.}~\bibnamefont {Vermersch}}, \bibinfo
  {author} {\bibfnamefont {T.}~\bibnamefont {Brydges}}, \bibinfo {author}
  {\bibfnamefont {C.}~\bibnamefont {Maier}}, \bibinfo {author} {\bibfnamefont
  {P.}~\bibnamefont {Zoller}}, \bibinfo {author} {\bibfnamefont
  {R.}~\bibnamefont {Blatt}}, \ and\ \bibinfo {author} {\bibfnamefont {C.~F.}\
  \bibnamefont {Roos}},\ }\href {\doibase 10.1103/PhysRevLett.124.240505}
  {\bibfield  {journal} {\bibinfo  {journal} {Physical Review Letters}\
  }\textbf {\bibinfo {volume} {124}},\ \bibinfo {pages} {240505} (\bibinfo
  {year} {2020})}\BibitemShut {NoStop}%
\bibitem [{\citenamefont {Meier}\ \emph {et~al.}(2019)\citenamefont {Meier},
  \citenamefont {Ang'ong'a}, \citenamefont {An},\ and\ \citenamefont
  {Gadway}}]{meier_exploring_2019}%
  \BibitemOpen
  \bibfield  {author} {\bibinfo {author} {\bibfnamefont {E.~J.}\ \bibnamefont
  {Meier}}, \bibinfo {author} {\bibfnamefont {J.}~\bibnamefont {Ang'ong'a}},
  \bibinfo {author} {\bibfnamefont {F.~A.}\ \bibnamefont {An}}, \ and\ \bibinfo
  {author} {\bibfnamefont {B.}~\bibnamefont {Gadway}},\ }\href {\doibase
  10.1103/PhysRevA.100.013623} {\bibfield  {journal} {\bibinfo  {journal}
  {Physical Review A}\ }\textbf {\bibinfo {volume} {100}},\ \bibinfo {pages}
  {013623} (\bibinfo {year} {2019})}\BibitemShut {NoStop}%
\bibitem [{\citenamefont {Chen}\ \emph {et~al.}(2020)\citenamefont {Chen},
  \citenamefont {Hou}, \citenamefont {Zhou}, \citenamefont {Qian},
  \citenamefont {Shen},\ and\ \citenamefont {Xu}}]{chen_detecting_2020}%
  \BibitemOpen
  \bibfield  {author} {\bibinfo {author} {\bibfnamefont {B.}~\bibnamefont
  {Chen}}, \bibinfo {author} {\bibfnamefont {X.}~\bibnamefont {Hou}}, \bibinfo
  {author} {\bibfnamefont {F.}~\bibnamefont {Zhou}}, \bibinfo {author}
  {\bibfnamefont {P.}~\bibnamefont {Qian}}, \bibinfo {author} {\bibfnamefont
  {H.}~\bibnamefont {Shen}}, \ and\ \bibinfo {author} {\bibfnamefont
  {N.}~\bibnamefont {Xu}},\ }\href {\doibase 10.1063/5.0004152} {\bibfield
  {journal} {\bibinfo  {journal} {Applied Physics Letters}\ }\textbf {\bibinfo
  {volume} {116}},\ \bibinfo {pages} {194002} (\bibinfo {year}
  {2020})}\BibitemShut {NoStop}%
\bibitem [{\citenamefont {Landsman}\ \emph {et~al.}(2019)\citenamefont
  {Landsman}, \citenamefont {Figgatt}, \citenamefont {Schuster}, \citenamefont
  {Linke}, \citenamefont {Yoshida}, \citenamefont {Yao},\ and\ \citenamefont
  {Monroe}}]{landsman_verified_2019}%
  \BibitemOpen
  \bibfield  {author} {\bibinfo {author} {\bibfnamefont {K.~A.}\ \bibnamefont
  {Landsman}}, \bibinfo {author} {\bibfnamefont {C.}~\bibnamefont {Figgatt}},
  \bibinfo {author} {\bibfnamefont {T.}~\bibnamefont {Schuster}}, \bibinfo
  {author} {\bibfnamefont {N.~M.}\ \bibnamefont {Linke}}, \bibinfo {author}
  {\bibfnamefont {B.}~\bibnamefont {Yoshida}}, \bibinfo {author} {\bibfnamefont
  {N.~Y.}\ \bibnamefont {Yao}}, \ and\ \bibinfo {author} {\bibfnamefont
  {C.}~\bibnamefont {Monroe}},\ }\href {\doibase 10.1038/s41586-019-0952-6}
  {\bibfield  {journal} {\bibinfo  {journal} {Nature}\ }\textbf {\bibinfo
  {volume} {567}},\ \bibinfo {pages} {61} (\bibinfo {year} {2019})}\BibitemShut
  {NoStop}%
\bibitem [{\citenamefont {Styliaris}\ \emph {et~al.}(2021)\citenamefont
  {Styliaris}, \citenamefont {Anand},\ and\ \citenamefont
  {Zanardi}}]{styliaris_information_2021}%
  \BibitemOpen
  \bibfield  {author} {\bibinfo {author} {\bibfnamefont {G.}~\bibnamefont
  {Styliaris}}, \bibinfo {author} {\bibfnamefont {N.}~\bibnamefont {Anand}}, \
  and\ \bibinfo {author} {\bibfnamefont {P.}~\bibnamefont {Zanardi}},\ }\href
  {\doibase 10.1103/PhysRevLett.126.030601} {\bibfield  {journal} {\bibinfo
  {journal} {Physical Review Letters}\ }\textbf {\bibinfo {volume} {126}},\
  \bibinfo {pages} {030601} (\bibinfo {year} {2021})}\BibitemShut {NoStop}%
\bibitem [{\citenamefont {Zanardi}\ and\ \citenamefont
  {Anand}(2021)}]{zanardi_information_2021}%
  \BibitemOpen
  \bibfield  {author} {\bibinfo {author} {\bibfnamefont {P.}~\bibnamefont
  {Zanardi}}\ and\ \bibinfo {author} {\bibfnamefont {N.}~\bibnamefont
  {Anand}},\ }\href {\doibase 10.1103/PhysRevA.103.062214} {\bibfield
  {journal} {\bibinfo  {journal} {Physical Review A}\ }\textbf {\bibinfo
  {volume} {103}},\ \bibinfo {pages} {062214} (\bibinfo {year} {2021})},\
  \bibinfo {note} {publisher: American Physical Society}\BibitemShut {NoStop}%
\bibitem [{\citenamefont {Anand}\ and\ \citenamefont
  {Zanardi}(2022)}]{anand_brotocs_2021}%
  \BibitemOpen
  \bibfield  {author} {\bibinfo {author} {\bibfnamefont {N.}~\bibnamefont
  {Anand}}\ and\ \bibinfo {author} {\bibfnamefont {P.}~\bibnamefont
  {Zanardi}},\ }\href {\doibase 10.22331/q-2022-06-27-746} {\bibfield
  {journal} {\bibinfo  {journal} {{Quantum}}\ }\textbf {\bibinfo {volume}
  {6}},\ \bibinfo {pages} {746} (\bibinfo {year} {2022})}\BibitemShut {NoStop}%
\bibitem [{\citenamefont {Anand}\ \emph {et~al.}(2021)\citenamefont {Anand},
  \citenamefont {Styliaris}, \citenamefont {Kumari},\ and\ \citenamefont
  {Zanardi}}]{anand_quantum_2021}%
  \BibitemOpen
  \bibfield  {author} {\bibinfo {author} {\bibfnamefont {N.}~\bibnamefont
  {Anand}}, \bibinfo {author} {\bibfnamefont {G.}~\bibnamefont {Styliaris}},
  \bibinfo {author} {\bibfnamefont {M.}~\bibnamefont {Kumari}}, \ and\ \bibinfo
  {author} {\bibfnamefont {P.}~\bibnamefont {Zanardi}},\ }\href {\doibase
  10.1103/PhysRevResearch.3.023214} {\bibfield  {journal} {\bibinfo  {journal}
  {Physical Review Research}\ }\textbf {\bibinfo {volume} {3}},\ \bibinfo
  {pages} {023214} (\bibinfo {year} {2021})}\BibitemShut {NoStop}%
\bibitem [{\citenamefont {Yan}\ \emph {et~al.}(2020)\citenamefont {Yan},
  \citenamefont {Cincio},\ and\ \citenamefont {Zurek}}]{yan_information_2020}%
  \BibitemOpen
  \bibfield  {author} {\bibinfo {author} {\bibfnamefont {B.}~\bibnamefont
  {Yan}}, \bibinfo {author} {\bibfnamefont {L.}~\bibnamefont {Cincio}}, \ and\
  \bibinfo {author} {\bibfnamefont {W.~H.}\ \bibnamefont {Zurek}},\ }\href
  {\doibase 10.1103/PhysRevLett.124.160603} {\bibfield  {journal} {\bibinfo
  {journal} {Physical Review Letters}\ }\textbf {\bibinfo {volume} {124}},\
  \bibinfo {pages} {160603} (\bibinfo {year} {2020})}\BibitemShut {NoStop}%
\bibitem [{\citenamefont {Yunger~Halpern}\ \emph {et~al.}(2018)\citenamefont
  {Yunger~Halpern}, \citenamefont {Swingle},\ and\ \citenamefont
  {Dressel}}]{PhysRevA.97.042105}%
  \BibitemOpen
  \bibfield  {author} {\bibinfo {author} {\bibfnamefont {N.}~\bibnamefont
  {Yunger~Halpern}}, \bibinfo {author} {\bibfnamefont {B.}~\bibnamefont
  {Swingle}}, \ and\ \bibinfo {author} {\bibfnamefont {J.}~\bibnamefont
  {Dressel}},\ }\href {\doibase 10.1103/PhysRevA.97.042105} {\bibfield
  {journal} {\bibinfo  {journal} {Phys. Rev. A}\ }\textbf {\bibinfo {volume}
  {97}},\ \bibinfo {pages} {042105} (\bibinfo {year} {2018})}\BibitemShut
  {NoStop}%
\bibitem [{\citenamefont {G\"arttner}\ \emph {et~al.}(2018)\citenamefont
  {G\"arttner}, \citenamefont {Hauke},\ and\ \citenamefont
  {Rey}}]{PhysRevLett.120.040402}%
  \BibitemOpen
  \bibfield  {author} {\bibinfo {author} {\bibfnamefont {M.}~\bibnamefont
  {G\"arttner}}, \bibinfo {author} {\bibfnamefont {P.}~\bibnamefont {Hauke}}, \
  and\ \bibinfo {author} {\bibfnamefont {A.~M.}\ \bibnamefont {Rey}},\ }\href
  {\doibase 10.1103/PhysRevLett.120.040402} {\bibfield  {journal} {\bibinfo
  {journal} {Phys. Rev. Lett.}\ }\textbf {\bibinfo {volume} {120}},\ \bibinfo
  {pages} {040402} (\bibinfo {year} {2018})}\BibitemShut {NoStop}%
\bibitem [{\citenamefont {Borgonovi}\ \emph {et~al.}(2019)\citenamefont
  {Borgonovi}, \citenamefont {Izrailev},\ and\ \citenamefont
  {Santos}}]{PhysRevE.99.052143}%
  \BibitemOpen
  \bibfield  {author} {\bibinfo {author} {\bibfnamefont {F.}~\bibnamefont
  {Borgonovi}}, \bibinfo {author} {\bibfnamefont {F.~M.}\ \bibnamefont
  {Izrailev}}, \ and\ \bibinfo {author} {\bibfnamefont {L.~F.}\ \bibnamefont
  {Santos}},\ }\href {\doibase 10.1103/PhysRevE.99.052143} {\bibfield
  {journal} {\bibinfo  {journal} {Phys. Rev. E}\ }\textbf {\bibinfo {volume}
  {99}},\ \bibinfo {pages} {052143} (\bibinfo {year} {2019})}\BibitemShut
  {NoStop}%
\bibitem [{\citenamefont {Yoshida}\ and\ \citenamefont
  {Yao}(2019)}]{yoshida_disentangling_2019}%
  \BibitemOpen
  \bibfield  {author} {\bibinfo {author} {\bibfnamefont {B.}~\bibnamefont
  {Yoshida}}\ and\ \bibinfo {author} {\bibfnamefont {N.~Y.}\ \bibnamefont
  {Yao}},\ }\href {\doibase 10.1103/PhysRevX.9.011006} {\bibfield  {journal}
  {\bibinfo  {journal} {Physical Review X}\ }\textbf {\bibinfo {volume} {9}},\
  \bibinfo {pages} {011006} (\bibinfo {year} {2019})}\BibitemShut {NoStop}%
\bibitem [{\citenamefont {Touil}\ and\ \citenamefont
  {Deffner}(2021)}]{touil_information_2021}%
  \BibitemOpen
  \bibfield  {author} {\bibinfo {author} {\bibfnamefont {A.}~\bibnamefont
  {Touil}}\ and\ \bibinfo {author} {\bibfnamefont {S.}~\bibnamefont
  {Deffner}},\ }\href {\doibase 10.1103/PRXQuantum.2.010306} {\bibfield
  {journal} {\bibinfo  {journal} {PRX Quantum}\ }\textbf {\bibinfo {volume}
  {2}},\ \bibinfo {pages} {010306} (\bibinfo {year} {2021})}\BibitemShut
  {NoStop}%
\bibitem [{\citenamefont {Zanardi}(2022)}]{zanardi_quantum_nodate}%
  \BibitemOpen
  \bibfield  {author} {\bibinfo {author} {\bibfnamefont {P.}~\bibnamefont
  {Zanardi}},\ }\href {\doibase 10.22331/q-2022-03-11-666} {\bibfield
  {journal} {\bibinfo  {journal} {Quantum}\ }\textbf {\bibinfo {volume} {6}},\
  \bibinfo {pages} {666} (\bibinfo {year} {2022})}\BibitemShut {NoStop}%
\bibitem [{\citenamefont {Swingle}\ and\ \citenamefont
  {Yunger~Halpern}(2018)}]{swingle_resilience_2018}%
  \BibitemOpen
  \bibfield  {author} {\bibinfo {author} {\bibfnamefont {B.}~\bibnamefont
  {Swingle}}\ and\ \bibinfo {author} {\bibfnamefont {N.}~\bibnamefont
  {Yunger~Halpern}},\ }\href {\doibase 10.1103/PhysRevA.97.062113} {\bibfield
  {journal} {\bibinfo  {journal} {Physical Review A}\ }\textbf {\bibinfo
  {volume} {97}},\ \bibinfo {pages} {062113} (\bibinfo {year}
  {2018})}\BibitemShut {NoStop}%
\bibitem [{\citenamefont {Davidson}(1996)}]{davidson_c-algebras_1996}%
  \BibitemOpen
  \bibfield  {author} {\bibinfo {author} {\bibfnamefont {K.}~\bibnamefont
  {Davidson}},\ }\href {\doibase 10.1090/fim/006} {{\selectlanguage {en}\emph
  {\bibinfo {title} {C*-{Algebras} by {Example}}}}},\ \bibinfo {series} {Fields
  {Institute} {Monographs}}, Vol.~\bibinfo {volume} {6}\ (\bibinfo  {publisher}
  {American Mathematical Society},\ \bibinfo {year} {1996})\ \bibinfo {note}
  {iSSN: 1069-5273, 2472-4173}\BibitemShut {NoStop}%
\bibitem [{\citenamefont {Zanardi}(2001{\natexlab{a}})}]{zanardi_virtual_2001}%
  \BibitemOpen
  \bibfield  {author} {\bibinfo {author} {\bibfnamefont {P.}~\bibnamefont
  {Zanardi}},\ }\href {\doibase 10.1103/PhysRevLett.87.077901} {\bibfield
  {journal} {\bibinfo  {journal} {Physical Review Letters}\ }\textbf {\bibinfo
  {volume} {87}},\ \bibinfo {pages} {077901} (\bibinfo {year}
  {2001}{\natexlab{a}})}\BibitemShut {NoStop}%
\bibitem [{\citenamefont {Zanardi}\ \emph {et~al.}(2004)\citenamefont
  {Zanardi}, \citenamefont {Lidar},\ and\ \citenamefont
  {Lloyd}}]{zanardi_quantum_2004}%
  \BibitemOpen
  \bibfield  {author} {\bibinfo {author} {\bibfnamefont {P.}~\bibnamefont
  {Zanardi}}, \bibinfo {author} {\bibfnamefont {D.~A.}\ \bibnamefont {Lidar}},
  \ and\ \bibinfo {author} {\bibfnamefont {S.}~\bibnamefont {Lloyd}},\ }\href
  {\doibase 10.1103/PhysRevLett.92.060402} {\bibfield  {journal} {\bibinfo
  {journal} {Physical Review Letters}\ }\textbf {\bibinfo {volume} {92}},\
  \bibinfo {pages} {060402} (\bibinfo {year} {2004})}\BibitemShut {NoStop}%
\bibitem [{\citenamefont {Giulini}(2000)}]{giulini_decoherence_2000}%
  \BibitemOpen
  \bibfield  {author} {\bibinfo {author} {\bibfnamefont {D.}~\bibnamefont
  {Giulini}},\ }\href {http://arxiv.org/abs/quant-ph/0010090} {\emph {\bibinfo
  {title} {Decoherence: {A} dynamical approach to superselection rules?}}},\
  \bibinfo {type} {Tech. Rep.}\ \bibinfo {number} {arXiv:quant-ph/0010090}\
  (\bibinfo  {institution} {arXiv},\ \bibinfo {year} {2000})\ \bibinfo {note}
  {arXiv:quant-ph/0010090 type: article}\BibitemShut {NoStop}%
\bibitem [{\citenamefont {Zanardi}\ \emph
  {et~al.}(2017{\natexlab{a}})\citenamefont {Zanardi}, \citenamefont
  {Styliaris},\ and\ \citenamefont
  {Campos~Venuti}}]{zanardi_coherence-generating_2017}%
  \BibitemOpen
  \bibfield  {author} {\bibinfo {author} {\bibfnamefont {P.}~\bibnamefont
  {Zanardi}}, \bibinfo {author} {\bibfnamefont {G.}~\bibnamefont {Styliaris}},
  \ and\ \bibinfo {author} {\bibfnamefont {L.}~\bibnamefont {Campos~Venuti}},\
  }\href {\doibase 10.1103/PhysRevA.95.052306} {\bibfield  {journal} {\bibinfo
  {journal} {Physical Review A}\ }\textbf {\bibinfo {volume} {95}},\ \bibinfo
  {pages} {052306} (\bibinfo {year} {2017}{\natexlab{a}})}\BibitemShut
  {NoStop}%
\bibitem [{\citenamefont {Zanardi}\ and\ \citenamefont
  {Campos~Venuti}(2018)}]{zanardi_quantum_2018}%
  \BibitemOpen
  \bibfield  {author} {\bibinfo {author} {\bibfnamefont {P.}~\bibnamefont
  {Zanardi}}\ and\ \bibinfo {author} {\bibfnamefont {L.}~\bibnamefont
  {Campos~Venuti}},\ }\href {\doibase 10.1063/1.4997146} {\bibfield  {journal}
  {\bibinfo  {journal} {Journal of Mathematical Physics}\ }\textbf {\bibinfo
  {volume} {59}},\ \bibinfo {pages} {012203} (\bibinfo {year}
  {2018})}\BibitemShut {NoStop}%
\bibitem [{\citenamefont
  {Zanardi}(2001{\natexlab{b}})}]{zanardi_entanglement_2001}%
  \BibitemOpen
  \bibfield  {author} {\bibinfo {author} {\bibfnamefont {P.}~\bibnamefont
  {Zanardi}},\ }\href {\doibase 10.1103/PhysRevA.63.040304} {\bibfield
  {journal} {\bibinfo  {journal} {Phys. Rev. A}\ }\textbf {\bibinfo {volume}
  {63}},\ \bibinfo {pages} {040304} (\bibinfo {year}
  {2001}{\natexlab{b}})}\BibitemShut {NoStop}%
\bibitem [{\citenamefont {Wang}\ and\ \citenamefont
  {Zanardi}(2002)}]{wang_quantum_2002}%
  \BibitemOpen
  \bibfield  {author} {\bibinfo {author} {\bibfnamefont {X.}~\bibnamefont
  {Wang}}\ and\ \bibinfo {author} {\bibfnamefont {P.}~\bibnamefont {Zanardi}},\
  }\href {\doibase 10.1103/PhysRevA.66.044303} {\bibfield  {journal} {\bibinfo
  {journal} {Physical Review A}\ }\textbf {\bibinfo {volume} {66}},\ \bibinfo
  {pages} {044303} (\bibinfo {year} {2002})}\BibitemShut {NoStop}%
\bibitem [{\citenamefont {Zhang}\ \emph {et~al.}(2019)\citenamefont {Zhang},
  \citenamefont {Huang},\ and\ \citenamefont {Chen}}]{zhang_information_2019}%
  \BibitemOpen
  \bibfield  {author} {\bibinfo {author} {\bibfnamefont {Y.-L.}\ \bibnamefont
  {Zhang}}, \bibinfo {author} {\bibfnamefont {Y.}~\bibnamefont {Huang}}, \ and\
  \bibinfo {author} {\bibfnamefont {X.}~\bibnamefont {Chen}},\ }\href {\doibase
  10.1103/PhysRevB.99.014303} {\bibfield  {journal} {\bibinfo  {journal}
  {Physical Review B}\ }\textbf {\bibinfo {volume} {99}},\ \bibinfo {pages}
  {014303} (\bibinfo {year} {2019})}\BibitemShut {NoStop}%
\bibitem [{\citenamefont {Dom\'{\i}nguez}\ and\ \citenamefont
  {\'Alvarez}(2021)}]{dominguez_dynamics_2021}%
  \BibitemOpen
  \bibfield  {author} {\bibinfo {author} {\bibfnamefont {F.~D.}\ \bibnamefont
  {Dom\'{\i}nguez}}\ and\ \bibinfo {author} {\bibfnamefont {G.~A.}\
  \bibnamefont {\'Alvarez}},\ }\href {\doibase 10.1103/PhysRevA.104.062406}
  {\bibfield  {journal} {\bibinfo  {journal} {Phys. Rev. A}\ }\textbf {\bibinfo
  {volume} {104}},\ \bibinfo {pages} {062406} (\bibinfo {year}
  {2021})}\BibitemShut {NoStop}%
\bibitem [{\citenamefont {Mehta}(2004)}]{mehta_random_2004}%
  \BibitemOpen
  \bibfield  {author} {\bibinfo {author} {\bibfnamefont {M.~L.}\ \bibnamefont
  {Mehta}},\ }\href@noop {} {{\selectlanguage {en}\emph {\bibinfo {title}
  {Random Matrices}}}},\ \bibinfo {edition} {3rd}\ ed.,\ \bibinfo {series}
  {Pure and Applied Mathematics Series}\ No.\ \bibinfo {number} {142}\
  (\bibinfo  {publisher} {{Elsevier}},\ \bibinfo {address} {{Amsterdam}},\
  \bibinfo {year} {2004})\BibitemShut {NoStop}%
\bibitem [{\citenamefont {Guhr}\ \emph {et~al.}(1998)\citenamefont {Guhr},
  \citenamefont {{M{\"u}ller{\textendash}Groeling}},\ and\ \citenamefont
  {Weidenm{\"u}ller}}]{guhr_random-matrix_1998}%
  \BibitemOpen
  \bibfield  {author} {\bibinfo {author} {\bibfnamefont {T.}~\bibnamefont
  {Guhr}}, \bibinfo {author} {\bibfnamefont {A.}~\bibnamefont
  {{M{\"u}ller{\textendash}Groeling}}}, \ and\ \bibinfo {author} {\bibfnamefont
  {H.~A.}\ \bibnamefont {Weidenm{\"u}ller}},\ }\href {\doibase
  10.1016/S0370-1573(97)00088-4} {\bibfield  {journal} {\bibinfo  {journal}
  {Physics Reports}\ }\textbf {\bibinfo {volume} {299}},\ \bibinfo {pages}
  {189} (\bibinfo {year} {1998})}\BibitemShut {NoStop}%
\bibitem [{\citenamefont {Styliaris}\ \emph {et~al.}(2018)\citenamefont
  {Styliaris}, \citenamefont {Campos~Venuti},\ and\ \citenamefont
  {Zanardi}}]{styliaris_coherence-generating_2018}%
  \BibitemOpen
  \bibfield  {author} {\bibinfo {author} {\bibfnamefont {G.}~\bibnamefont
  {Styliaris}}, \bibinfo {author} {\bibfnamefont {L.}~\bibnamefont
  {Campos~Venuti}}, \ and\ \bibinfo {author} {\bibfnamefont {P.}~\bibnamefont
  {Zanardi}},\ }\href {\doibase 10.1103/PhysRevA.97.032304} {\bibfield
  {journal} {\bibinfo  {journal} {Physical Review A}\ }\textbf {\bibinfo
  {volume} {97}},\ \bibinfo {pages} {032304} (\bibinfo {year}
  {2018})}\BibitemShut {NoStop}%
\bibitem [{\citenamefont {Styliaris}\ \emph {et~al.}(2019)\citenamefont
  {Styliaris}, \citenamefont {Anand}, \citenamefont {Campos~Venuti},\ and\
  \citenamefont {Zanardi}}]{styliaris_quantum_2019}%
  \BibitemOpen
  \bibfield  {author} {\bibinfo {author} {\bibfnamefont {G.}~\bibnamefont
  {Styliaris}}, \bibinfo {author} {\bibfnamefont {N.}~\bibnamefont {Anand}},
  \bibinfo {author} {\bibfnamefont {L.}~\bibnamefont {Campos~Venuti}}, \ and\
  \bibinfo {author} {\bibfnamefont {P.}~\bibnamefont {Zanardi}},\ }\href
  {\doibase 10.1103/PhysRevB.100.224204} {\bibfield  {journal} {\bibinfo
  {journal} {Physical Review B}\ }\textbf {\bibinfo {volume} {100}},\ \bibinfo
  {pages} {224204} (\bibinfo {year} {2019})}\BibitemShut {NoStop}%
\bibitem [{\citenamefont {Zanardi}\ \emph
  {et~al.}(2017{\natexlab{b}})\citenamefont {Zanardi}, \citenamefont
  {Styliaris},\ and\ \citenamefont {Campos~Venuti}}]{zanardi_measures_2017}%
  \BibitemOpen
  \bibfield  {author} {\bibinfo {author} {\bibfnamefont {P.}~\bibnamefont
  {Zanardi}}, \bibinfo {author} {\bibfnamefont {G.}~\bibnamefont {Styliaris}},
  \ and\ \bibinfo {author} {\bibfnamefont {L.}~\bibnamefont {Campos~Venuti}},\
  }\href {\doibase 10.1103/PhysRevA.95.052307} {\bibfield  {journal} {\bibinfo
  {journal} {Physical Review A}\ }\textbf {\bibinfo {volume} {95}},\ \bibinfo
  {pages} {052307} (\bibinfo {year} {2017}{\natexlab{b}})}\BibitemShut
  {NoStop}%
\bibitem [{\citenamefont {Durt}\ \emph {et~al.}(2010)\citenamefont {Durt},
  \citenamefont {Englert}, \citenamefont {Bengtsson},\ and\ \citenamefont
  {{\.Z}yczkowski}}]{durt2010mutually}%
  \BibitemOpen
  \bibfield  {author} {\bibinfo {author} {\bibfnamefont {T.}~\bibnamefont
  {Durt}}, \bibinfo {author} {\bibfnamefont {B.-G.}\ \bibnamefont {Englert}},
  \bibinfo {author} {\bibfnamefont {I.}~\bibnamefont {Bengtsson}}, \ and\
  \bibinfo {author} {\bibfnamefont {K.}~\bibnamefont {{\.Z}yczkowski}},\ }\href
  {\doibase 10.1142/S0219749910006502} {\bibfield  {journal} {\bibinfo
  {journal} {International journal of quantum information}\ }\textbf {\bibinfo
  {volume} {8}},\ \bibinfo {pages} {535} (\bibinfo {year} {2010})}\BibitemShut
  {NoStop}%
\bibitem [{\citenamefont {Turner}\ \emph
  {et~al.}(2018{\natexlab{a}})\citenamefont {Turner}, \citenamefont
  {Michailidis}, \citenamefont {Abanin}, \citenamefont {Serbyn},\ and\
  \citenamefont {Papić}}]{turner_weak_2018}%
  \BibitemOpen
  \bibfield  {author} {\bibinfo {author} {\bibfnamefont {C.~J.}\ \bibnamefont
  {Turner}}, \bibinfo {author} {\bibfnamefont {A.~A.}\ \bibnamefont
  {Michailidis}}, \bibinfo {author} {\bibfnamefont {D.~A.}\ \bibnamefont
  {Abanin}}, \bibinfo {author} {\bibfnamefont {M.}~\bibnamefont {Serbyn}}, \
  and\ \bibinfo {author} {\bibfnamefont {Z.}~\bibnamefont {Papić}},\ }\href
  {\doibase 10.1038/s41567-018-0137-5} {\bibfield  {journal} {\bibinfo
  {journal} {Nature Physics}\ }\textbf {\bibinfo {volume} {14}},\ \bibinfo
  {pages} {745} (\bibinfo {year} {2018}{\natexlab{a}})}\BibitemShut {NoStop}%
\bibitem [{\citenamefont {Serbyn}\ \emph {et~al.}(2021)\citenamefont {Serbyn},
  \citenamefont {Abanin},\ and\ \citenamefont {Papić}}]{serbyn_quantum_2021}%
  \BibitemOpen
  \bibfield  {author} {\bibinfo {author} {\bibfnamefont {M.}~\bibnamefont
  {Serbyn}}, \bibinfo {author} {\bibfnamefont {D.~A.}\ \bibnamefont {Abanin}},
  \ and\ \bibinfo {author} {\bibfnamefont {Z.}~\bibnamefont {Papić}},\ }\href
  {\doibase 10.1038/s41567-021-01230-2} {\bibfield  {journal} {\bibinfo
  {journal} {Nature Physics}\ }\textbf {\bibinfo {volume} {17}},\ \bibinfo
  {pages} {675} (\bibinfo {year} {2021})}\BibitemShut {NoStop}%
\bibitem [{\citenamefont {Lidar}\ and\ \citenamefont
  {Brun}(2013)}]{lidar_quantum_2013}%
  \BibitemOpen
  \bibinfo {editor} {\bibfnamefont {D.~A.}\ \bibnamefont {Lidar}}\ and\
  \bibinfo {editor} {\bibfnamefont {T.~A.}\ \bibnamefont {Brun}},\ eds.,\ \href
  {\doibase 10.1017/CBO9781139034807} {\emph {\bibinfo {title} {Quantum {Error}
  {Correction}}}}\ (\bibinfo  {publisher} {Cambridge University Press},\
  \bibinfo {address} {Cambridge},\ \bibinfo {year} {2013})\BibitemShut
  {NoStop}%
\bibitem [{\citenamefont {Breuer}\ and\ \citenamefont
  {Petruccione}(2007)}]{breuer_theory_2007}%
  \BibitemOpen
  \bibfield  {author} {\bibinfo {author} {\bibfnamefont {H.-P.}\ \bibnamefont
  {Breuer}}\ and\ \bibinfo {author} {\bibfnamefont {F.}~\bibnamefont
  {Petruccione}},\ }\href {\doibase 10.1093/acprof:oso/9780199213900.001.0001}
  {{\selectlanguage {eng}\emph {\bibinfo {title} {The {Theory} of {Open}
  {Quantum} {Systems}}}}}\ (\bibinfo  {publisher} {Oxford University Press},\
  \bibinfo {address} {Oxford},\ \bibinfo {year} {2007})\BibitemShut {NoStop}%
\bibitem [{\citenamefont {Am-Shallem}\ \emph {et~al.}(2015)\citenamefont
  {Am-Shallem}, \citenamefont {Levy}, \citenamefont {Schaefer},\ and\
  \citenamefont {Kosloff}}]{am-shallem_three_2015}%
  \BibitemOpen
  \bibfield  {author} {\bibinfo {author} {\bibfnamefont {M.}~\bibnamefont
  {Am-Shallem}}, \bibinfo {author} {\bibfnamefont {A.}~\bibnamefont {Levy}},
  \bibinfo {author} {\bibfnamefont {I.}~\bibnamefont {Schaefer}}, \ and\
  \bibinfo {author} {\bibfnamefont {R.}~\bibnamefont {Kosloff}},\ }\href
  {http://arxiv.org/abs/1510.08634} {\emph {\bibinfo {title} {Three approaches
  for representing {Lindblad} dynamics by a matrix-vector notation}}},\
  \bibinfo {type} {Tech. Rep.}\ \bibinfo {number} {arXiv:1510.08634}\ (\bibinfo
   {institution} {arXiv},\ \bibinfo {year} {2015})\ \bibinfo {note}
  {arXiv:1510.08634 [quant-ph] type: article}\BibitemShut {NoStop}%
\bibitem [{\citenamefont {Bernien}\ \emph {et~al.}(2017)\citenamefont
  {Bernien}, \citenamefont {Schwartz}, \citenamefont {Keesling}, \citenamefont
  {Levine}, \citenamefont {Omran}, \citenamefont {Pichler}, \citenamefont
  {Choi}, \citenamefont {Zibrov}, \citenamefont {Endres}, \citenamefont
  {Greiner}, \citenamefont {Vuletić},\ and\ \citenamefont
  {Lukin}}]{bernien_probing_2017}%
  \BibitemOpen
  \bibfield  {author} {\bibinfo {author} {\bibfnamefont {H.}~\bibnamefont
  {Bernien}}, \bibinfo {author} {\bibfnamefont {S.}~\bibnamefont {Schwartz}},
  \bibinfo {author} {\bibfnamefont {A.}~\bibnamefont {Keesling}}, \bibinfo
  {author} {\bibfnamefont {H.}~\bibnamefont {Levine}}, \bibinfo {author}
  {\bibfnamefont {A.}~\bibnamefont {Omran}}, \bibinfo {author} {\bibfnamefont
  {H.}~\bibnamefont {Pichler}}, \bibinfo {author} {\bibfnamefont
  {S.}~\bibnamefont {Choi}}, \bibinfo {author} {\bibfnamefont {A.~S.}\
  \bibnamefont {Zibrov}}, \bibinfo {author} {\bibfnamefont {M.}~\bibnamefont
  {Endres}}, \bibinfo {author} {\bibfnamefont {M.}~\bibnamefont {Greiner}},
  \bibinfo {author} {\bibfnamefont {V.}~\bibnamefont {Vuletić}}, \ and\
  \bibinfo {author} {\bibfnamefont {M.~D.}\ \bibnamefont {Lukin}},\ }\href
  {\doibase 10.1038/nature24622} {\bibfield  {journal} {\bibinfo  {journal}
  {Nature}\ }\textbf {\bibinfo {volume} {551}},\ \bibinfo {pages} {579}
  (\bibinfo {year} {2017})}\BibitemShut {NoStop}%
\bibitem [{\citenamefont {Lesanovsky}\ and\ \citenamefont
  {Katsura}(2012)}]{lesanovsky_interacting_2012}%
  \BibitemOpen
  \bibfield  {author} {\bibinfo {author} {\bibfnamefont {I.}~\bibnamefont
  {Lesanovsky}}\ and\ \bibinfo {author} {\bibfnamefont {H.}~\bibnamefont
  {Katsura}},\ }\href {\doibase 10.1103/PhysRevA.86.041601} {\bibfield
  {journal} {\bibinfo  {journal} {Phys. Rev. A}\ }\textbf {\bibinfo {volume}
  {86}},\ \bibinfo {pages} {041601} (\bibinfo {year} {2012})}\BibitemShut
  {NoStop}%
\bibitem [{\citenamefont {Saffman}\ \emph {et~al.}(2010)\citenamefont
  {Saffman}, \citenamefont {Walker},\ and\ \citenamefont
  {M\o{}lmer}}]{saffman_quantum_2010}%
  \BibitemOpen
  \bibfield  {author} {\bibinfo {author} {\bibfnamefont {M.}~\bibnamefont
  {Saffman}}, \bibinfo {author} {\bibfnamefont {T.~G.}\ \bibnamefont {Walker}},
  \ and\ \bibinfo {author} {\bibfnamefont {K.}~\bibnamefont {M\o{}lmer}},\
  }\href {\doibase 10.1103/RevModPhys.82.2313} {\bibfield  {journal} {\bibinfo
  {journal} {Rev. Mod. Phys.}\ }\textbf {\bibinfo {volume} {82}},\ \bibinfo
  {pages} {2313} (\bibinfo {year} {2010})}\BibitemShut {NoStop}%
\bibitem [{\citenamefont {Yuan}\ \emph {et~al.}(2022)\citenamefont {Yuan},
  \citenamefont {Zhang}, \citenamefont {Wang}, \citenamefont {Duan},\ and\
  \citenamefont {Deng}}]{yuan_quantum_2022}%
  \BibitemOpen
  \bibfield  {author} {\bibinfo {author} {\bibfnamefont {D.}~\bibnamefont
  {Yuan}}, \bibinfo {author} {\bibfnamefont {S.-Y.}\ \bibnamefont {Zhang}},
  \bibinfo {author} {\bibfnamefont {Y.}~\bibnamefont {Wang}}, \bibinfo {author}
  {\bibfnamefont {L.-M.}\ \bibnamefont {Duan}}, \ and\ \bibinfo {author}
  {\bibfnamefont {D.-L.}\ \bibnamefont {Deng}},\ }\href {\doibase
  10.1103/PhysRevResearch.4.023095} {\bibfield  {journal} {\bibinfo  {journal}
  {Physical Review Research}\ }\textbf {\bibinfo {volume} {4}},\ \bibinfo
  {pages} {023095} (\bibinfo {year} {2022})},\ \bibinfo {note}
  {arXiv:2201.01777 [cond-mat, physics:quant-ph]}\BibitemShut {NoStop}%
\bibitem [{\citenamefont {Turner}\ \emph
  {et~al.}(2018{\natexlab{b}})\citenamefont {Turner}, \citenamefont
  {Michailidis}, \citenamefont {Abanin}, \citenamefont {Serbyn},\ and\
  \citenamefont {Papi\ifmmode~\acute{c}\else
  \'{c}\fi{}}}]{turner_quantum_2018}%
  \BibitemOpen
  \bibfield  {author} {\bibinfo {author} {\bibfnamefont {C.~J.}\ \bibnamefont
  {Turner}}, \bibinfo {author} {\bibfnamefont {A.~A.}\ \bibnamefont
  {Michailidis}}, \bibinfo {author} {\bibfnamefont {D.~A.}\ \bibnamefont
  {Abanin}}, \bibinfo {author} {\bibfnamefont {M.}~\bibnamefont {Serbyn}}, \
  and\ \bibinfo {author} {\bibfnamefont {Z.}~\bibnamefont
  {Papi\ifmmode~\acute{c}\else \'{c}\fi{}}},\ }\href {\doibase
  10.1103/PhysRevB.98.155134} {\bibfield  {journal} {\bibinfo  {journal} {Phys.
  Rev. B}\ }\textbf {\bibinfo {volume} {98}},\ \bibinfo {pages} {155134}
  (\bibinfo {year} {2018}{\natexlab{b}})}\BibitemShut {NoStop}%
\bibitem [{\citenamefont {Deutsch}(1991)}]{deutsch_quantum_1991}%
  \BibitemOpen
  \bibfield  {author} {\bibinfo {author} {\bibfnamefont {J.~M.}\ \bibnamefont
  {Deutsch}},\ }\href {\doibase 10.1103/PhysRevA.43.2046} {\bibfield  {journal}
  {\bibinfo  {journal} {Physical Review A}\ }\textbf {\bibinfo {volume} {43}},\
  \bibinfo {pages} {2046} (\bibinfo {year} {1991})}\BibitemShut {NoStop}%
\bibitem [{\citenamefont {Srednicki}(1994)}]{srednicki_chaos_1994}%
  \BibitemOpen
  \bibfield  {author} {\bibinfo {author} {\bibfnamefont {M.}~\bibnamefont
  {Srednicki}},\ }\href {\doibase 10.1103/PhysRevE.50.888} {\bibfield
  {journal} {\bibinfo  {journal} {Physical Review E}\ }\textbf {\bibinfo
  {volume} {50}},\ \bibinfo {pages} {888} (\bibinfo {year} {1994})}\BibitemShut
  {NoStop}%
\bibitem [{\citenamefont {Fendley}\ \emph {et~al.}(2004)\citenamefont
  {Fendley}, \citenamefont {Sengupta},\ and\ \citenamefont
  {Sachdev}}]{fendley_competing_2004}%
  \BibitemOpen
  \bibfield  {author} {\bibinfo {author} {\bibfnamefont {P.}~\bibnamefont
  {Fendley}}, \bibinfo {author} {\bibfnamefont {K.}~\bibnamefont {Sengupta}}, \
  and\ \bibinfo {author} {\bibfnamefont {S.}~\bibnamefont {Sachdev}},\ }\href
  {\doibase 10.1103/PhysRevB.69.075106} {\bibfield  {journal} {\bibinfo
  {journal} {Physical Review B}\ }\textbf {\bibinfo {volume} {69}},\ \bibinfo
  {pages} {075106} (\bibinfo {year} {2004})}\BibitemShut {NoStop}%
\bibitem [{\citenamefont {Mondragon-Shem}\ \emph {et~al.}(2021)\citenamefont
  {Mondragon-Shem}, \citenamefont {Vavilov},\ and\ \citenamefont
  {Martin}}]{mondragon-shem_fate_2021}%
  \BibitemOpen
  \bibfield  {author} {\bibinfo {author} {\bibfnamefont {I.}~\bibnamefont
  {Mondragon-Shem}}, \bibinfo {author} {\bibfnamefont {M.~G.}\ \bibnamefont
  {Vavilov}}, \ and\ \bibinfo {author} {\bibfnamefont {I.}~\bibnamefont
  {Martin}},\ }\href {\doibase 10.1103/PRXQuantum.2.030349} {\bibfield
  {journal} {\bibinfo  {journal} {PRX Quantum}\ }\textbf {\bibinfo {volume}
  {2}},\ \bibinfo {pages} {030349} (\bibinfo {year} {2021})},\ \bibinfo {note}
  {arXiv:2010.10535 [cond-mat, physics:quant-ph]}\BibitemShut {NoStop}%
\bibitem [{\citenamefont {Zanardi}\ and\ \citenamefont
  {Rasetti}(1997{\natexlab{a}})}]{zanardi_noiseless_1997}%
  \BibitemOpen
  \bibfield  {author} {\bibinfo {author} {\bibfnamefont {P.}~\bibnamefont
  {Zanardi}}\ and\ \bibinfo {author} {\bibfnamefont {M.}~\bibnamefont
  {Rasetti}},\ }\href {\doibase 10.1103/PhysRevLett.79.3306} {\bibfield
  {journal} {\bibinfo  {journal} {Physical Review Letters}\ }\textbf {\bibinfo
  {volume} {79}},\ \bibinfo {pages} {3306} (\bibinfo {year}
  {1997}{\natexlab{a}})}\BibitemShut {NoStop}%
\bibitem [{\citenamefont {Zanardi}\ and\ \citenamefont
  {Rasetti}(1997{\natexlab{b}})}]{zanardi_error_1997}%
  \BibitemOpen
  \bibfield  {author} {\bibinfo {author} {\bibfnamefont {P.}~\bibnamefont
  {Zanardi}}\ and\ \bibinfo {author} {\bibfnamefont {M.}~\bibnamefont
  {Rasetti}},\ }\href {\doibase 10.1142/S0217984997001304} {\bibfield
  {journal} {\bibinfo  {journal} {Modern Physics Letters B}\ }\textbf {\bibinfo
  {volume} {11}},\ \bibinfo {pages} {1085} (\bibinfo {year}
  {1997}{\natexlab{b}})},\ \bibinfo {note} {publisher: World Scientific
  Publishing Co.}\BibitemShut {Stop}%
\bibitem [{\citenamefont {Lidar}\ \emph {et~al.}(1998)\citenamefont {Lidar},
  \citenamefont {Chuang},\ and\ \citenamefont
  {Whaley}}]{lidar_decoherence-free_1998}%
  \BibitemOpen
  \bibfield  {author} {\bibinfo {author} {\bibfnamefont {D.~A.}\ \bibnamefont
  {Lidar}}, \bibinfo {author} {\bibfnamefont {I.~L.}\ \bibnamefont {Chuang}}, \
  and\ \bibinfo {author} {\bibfnamefont {K.~B.}\ \bibnamefont {Whaley}},\
  }\href {\doibase 10.1103/PhysRevLett.81.2594} {\bibfield  {journal} {\bibinfo
   {journal} {Physical Review Letters}\ }\textbf {\bibinfo {volume} {81}},\
  \bibinfo {pages} {2594} (\bibinfo {year} {1998})}\BibitemShut {NoStop}%
\bibitem [{\citenamefont {Moudgalya}\ and\ \citenamefont
  {Motrunich}(2022)}]{moudgalya_hilbert_2021}%
  \BibitemOpen
  \bibfield  {author} {\bibinfo {author} {\bibfnamefont {S.}~\bibnamefont
  {Moudgalya}}\ and\ \bibinfo {author} {\bibfnamefont {O.~I.}\ \bibnamefont
  {Motrunich}},\ }\href {\doibase 10.1103/PhysRevX.12.011050} {\bibfield
  {journal} {\bibinfo  {journal} {Phys. Rev. X}\ }\textbf {\bibinfo {volume}
  {12}},\ \bibinfo {pages} {011050} (\bibinfo {year} {2022})}\BibitemShut
  {NoStop}%
\bibitem [{\citenamefont {Bhatia}(1997)}]{bhatia_matrix_1997}%
  \BibitemOpen
  \bibfield  {author} {\bibinfo {author} {\bibfnamefont {R.}~\bibnamefont
  {Bhatia}},\ }\href {\doibase 10.1007/978-1-4612-0653-8} {{\selectlanguage
  {en}\emph {\bibinfo {title} {Matrix {Analysis}}}}},\ \bibinfo {series}
  {Graduate {Texts} in {Mathematics}}, Vol.\ \bibinfo {volume} {169}\ (\bibinfo
   {publisher} {Springer New York},\ \bibinfo {address} {New York, NY},\
  \bibinfo {year} {1997})\BibitemShut {NoStop}%
\bibitem [{\citenamefont {Pérez-García}\ \emph {et~al.}(2006)\citenamefont
  {Pérez-García}, \citenamefont {Wolf}, \citenamefont {Petz},\ and\
  \citenamefont {Ruskai}}]{perez-garcia_contractivity_2006}%
  \BibitemOpen
  \bibfield  {author} {\bibinfo {author} {\bibfnamefont {D.}~\bibnamefont
  {Pérez-García}}, \bibinfo {author} {\bibfnamefont {M.~M.}\ \bibnamefont
  {Wolf}}, \bibinfo {author} {\bibfnamefont {D.}~\bibnamefont {Petz}}, \ and\
  \bibinfo {author} {\bibfnamefont {M.~B.}\ \bibnamefont {Ruskai}},\ }\href
  {\doibase 10.1063/1.2218675} {\bibfield  {journal} {\bibinfo  {journal}
  {Journal of Mathematical Physics}\ }\textbf {\bibinfo {volume} {47}},\
  \bibinfo {pages} {083506} (\bibinfo {year} {2006})},\ \bibinfo {note}
  {publisher: American Institute of Physics}\BibitemShut {NoStop}%
\bibitem [{\citenamefont {Goodman}\ and\ \citenamefont
  {Wallach}(2009)}]{goodman_symmetry_2009}%
  \BibitemOpen
  \bibfield  {author} {\bibinfo {author} {\bibfnamefont {R.}~\bibnamefont
  {Goodman}}\ and\ \bibinfo {author} {\bibfnamefont {N.~R.}\ \bibnamefont
  {Wallach}},\ }\href {\doibase 10.1007/978-0-387-79852-3} {{\selectlanguage
  {en}\emph {\bibinfo {title} {Symmetry, {Representations}, and
  {Invariants}}}}},\ \bibinfo {series} {Graduate {Texts} in {Mathematics}},
  Vol.\ \bibinfo {volume} {255}\ (\bibinfo  {publisher} {Springer New York},\
  \bibinfo {address} {New York, NY},\ \bibinfo {year} {2009})\BibitemShut
  {NoStop}%
\end{thebibliography}%
\onecolumngrid
\newpage
\appendix
\section{Supplemental Material} \label{append}
\subsection{Proof of \autoref{prop1}}
Note that for a CPTP map $\mathcal{E}(Y)^\dagger = \mathcal{E}(Y^\dagger)$. Then, \cref{eq4} can be rewritten as:
\begin{equation} \label{eqa1}
{G}_{\mathcal{A}}(\mathcal{E})= \frac{1}{d} \, {\mathlarger{\mathbb{E}}}_{X_\mathcal{A},Y_{\mathcal{A}^\prime}} \left[ \Tr \left[ \mathcal{E} (Y_{\mathcal{A}^\prime}^\dagger ) \,  \mathcal{E} (Y_{\mathcal{A}^\prime}) \right] - \Re\left( \Tr \left[ X_{\mathcal{A}} \, \mathcal{E} (Y_{\mathcal{A}^\prime}) \, X_{\mathcal{A}}^\dagger \, \mathcal{E}(Y_{\mathcal{A}^\prime}^\dagger ) \right] \right) \right]
\end{equation}
where $X_\mathcal{A} \in \mathcal{A}$, $Y_{\mathcal{A}^\prime}\in \mathcal{A}^\prime$ and ${\mathlarger{\mathbb{E}}}_{X_\mathcal{A},Y_{\mathcal{A}^\prime}}$ denotes averaging over the Haar measures on the unitary subgroups of operators in $\mathcal{A}$ and $\mathcal{A}^\prime$. Letting $S$ denote the swap operator in the replica space $\mathcal{H}^{\otimes 2}=\mathcal{H} \otimes \mathcal{H}$ and recalling the ``replica trick'':
\begin{equation} \label{eqa2}
\Tr \left[ S \, (M \otimes N ) \right] = \Tr \left[ M N \right]
\end{equation}
we have further
\begin{align}
{G}_{\mathcal{A}}(\mathcal{E})&=\frac{1}{d} \left( \Tr \left[ S \; \mathcal{E}^{\otimes 2} \left( {\mathlarger{\mathbb{E}}}_{Y_{\mathcal{A}^\prime}} \left[ Y_{\mathcal{A}^\prime} \otimes Y_{\mathcal{A}^\prime}^\dagger \right] \right) \right] -\Tr \left[ S \, {\mathlarger{\mathbb{E}}}_{X_\mathcal{A}} \left[ X_{\mathcal{A}} \otimes X_{\mathcal{A}}^\dagger \right]\; \mathcal{E}^{\otimes 2} \left( {\mathlarger{\mathbb{E}}}_{Y_{\mathcal{A}^\prime}} \left[ Y_{\mathcal{A}^\prime} \otimes Y_{\mathcal{A}^\prime}^\dagger \right] \right) \right] \right) \nonumber \\
&\equiv \frac{1}{d} \Tr\left[ S(\mathds{1}_{d^2}-\Omega_{\mathcal{A}})\; \mathcal{E}^{\otimes 2}(\Omega_{\mathcal{A}^\prime})\right] \label{eqa3}
\end{align}
where $\Omega_{\mathcal{A}}:={\mathlarger{\mathbb{E}}}_{X_\mathcal{A}} \left[ X_{\mathcal{A}} \otimes X_{\mathcal{A}}^\dagger \right]$, $\Omega_{\mathcal{A}^\prime}:={\mathlarger{\mathbb{E}}}_{Y_{\mathcal{A}^\prime}} \left[ Y_{\mathcal{A}^\prime} \otimes Y_{\mathcal{A}^\prime}^\dagger \right] $. We now use the following result:
\begin{align}
{\mathlarger{\mathbb{E}}}_{Q} \, \left[ Q \otimes Q^\dagger \right] = \frac{S}{d} \label{eqa4}
\end{align}
A proof of this result is as follows. Left invariance of the Haar measure implies that for any linear operators $M,U \in \mathcal{L}(\mathcal{H})$ where $U$ is unitary, we have that $\left[ U, {\mathlarger{\mathbb{E}}}_{Q} \, QMQ^\dagger \right] = 0$ and as a consequence of Schur's lemma:
\begin{align}
{\mathlarger{\mathbb{E}}}_{Q} \, QMQ^\dagger = \frac{\Tr [M]}{d} \mathds{1}_d \label{eqa5}
\end{align}
By direct computation one can also show that:
\begin{align}
&\Tr_2 \left[ S \; (M\otimes \mathds{1}_d )\; S \right] = \Tr_2\left[ \mathds{1}_d \otimes M \right] = \Tr [M] \; \mathds{1}_d  \label{eqa6} \\
&\Tr_2 \left[ (Q \otimes Q^\dagger ) (M\otimes \mathds{1}_d)\; S \right] = QMQ^\dagger \label{eqa7}
\end{align}
where $\Tr_2$ denotes the partial trace over the second copy of $\mathcal{H}\otimes\mathcal{H}$. \cref{eqa4} then follows by combining \cref{eqa5}, (\ref{eqa6}), (\ref{eqa7}). Note that left invariance of the Haar measure also implies that:
\begin{align}
{\mathlarger{\mathbb{E}}}_{Q} \, Q = 0 \label{eqa8}
\end{align}
In our case, we have $X_\mathcal{A}\in \mathcal{A}$, which means that $X_\mathcal{A}=\oplus_{J=1}^{d_Z} \mathds{1}_{n_J} \otimes X_{d_J}$. So:
\begin{align}
\Omega_\mathcal{A}&= {\mathlarger{\mathbb{E}}}_{X_\mathcal{A}} \left[ X_\mathcal{A} \otimes X_\mathcal{A}^\dagger \right] =
{\mathlarger{\mathbb{E}}}_{X_\mathcal{A}}  \left[ \oplus_{J,J^\prime=1}^{d_Z} \mathds{1}_{n_J} \otimes X_{d_J} \otimes \mathds{1}_{n_{J^\prime}} \otimes X_{d_{J^\prime}}^\dagger \right] \nonumber \\
&= \oplus_{J=1}^{d_Z} \; {\mathlarger{\mathbb{E}}}_{X_{d_J}} \left[ \mathds{1}_{n_J} \otimes X_{d_J} \otimes \mathds{1}_{n_J} \otimes X_{d_{J}}^\dagger \right] \oplus_{J\neq J^\prime=1}^{d_Z} \; {\mathlarger{\mathbb{E}}}_{X_{d_J},X_{d_{J^\prime}}} \left[ \mathds{1}_{n_J} \otimes X_{d_J} \otimes \mathds{1}_{n_{J^\prime}} \otimes X_{d_{J^\prime}}^\dagger \right]  \nonumber \\
&\cong \oplus_{J=1}^{d_Z} \mathds{1}_{n_J}^{\otimes 2} \; {\mathlarger{\mathbb{E}}}_{X_{d_J}} \left[ X_{d_J} \otimes X_{d_{J}}^\dagger \right] \oplus_{J\neq J^\prime=1}^{d_Z} \mathds{1}_{n_J} \otimes \mathds{1}_{n_{J^\prime}} \; {\mathlarger{\mathbb{E}}}_{X_{d_{J^\prime}}} \left[\left( {\mathlarger{\mathbb{E}}}_{X_{d_{J}}} X_{d_J} \right) \otimes X_{d_{J^\prime}}^\dagger  \right] \nonumber \\
&\mathrel{\overset{\smash{\scriptscriptstyle (\ref{eqa4}),(\ref{eqa8})}}{=\joinrel = \joinrel =}}\oplus_{J=1}^{d_Z} \mathds{1}_{n_J}^{\otimes 2} \otimes \frac{S_{d_J}}{d_J} \label{eqa9}
\end{align}
By virtue of the structure theorem \cref{eq3} we can choose the following orthogonal basis of $\mathcal{A}$
\begin{align}
e_\alpha = \frac{\mathds{1}_{n_J}}{\sqrt{d_J}} \otimes \ket{k}\bra{l}, \quad \alpha:=(J,l,m),\;  l,m=1,\dots ,d_J, \; J=1,\dots ,d_{\mathcal{Z}}\label{eqa10}
\end{align}
Then,
\begin{align}
\sum_{\alpha =1}^{d(\mathcal{A})} e_\alpha \otimes e_\alpha^\dagger = \oplus_{J=1}^{d_\mathcal{Z}} \sum_{k,l=1}^{d_J} \frac{\mathds{1}_{n_J}}{\sqrt{d_J}} \otimes \ket{k}\bra{l} \otimes \frac{\mathds{1}_{n_J}}{\sqrt{d_J}} \otimes \ket{l}\bra{k}\cong \oplus_{J=1}^{d_\mathcal{Z}} \mathds{1}_{n_J}^{\otimes 2} \otimes \frac{S_{d_J}}{d_J} \label{eqa11}
\end{align}
Comparing \cref{eqa9}, (\ref{eqa11}) we get 
\begin{align}
\Omega_\mathcal{A} =\sum_{\alpha =1}^{d(\mathcal{A})} e_\alpha \otimes e_\alpha^\dagger\cong\oplus_{J=1}^{d_\mathcal{Z}} \mathds{1}_{n_J}^{\otimes 2} \otimes \frac{S_{d_J}}{d_J} \label{eqa12}
\end{align}
Similarly, 
\begin{align}
\Omega_\mathcal{A^\prime} = \sum_{\gamma =1}^{d(\mathcal{A^\prime})} f_\gamma \otimes f_\gamma^\dagger\cong\oplus_{J=1}^{d_\mathcal{Z}} \frac{S_{n_J}}{n_J} \otimes \mathds{1}_{d_J}^{\otimes 2} \label{eqa13}
\end{align}
where $f_\gamma$ is an orthogonal basis of $\mathcal{A}^\prime$ given as
\begin{align}
f_\gamma = \ket{p}\bra{q} \otimes \frac{\mathds{1}_{d_J}}{\sqrt{n_J}}, \quad \gamma:=(J,p,q),\;  p,q=1,\dots ,n_J, \; J=1,\dots ,d_{\mathcal{Z}} \label{eqa14}
\end{align}
Note that the orthogonal bases in \cref{eqa10}, (\ref{eqa13}) are defined up to unitary transformations and are suitable for expressing the projectors on $\mathcal{A}^\prime$, $\mathcal{A}$ in an OSR as $\mathbb{P}_{\mathcal{A}^\prime} [\bullet ] = \sum_{\alpha =1}^{d(\mathcal{A} )}e_\alpha [\bullet ] e_\alpha^\dagger$, $\mathbb{P}_{\mathcal{A}} [\bullet ] = \sum_{\gamma =1}^{d(\mathcal{A}^\prime )}f_\gamma [\bullet ] f_\gamma^\dagger$.
\subsection{Proof of \autoref{cor1}}
Using \cref{eqa12} we have
\begin{align}
\Omega_\mathcal{A} \cong \oplus_{J=1}^{d_\mathcal{Z}} \mathds{1}_{n_J}^{\otimes 2} \otimes \frac{S_{d_J}}{d_J} = S \, \oplus_{J=1}^{d_\mathcal{Z}} \frac{S_{n_J}}{d_J} \otimes \mathds{1}_{d_J}^{\otimes 2} \cong S \,\sum_{\gamma =1}^{d(\mathcal{A^\prime})} \tilde{f}_\gamma \otimes \tilde{f}_\gamma^\dagger \label{eqa15}
\end{align}
where $\tilde{f}_\gamma := \frac{f_\gamma}{\lVert f_\gamma \rVert_2}$. Then, from \cref{eq5} we have
\begin{align}
{G}_{\mathcal{A}} (\mathcal{E} ) &=\frac{1}{d} \Tr \left[ S \left( 1 - S \sum_{\gamma =1}^{d({\mathcal{A}^\prime})} \tilde{f}_\gamma \otimes \tilde{f}_\gamma^\dagger \right) \sum_{\gamma^\prime =1}^{d({\mathcal{A}^\prime})} \mathcal{E} (f_{\gamma^\prime} ) \otimes \mathcal{E} (f_{\gamma^\prime}^\dagger ) \right] \nonumber\\
&= \frac{1}{d} \sum_{\gamma^\prime =1}^{d({\mathcal{A}^\prime})} \left( \Tr \left[ S \, \mathcal{E} (f_{\gamma^\prime}) \otimes \mathcal{E} (f_{\gamma^\prime}^\dagger ) \right] - \sum_{\gamma =1}^{d({\mathcal{A}^\prime})} \Tr \left[ \tilde{f}_\gamma \, \mathcal{E} (f_{\gamma^\prime}) \otimes \tilde{f}_\gamma^\dagger \, \mathcal{E} (f_{\gamma^\prime}^\dagger) \right]\right) \nonumber\\
&=\frac{1}{d}\sum_{\gamma^\prime =1}^{d(\mathcal{A}^\prime )} \lVert \mathcal{E} (f_{\gamma^\prime} ) \rVert_2^2 - \frac{1}{d} \sum_{\gamma , \gamma^\prime =1}^{d(\mathcal{A}^\prime )}  
\left\lvert \langle \tilde{f}_\gamma^\dagger , \mathcal{E} (f_{\gamma^\prime })\rangle\right\rvert^2 \label{eqa16}
\end{align}
\subsection{Proof of \autoref{prop2}}
Using the definition \cref{eq4}:
\begin{align}
{G}_{\mathcal{A}}(\mathcal{E})=0 &\Leftrightarrow {\mathlarger{\mathbb{E}}}_{X_\mathcal{A},Y_{\mathcal{A}^\prime}}\left\lVert \left[ X_\mathcal{A}, \mathcal{E}(Y_{\mathcal{A}^\prime}) \right] \right\rVert_2^2 =0 \Leftrightarrow \lVert \left[ X , \mathcal{E}(Y) \right] \rVert_2^2 = 0 \nonumber \\
&\Leftrightarrow \left[ X , \mathcal{E}(Y) \right] = 0 \; \; \forall \; X\in U(\mathcal{A}), \; Y\in U(\mathcal{A}^\prime) \label{eqa17}
\end{align}
Since every finite dimensional $C^*$-algebra is $*$-isomorphic to the direct sum of full matrix algebras \cite{davidson_c-algebras_1996}, it follows that one can always find a unitary basis for $\mathcal{A}$, $\mathcal{A}^\prime$, thus \cref{eqa17} is equivalent to
\begin{align}
\begin{aligned}\label{eqa18}
{G}_{\mathcal{A}}(\mathcal{E})=0 &\Leftrightarrow \left[ M , \mathcal{E}(N) \right] = 0 \; \; \forall \; M \in \mathcal{A},N \in \mathcal{A}^\prime \\
&\Leftrightarrow \mathcal{E}(\mathcal{A}^\prime ) \subseteq \mathcal{A}^\prime
\end{aligned}
\end{align}
\subsection{Proof of \autoref{prop3}}
Using \cref{eqa12}, (\ref{eqa13}) in \cref{eq5} we have
\begin{align}
{G}_{\mathcal{A}}(\mathcal{E})&=\frac{1}{d} \sum_{\gamma = 1}^{d(\mathcal{A}^\prime )} \Tr \left[ S \left( \mathds{1}_{d^2} - \sum_{\alpha =1}^{d(\mathcal{A})} e_\alpha \otimes e_\alpha^\dagger \right) \mathcal{E}^{\otimes 2} \left( \sum_{\gamma =1}^{d(\mathcal{A}^\prime )} f_\gamma \otimes f_\gamma^\dagger \right) \right] \nonumber\\
&=\frac{1}{d}  \sum_{\gamma = 1}^{d(\mathcal{A}^\prime )} \left( \Tr \left[S \; \mathcal{E} (f_\gamma ) \otimes \mathcal{E} (f_\gamma ^\dagger ) \right] - \sum_{\alpha =1}^{d(\mathcal{A} )} \Tr \left[S \; e_\alpha \mathcal{E}(f_\gamma ) \otimes e_\alpha^\dagger \mathcal{E} (f_\gamma)^\dagger \right] \right) \nonumber \\
&= \frac{1}{d}  \sum_{\gamma = 1}^{d(\mathcal{A}^\prime )} \left(  \langle \mathcal{E} (f_\gamma ), \mathcal{E} (f_\gamma ) \rangle -  \langle \mathcal{E} (f_\gamma ), \mathbb{P}_{\mathcal{A}^\prime }\, \mathcal{E} (f_\gamma ) \rangle \right) \label{eqa19}
\end{align}
Since $\mathbb{P}_{\mathcal{A}^\prime}$ is an orthogonal projector $\langle (\mathcal{I} - \mathbb{P}_{\mathcal{A}^\prime} ) \, \mathcal{E} (f_\gamma ), \mathbb{P}_{\mathcal{A}^\prime }\, \mathcal{E} (f_\gamma ) \rangle = 0$ and \cref{eqa19} becomes
\begin{align}
{G}_{\mathcal{A}}(\mathcal{E})=\frac{1}{d} \sum_{\gamma =1}^{d(\mathcal{A}^\prime )} \left( \lVert \mathcal{E}(f_\gamma ) \rVert_2^2 -\lVert \mathbb{P}_{\mathcal{A}^\prime}\, \mathcal{E} (f_\gamma )\rVert_2^2 \right) \label{eqa20}
\end{align}
On the other hand \cref{eqa19} also yields
\begin{align}
{G}_{\mathcal{A}}(\mathcal{E}) &= \frac{1}{d} \sum_{\gamma = 1}^{d(\mathcal{A}^\prime )} \langle \mathcal{E} (f_\gamma ) , \left( \mathcal{I} - \mathbb{P}_{\mathcal{A}^\prime} \right) \, \mathcal{E} (f_\gamma ) \rangle \nonumber \\
&= \frac{1}{d} \sum_{\gamma=1}^{d(\mathcal{A}^\prime )} \lVert (\mathcal{I}-\mathbb{P}_{\mathcal{A}^\prime }) \, \mathcal{E} (f_\gamma ) \rVert_2^2 \label{eqa21}
\end{align}
\subsection{Proof of \autoref{cor2}}
From \cref{eqa20} we have
\begin{align}
{G}_{\mathcal{A}}^{(2)}(\mathcal{E}) &= \frac{1}{d} \sum_{\gamma =1}^{d(\mathcal{A}^\prime )} \lVert \mathbb{P}_{\mathcal{A}^\prime}\, \mathcal{E} (f_\gamma )\rVert_2^2 = \frac{1}{d}  \sum_{\gamma = 1}^{d(\mathcal{A}^\prime )}\langle \mathcal{E} (f_\gamma ), \mathbb{P}_{\mathcal{A}^\prime }\, \mathcal{E} (f_\gamma ) \rangle \nonumber \\
&= \frac{1}{d} \sum_{\gamma =1}^{d(\mathcal{A}^\prime )}  \sum_{\alpha =1}^{d(\mathcal{A} )} \Tr \left[ \mathcal{E} (f_\gamma )^\dagger \, e_\alpha^\dagger \, \mathcal{E} (f_\gamma ) \, e_\alpha \right] \label{eqa22}
\end{align}
\subsection{Proof of \autoref{prop4}}
Recall that any unital, positive, trace-preserving map $\mathcal{T}$ is contractive for the $p$-norm \footnote{The (Schatten) $p$-norm \cite{bhatia_matrix_1997} is defined as $\lVert X \rVert_p := \left( \sum_i s_i^k \right)^{1/k}$, where $\{s_i\}_i$ are the singular values of $X$.} $\forall \; p\in \left[ 1,\infty \right]$ in the sense that $\sup_X \frac{\lVert \mathcal{T} (X) \rVert_p}{\lVert X \rVert_p} \leq 1$ \cite{perez-garcia_contractivity_2006}. Since $\mathcal{E}$ is a unital CPTP map, this in particular implies that 
\begin{equation}
\lVert \mathcal{E} (X) \rVert_2 \leq \lVert X \rVert_2 \; \forall \; X\in \mathcal{L}(\mathcal{H}) \label{eqa23}
\end{equation}
Moreover, as a direct consequence of $\left\lVert X - \frac{\Tr X}{d} \mathds{1}_d \right\rVert_2^2 \geq 0$ we have
\begin{equation}
\lVert X \rVert_2^2 \geq \frac{\left\lvert \Tr X \right\rvert^2}{d} \quad  \forall \; X\in\mathcal{L}(\mathcal{H}) \label{eqa24}
\end{equation}
Finally, recall that the Cauchy-Schwarz inequality implies that $d^2 \leq d(\mathcal{A}^\prime) \, d(\mathcal{A})$. Using the above observations in \cref{eqa20}
\begin{align}
{G}_{\mathcal{A}}(\mathcal{E}) &\leq \frac{1}{d} \sum_{\gamma = 1}^{d(\mathcal{A}^\prime )} \lVert f_\gamma \rVert_2^2 - \frac{1}{d} \sum_{\gamma = 1}^{d(\mathcal{A}^\prime )}\frac{\left\lvert \Tr \left[ \mathbb{P}_{\mathcal{A}^\prime} \, \mathcal{E} (f_\gamma ) \right] \right\rvert^2}{d} \nonumber\\
&= \frac{1}{d} \sum_{\gamma = 1}^{d(\mathcal{A}^\prime )}  \Tr [f_\gamma^\dagger f_\gamma ] - \frac{1}{d} \sum_{\gamma = 1}^{d(\mathcal{A}^\prime )} \frac{\left\lvert \Tr \left[ \sum_{\alpha =1}^{d(\mathcal{A})} e_\alpha \mathcal{E} (f_\gamma ) e_\alpha^\dagger \right] \right\rvert^2}{d} \nonumber\\
&= \frac{1}{d} \Tr [ \sum_{J=1}^{d_\mathcal{Z}}\sum_{p,q = 1}^{n_J}\frac{1}{\sqrt{n_J}}f_{(Jqq)}] -  \frac{1}{d} \sum_{\gamma = 1}^{d(\mathcal{A}^\prime )} \frac{\left\lvert \Tr \left[ \sum_{J=1}^{d_{\mathcal{Z}}}\sum_{k,l =1}^{d_J}  \mathcal{E} (f_\gamma ) \frac{1}{\sqrt{d_{J}}} e_{(Jll)} \right] \right\rvert^2}{d} \nonumber\\
&= \frac{1}{d} \Tr [\mathds{1}_d] - \frac{1}{d} \sum_{\gamma = 1}^{d(\mathcal{A}^\prime )} \frac{\left\lvert \Tr \left[  \mathcal{E} (f_\gamma ) \mathds{1}_d \right] \right\rvert^2}{d} = 1 - \frac{1}{d} \sum_{\gamma = 1}^{d(\mathcal{A}^\prime )} \frac{\left\lvert \Tr f_\gamma \right\rvert^2}{d} = 1- \frac{d(\mathcal{A} )}{d^2} \leq 1 - \frac{1}{d(\mathcal{A}^\prime )} \label{eqa25}
\end{align}
On the other hand consider a unitary transformation $\hat{e}_\beta := \sum_{\alpha =1}^{d( \mathcal{A} )} (U)_{\alpha \beta} \, e_\alpha$ of the basis in \cref{eqa10}, given by a unitary $U$ such that $(U)_{(Jkl)1}=\sqrt{\frac{d_J}{d( \mathcal{A} )}} \delta_{k l}$. This is a valid choice since $\sum_{(Jkl)} (U^\dagger )_{1(Jkl)} (U)_{(Jkl)1}=1$, as it should for a unitary matrix. Then, 
\begin{align}
\hat{e}_1=\sum_{J=1}^{d_Z} \sum_{k,l=1}^{d_J} \sqrt{\frac{d_J}{d( \mathcal{A} )}} \delta_{k l} \frac{\mathds{1}_{n_J}}{\sqrt{d_J}} \otimes \ket{k}\bra{l} =\frac{\mathds{1}_{d}}{\sqrt{d(\mathcal{A} )}} \label{eqa26}
\end{align}
Moreover, expressing
\begin{equation*}
\mathds{1}_{n_J}=\sum_{n^{(J)}=1}^{n_J} \ket{n^{(J)}}\bra{n^{(J)}}, \; \mathds{1}_{d_J}=\sum_{d^{(J)}=1}^{d_J} \ket{d^{(J)}}\bra{d^{(J)}}, \; \mathcal{E} [\bullet ] = \sum_\delta K_\delta \bullet  K_\delta^\dagger
\end{equation*} and using \cref{eqa10}, (\ref{eqa14}) one can find after some algebra that
\begin{align*}
\begin{split}
\sum_{\gamma =1}^{d(\mathcal{A}^\prime)}& \Tr[ \hat{e}_\beta \mathcal{E}(f_\gamma ) \hat{e}_\beta^\dagger \mathcal{E} (f_\gamma^\dagger)] = \\
&= \sum_{J=1}^{d_\mathcal{Z}} \sum_{\delta , \delta^\prime} \sum_{d^{(J)},{d^\prime}^{(J)}=1}^{d_J} \frac{1}{n_J} \left\lvert \sum_{J^\prime =1}^{d_{\mathcal{Z}}}\sum_{k,l=1}^{d(\mathcal{A})} \sum_{p =1}^{d(\mathcal{A}^\prime)} \sum_{n^{(J^\prime)}=1}^{n_J} \frac{(U)_{(J^\prime kl) \beta}}{\sqrt{d_{J^\prime}}} \mel{n^{(J^\prime)},l}{K_{\delta}}{p,{d}^{(J)}} \mel{p,{d^\prime}^{(J)}}{K_{\delta^\prime}^\dagger}{n^{(J^\prime)},k} \right\rvert^2 \geq 0 \quad \forall \beta
\end{split}
\end{align*}
Then, using \cref{eqa20}, (\ref{eqa22}) with the basis $\hat{e}_\beta$ and \cref{eqa23} we obtain
\begin{align}
{G}_{\mathcal{A}}(\mathcal{E} ) &\leq \frac{1}{d} \sum_{\gamma =1}^{d(\mathcal{A}^\prime )} \lVert \mathcal{E} (f_\gamma ) \rVert_2^2 - \frac{1}{d} \sum_{\gamma =1} Tr[ \mathcal{E}(f_\gamma )^\dagger \hat{e}_1^\dagger \mathcal{E} (f_\gamma ) \hat{e}_1 ]  \nonumber \\
&=\frac{1}{d} \sum_{\gamma =1}^{d(\mathcal{A}^\prime )} \lVert \mathcal{E} (f_\gamma ) \rVert_2^2 - \frac{1}{d \, d(\mathcal{A} )} \sum_{\gamma =1} \Tr[ \mathcal{E} (f_\gamma )^\dagger \mathcal{E}(f_\gamma )] = \left( \frac{1}{d} - \frac{1}{d \, d(\mathcal{A} )} \right) \sum_{\gamma =1}^{d(\mathcal{A}^\prime)} \lVert \mathcal{E} (f_\gamma ) \rVert_2^2 \nonumber \\
&\leq \left( \frac{1}{d} - \frac{1}{d \, d(\mathcal{A} )} \right) \sum_{\gamma =1}^{d(\mathcal{A}^\prime)} \lVert f_\gamma \rVert_2^2 = 1 - \frac{1}{d(\mathcal{A})} \label{eqa27}
\end{align}
From \cref{eqa25}, (\ref{eqa27}) it follows that ${G}_{\mathcal{A}}(\mathcal{E}) \leq \min \left\{ 1-\frac{1}{d(\mathcal{A})}, \; 1- \frac{1}{d(\mathcal{A}^\prime)} \right\}$.
\subsection{Proof of \autoref{prop5}}
This follows from the fact that $[X_\mathcal{A},\mathcal{U}(Y_{\mathcal{A}^\prime})]= - \mathcal{U} \left( [Y_{\mathcal{A}^\prime},\mathcal{U}^\dagger (X_\mathcal{A} ) ] \right)$, the unitary invariance of the 2-norm and the double commutant theorem. Using the definition \cref{eq4}
\begin{align}
{G}_{\mathcal{A}}(\mathcal{U}) :&= \frac{1}{2d} \; {\mathlarger{\mathbb{E}}}_{X_\mathcal{A},Y_{\mathcal{A}^\prime}}\left\lVert \left[ X_\mathcal{A}, \mathcal{U}(Y_{\mathcal{A}^\prime}) \right] \right\rVert_2^2 =\frac{1}{2d} \; {\mathlarger{\mathbb{E}}}_{X_\mathcal{A},Y_{\mathcal{A}^\prime}}\left\lVert  - \mathcal{U} \left( [Y_{\mathcal{A}^\prime},\mathcal{U}^\dagger (X_\mathcal{A}) ] \right) \right\rVert_2^2 \nonumber\\
&=\frac{1}{2d} \; {\mathlarger{\mathbb{E}}}_{X_\mathcal{A},Y_{\mathcal{A}^\prime}}\left\lVert  [Y_{\mathcal{A}^\prime},\mathcal{U}^\dagger (X_\mathcal{A}) ] \right\rVert_2^2 = {G}_{\mathcal{A}^\prime}(\mathcal{U}^\dagger) \label{eqa28}
\end{align}
\subsection{Proof of \autoref{prop6}}
Note from \cref{eqa10} that $\{e_\alpha \}$ is $\dagger$-closed, which implies that $S \, \Omega_\mathcal{A} \, S = \Omega_\mathcal{A}^\dagger =\Omega_\mathcal{A} \Leftrightarrow [S, \Omega_\mathcal{A} ] =0$. Also, clearly $[S, U^{\otimes 2} ] =0$. Finally, for the collinear case and using \cref{eqa13}, (\ref{eqa15}) we have 
\begin{equation*}
\Omega_{\mathcal{A}^\prime} = \oplus_{J=1}^{d_\mathcal{Z}} \mathds{1}_{d_J}^{\otimes 2} \otimes \frac{S_{n_J}}{n_J}=\oplus_{J=1}^{d_\mathcal{Z}} \lambda \, \mathds{1}_{d_J}^{\otimes 2} \otimes \frac{S_{n_J}}{d_J} = \frac{d}{d(\mathcal{A}^\prime)} \, S \, \Omega_\mathcal{A}
\end{equation*}
Then, using (\ref{eqa3}) we find
\begin{align}
{G}_{\mathcal{A}}(\mathcal{U})&=\frac{1}{d} \Tr\left[ S(\mathds{1}_{d^2}-\Omega_{\mathcal{A}})\; \mathcal{U}^{\otimes 2}(\Omega_{\mathcal{A}^\prime})\right] \nonumber \\
&=\frac{1}{d} \Tr \left[ S \,(\mathds{1}_d-\Omega_\mathcal{A})\; \mathcal{U}^{\otimes 2} \left( \frac{d}{d(\mathcal{A}^\prime)} \, S\, \Omega_{\mathcal{A}} \right)\right] \nonumber \\
&=\frac{1}{d(\mathcal{A}^\prime)} \left( \Tr[ \mathcal{U}^{\otimes 2} ( \Omega_{\mathcal{A}}) ] - \Tr[ \Omega_{\mathcal{A}} \, \mathcal{U}^{\otimes 2} (\Omega_\mathcal{A} ) ] \right) \nonumber \\
&= 1-\frac{1}{d(\mathcal{A}^\prime)} \left\langle \Omega_{\mathcal{A}}, \mathcal{U}^{\otimes 2} ( \Omega_{\mathcal{A}} )\right\rangle \label{eqa29}
\end{align}
where in the last line we also used that from \cref{eqa12}
\begin{equation*}
\Tr \left[ \mathcal{U}^{\otimes 2}(\Omega_{\mathcal{A}} ) \right] =\Tr \Omega_{\mathcal{A}} = \Tr[ \oplus_{J=1}^{d_\mathcal{Z}} \mathds{1}_{n_J}^{\otimes 2} \otimes \frac{S_{d_J}}{d_J} ] = \sum_{J=1}^{d_\mathcal{Z}} n_J^2 = d(\mathcal{A}^\prime)
\end{equation*}
The last expression in \cref{eqa29} coincides with Eq. (4) for the GAAC in Ref. \cite{zanardi_quantum_nodate}.
\subsection{Proof of \autoref{prop7}}
By Schur-Weyl duality the commutant of the algebra $\mathcal{K}$ generated by $\{ M^{\otimes 2} \, | \, M \in \mathcal{L}(\mathcal{H}) \}$ is $\mathcal{K}^\prime=\mathbf{C}S_2$, where $S_2=\{\mathds{1}_d, \, S\}$ is the symmetric group over the copies in $\mathcal{H} \otimes \mathcal{H}$ \cite{goodman_symmetry_2009}. Since we can always find a unitary basis of $\mathcal{L}(\mathcal{H})$, it follows that $\mathcal{K}$ is equivalently generated by $\{ U^{\otimes 2} \, | \, U \in \mathcal{L} (\mathcal{H} ), \, U\,U^\dagger = \mathds{1}_d \}$. Also, note that $\mathbb{P}_{\mathcal{K}^\prime} [\bullet ] := \overline{\mathcal{U}^{\otimes 2} \bullet }^{\, \mathcal{U}}\equiv \overline{U^{\otimes 2} [\bullet ] {U^\dagger}^{\otimes 2}}^{\, {U}}$ is an orthogonal projector on $\mathcal{K^\prime}$ \footnote{It is not hard to check that $\mathbb{P}_{\mathcal{K}^\prime}^\dagger = \mathbb{P}_{\mathcal{K}^\prime}$, $\mathbb{P}_{\mathcal{K}^\prime}^2=\mathbb{P}_{\mathcal{K}^\prime}$, $\{\mathbb{P}_{\mathcal{K}^\prime}(M)\, | \, M \in \mathcal{L}(\mathcal{H}^{\otimes 2})\} \subseteq \mathcal{K}^\prime$, $\mathbb{P}_{\mathcal{K}^\prime}(\mathds{1}_d)=\mathds{1}_d$, $\mathbb{P}_{\mathcal{K}^\prime}(S)=S$.}. So, we can express $\mathbb{P}_{\mathcal{K}^\prime}$ in terms of the orthonormal basis $\left\{ \frac{\mathds{1}_d +S}{\sqrt{2d(d+1)}},\frac{\mathds{1}_d -S}{\sqrt{2d(d-1)}} \right\}$ of $\mathbf{C}S_2$:
\begin{align}
\mathbb{P}_{\mathcal{K}^\prime} [\bullet]= \overline{\mathcal{U}^{\otimes 2} (\cdot )}^{\, \mathcal{U}} = \sum_{\eta = \pm 1} \frac{\mathds{1}_d + \eta \, S }{2d (d+ \eta )} \langle \mathds{1}_d + \eta \, S , \bullet \rangle \label{eqa30}
\end{align}
Now, using \cref{eqa3}, (\ref{eqa12}), (\ref{eqa13}), (\ref{eqa30}) we have
\begin{align}
\overline{{G}_{\mathcal{A}} (\mathcal{U} )}^{\, \mathcal{U}}&=\frac{1}{d} \Tr\left[ S(\mathds{1}_{d^2}-\Omega_{\mathcal{A}})\; \overline{\mathcal{U}^{\otimes 2}(\Omega_{\mathcal{A}^\prime})}^{\, \mathcal{U}}\right] \nonumber \\
&= \frac{1}{d} \Tr\left[ S(\mathds{1}_{d^2}-\Omega_{\mathcal{A}})\; \sum_{\eta = \pm 1} \frac{\mathds{1}_d + \eta \, S }{2d (d+ \eta )} \langle \mathds{1}_d + \eta \, S , \sum_{\gamma =1}^{d(\mathcal{A}^\prime )} f_\gamma \otimes f_\gamma^\dagger \rangle \right] \nonumber \\
&=\frac{1}{d} \Tr\left[ S(\mathds{1}_{d^2}-\Omega_{\mathcal{A}})\; \sum_{\eta = \pm 1} \frac{\mathds{1}_d + \eta \, S }{2d (d+ \eta )} \sum_{\gamma =1}^{d(\mathcal{A}^\prime )} \Tr [ f_\gamma \otimes f_\gamma^\dagger ] + \eta \Tr [S ( f_\gamma \otimes f_\gamma^\dagger )] \right] \nonumber \\
&= \frac{1}{d} \Tr\left[ S(\mathds{1}_{d^2}-\Omega_{\mathcal{A}})\; \sum_{\eta = \pm 1} \frac{\mathds{1}_d + \eta \, S }{2d (d+ \eta )} \sum_{\gamma =1}^{d(\mathcal{A}^\prime )} \left\lvert \Tr f_\gamma \right\rvert^2 + \eta \Tr [f_\gamma f_\gamma^\dagger ] \right] \nonumber \\
&=\frac{1}{d} \Tr\left[ S(\mathds{1}_{d^2}-\Omega_{\mathcal{A}})\; \sum_{\eta = \pm 1} \frac{\mathds{1}_d + \eta \, S }{2d (d+ \eta )} \left( d(\mathcal{A} ) + \eta \, d \right) \right] \nonumber \\
&=\sum_{\eta=\pm 1} \frac{d(\mathcal{A} ) + \eta \, d}{2d^2 (d+\eta )} \Tr \left[ S+\eta \, \mathds{1}_d - S \, \Omega_{\mathcal{A}} - \eta \, S \, \Omega_{\mathcal{A}} \, S \right] \nonumber \\
&= \sum_{\eta=\pm 1} \frac{d(\mathcal{A} ) + \eta \, d}{2d^2 (d+\eta )} \left( d + \eta \, d^2 - \Tr \left[ S \sum_{\alpha =1}^{d(\mathcal{A} )} e_\alpha \otimes e_\alpha^\dagger  \right] - \eta \Tr \left[ \sum_{\alpha =1}^{d(\mathcal{A} )} e_\alpha \otimes e_\alpha^\dagger\right] \right) \nonumber \\
&= \sum_{\eta=\pm 1} \frac{d(\mathcal{A} ) + \eta \, d}{2d^2 (d+\eta )} \left( d + \eta \, d^2 - \sum_{\alpha =1}^{d(\mathcal{A} )} \Tr \left[ e_\alpha e_\alpha^\dagger  \right] - \eta \sum_{\alpha =1}^{d(\mathcal{A} )} \lvert \Tr e_\alpha \rvert^2 \right) \nonumber \\
&= \sum_{\eta=\pm 1} \frac{(d(\mathcal{A} ) + \eta \, d)}{2d^2 (d+\eta )} \, \eta \, \left( d^2 - d(\mathcal{A}^\prime ) \right) = \frac{\left( d^2 - d( \mathcal{A} ) \right) \left( d^2 - d( \mathcal{A}^\prime ) \right)}{d^2 (d^2 -1)} \label{eqa31}
\end{align}
\subsection{\autoref{exmp1} calculations}
Notice that since $M$ is a unitary involution $(\text{Ad}M)^2=\mathcal{I}$, so $\mathcal{L}_1^2=-2 \mathcal{L}_1$ and inductively $\mathcal{L}_1^n= (-2)^{n-1} \mathcal{L}_1 \; \forall \, n\geq 1, \; n\in\mathbb{Z}$. So,
\begin{align}
    \mathcal{E}_{1t} = \sum_{n=0}^\infty \frac{(t\mathcal{L}_1)^n}{n!} = \mathcal{I} + \mathcal{L}_1 \sum_{n=1}^\infty \frac{t^n(-2)^{n-1}}{n!} = \mathcal{I} + \left( \text{Ad}M - \mathcal{I} \right) \frac{1}{2}\left(1-e^{-2t}\right) = \alpha (t) \, \mathcal{I} + \beta (t) \, \text{Ad}M \label{eqa32} 
\end{align}
where $\alpha (t) \equiv \frac{1+e^{-2t}}{2} $, $\beta (t) \equiv  \frac{1-e^{-2t}}{2} $. Recalling the definition \cref{eq4}
\begin{align}
    G_{\mathcal{A}_B}(\mathcal{E}_{1t}) &= \frac{1}{2\, 2^n} \; {\mathlarger{\mathbb{E}}}_{X_{\mathcal{A}_B},Y_{\mathcal{A}_B^\prime}} \left[ \left\Vert \left[ X_{\mathcal{A}_B}, \alpha(t) \, Y_{\mathcal{A}^\prime_B} + \beta(t) \, \text{Ad}M (Y_{\mathcal{A}^\prime_B}) \right] \right\Vert_2^2 \right] \nonumber \\
    &= \beta^2(t) \frac{1}{2 \, 2^n} \; {\mathlarger{\mathbb{E}}}_{X_{\mathcal{A}_B},Y_{\mathcal{A}_B^\prime}} \left[ \left\Vert \left[ X_{\mathcal{A}_B}, \text{Ad}M (Y_{\mathcal{A}^\prime_B}) \right] \right\Vert_2^2 \right] = \beta^2(t) \, G_{\mathcal{A}_B}(\text{Ad}M) \label{eqa33}
\end{align}
Since $\text{Ad}M$ is a unitary channel with $\left\lvert \mel{\mu^\prime}{M}{\mu} \right\rvert = 2^{-n/2}$,
\begin{align}
   G_{\mathcal{A}_B}(\text{Ad}M)&=1-\frac{1}{2^n} \sum_{\mu,\mu^\prime=1}^{2^n} \lvert \mel{\mu^\prime}{M}{\mu}\rvert^4 = 1-\frac{1}{2^n} \label{eqa34}
\end{align}
so,
\begin{align}
    G_{\mathcal{A}_B}(\mathcal{E}_{1t})=\beta^2(t) \left( 1- \frac{1}{2^n} \right) \label{eqa35}
\end{align}
\subsection{\autoref{exmp2} calculations}
Notice that since $\Pi_i$ are eigenprojectors of $H$, $[\text{ad}H,\mathcal{D}_H]=0$ and $\text{ad}H \, \mathcal{D}_H =0$. Also, $\mathcal{D}_H$ is an orthogonal projector, so
\begin{align}
    \mathcal{E}_{2t}&= e^{it\, \text{ad}H} e^{\lambda t \mathcal{D}_H} e^{-\lambda t}=e^{it\, \text{ad}H} \left( \mathds{1} - \mathcal{D}_H + e^{\lambda t} \mathcal{D}_H \right) e^{-\lambda t} = e^{-\lambda t} \left( e^{it\, \text{ad}H} + \left( e^{\lambda t} -1 \right) \mathcal{D}_H \right) \nonumber \\
    &= a(t) \, e^{it\, \text{ad}H} + (1-a(t)) \, \mathcal{D}_H \label{eqa36}
\end{align}
where $a(t)\equiv e^{-\lambda t}$. Since $\Tr \left[ P_\mu \Pi_i \right] = 2^{-n}$,
\begin{align}
\mathcal{D}_H \left( P_\mu \right) = \sum_{i=1}^{2^n} \Pi_i P_{\mu} \Pi_i = 2^{-n} \sum_{i=1}^{2^n} \Pi_i = \frac{\mathds{1}}{2^n} \label{eqa37}
\end{align}
so,
\begin{align}
    G_{\mathcal{A}_B}(\mathcal{E}_{2t}) &= \frac{1}{2\, 2^n} \; {\mathlarger{\mathbb{E}}}_{X_{\mathcal{A}_B},Y_{\mathcal{A}_B^\prime}} \left[ \left\lVert \left[ X_{\mathcal{A}_B}, a(t) \, e^{it \, \text{ad}H} (Y_{\mathcal{A}^\prime_B}) + (1-a(t)) \, \mathcal{D}_H (Y_{\mathcal{A}^\prime_B}) \right] \right\rVert_2^2 \right] \nonumber \\
    &= a^2(t) \frac{1}{2 \, 2^n} \; {\mathlarger{\mathbb{E}}}_{X_{\mathcal{A}_B},Y_{\mathcal{A}_B^\prime}} \left[ \left\Vert \left[ X_{\mathcal{A}_B}, e^{it \, \text{ad}H} (Y_{\mathcal{A}^\prime_B}) \right] \right\Vert_2^2 \right] = a^2(t) \, G_{\mathcal{A}_B}(e^{it \, \text{ad}H}) \label{eqa38}
\end{align}
Note that $e^{it \, \text{ad}H} [\bullet ] = e^{it\, H} \bullet e^{-it \, H}$ is a unitary channel and since $H^2=\mathds{1}$, $e^{it\, H}=\cos t \, \mathds{1} + i \, \sin t \, H$. So,
\begin{align}
    G_{\mathcal{A}_B}(e^{it \, \text{ad}H})&=1-\frac{1}{2^n} \sum_{\mu,\mu^\prime=1}^{2^n} \lvert \mel{\mu^\prime}{\cos t \, \mathds{1} + i \, \sin t \, H}{\mu}\rvert^4 = 1-\frac{1}{2^n} \sum_{\mu,\mu^\prime=1}^{2^n} \lvert \delta_{\mu^\prime \mu} \, \cos t + i \, \delta_{\mu^\prime\bar{\mu}} \, \sin t \rvert^4 \nonumber\\
    &= 1-\frac{1}{2^n} \sum_{\mu,\mu^\prime=1}^{2^n} \delta_{\mu^\prime \mu} \, \cos^4t +  \delta_{\mu^\prime \bar{\mu}}  \, \sin^4 t + 2 \, \delta_{\mu^\prime \mu} \, \delta_{\mu^\prime \bar{\mu}} \, \cos^2t \, \sin^2t = 1 - \left( \cos^4 t + \sin^4 t \right) = \frac{\sin^2 (2t)}{2} \label{eqa39}
\end{align}
where we used that $\mel{\mu^\prime}{H}{\mu}=\delta_{\mu^\prime \bar{\mu}}$ and $\bar{\mu} \neq \mu \; \, \forall \mu$. So,
\begin{align}
    G_{\mathcal{A}_B}(\mathcal{E}_{2t})=a^2(t) \, \frac{\sin^2 (2t)}{2} \label{eqa40}
\end{align}
\subsection{Proof of \cref{eq15}}
Using $\{ f_\gamma \}_{\gamma =1}^{d(\mathcal{A}^\prime )}= \{ \Pi , \mathds{1} - \Pi \}$ in \cref{eq6} and the fact that $\mathcal{E}$ is CPTP, we have
\begin{align}
  G_{\mathcal{A}_{LE}}(\mathcal{E})&=\frac{1}{d} \left( \lVert \mathcal{E} (\Pi ) \rVert_2^2 + \lVert \mathcal{E} (\mathds{1}-\Pi ) \rVert_2^2 - \lvert \langle \Pi , \mathcal{E}(\Pi) \rangle \rvert^2 - \lvert \langle \Pi , \mathcal{E}(\mathds{1}-\Pi) \rangle \rvert^2 - \left\lvert \langle \frac{\mathds{1}-\Pi}{\sqrt{d-1}} , \mathcal{E}(\Pi) \rangle  \right\rvert^2- \left\lvert \langle \frac{\mathds{1}-\Pi}{\sqrt{d-1}} , \mathcal{E}(\mathds{1} -\Pi) \rangle  \right\rvert^2 \right) \nonumber \\
  &= \frac{1}{d} \left( \lVert \mathcal{E} (\Pi ) \rVert_2^2 + \Tr[ \mathds{1} +\mathcal{E}(\Pi ) \, \mathcal{E} (\Pi ) -2 \, \mathcal{E} (\Pi ) ] - \lvert\Tr[ \Pi \, \mathcal{E} (\Pi ) ]\rvert^2 - \Tr[ \Pi - \Pi \, \mathcal{E} (\Pi ) ]^2 - \frac{1}{d-1} \lvert\Tr[ \mathcal{E}(\Pi ) - \Pi \, \mathcal{E} (\Pi ) ]\rvert^2 \right. \nonumber \\
  &\hspace{40pt} \left.- \frac{1}{d-1} \lvert\Tr[ \mathds{1} - \mathcal{E} (\Pi ) - \Pi + \Pi \, \mathcal{E} (\Pi )]\rvert^2 \right) \nonumber \\
  &= \frac{1}{d} \left( 2 \lVert \mathcal{E} (\Pi ) \rVert_2^2 + d - 2 - \mathcal{L}_2^2 - \left(1 - \mathcal{L}_2 \right)^2 - \frac{1}{d-1} \left( 1- \mathcal{L}_2 \right)^2 - \frac{1}{d-1} \left( d -2 + \mathcal{L}_2 \right)^2 \right) \nonumber \\
  &= \frac{2}{d} \left( \lVert \mathcal{E} (\Pi ) \rVert_2^2 - \frac{\mathcal{L}_2 \left( d\mathcal{L}_2 -2 \right) +1}{d-1} \right) \label{eqa41}
\end{align}
where $\mathcal{L}_2 := \Tr [ \Pi \, \mathcal{E}(\Pi )]$.
\subsection{Proof of \cref{eq17}}
One can compute the $\mathcal{A}$-OTOC using either the basis $\{f_\gamma \}_{\gamma =1}^{d(\mathcal{A}^\prime)}= \{\mathds{1}_J \}_{J=1}^{2^{n-k}}$ or the basis  $\{\hat{f}_\delta \}_{\delta =1}^{d(\mathcal{A}^\prime)}=\{S_\mu/2^{(n-k)/2} \}_{\mu=1}^{2^{n-k}}$. Here, it is convenient to use the former. Let us now formalize the chosen rank-1 dephasing operators
\begin{align}
\begin{aligned} \label{eqa42}
    &\Pi_i^{(J)}=\ket{\psi_i^{(J)}} \bra{\psi_i^{(J)}} , \quad i=1,\dots,\chi, \;  J=1,\dots,2^{n-k}\\
    &\Pi_{j\alpha} = \left( \sum_{J=1}^{2^{n-k}} \lambda_{j\alpha}^{(J)} \ket{\psi_j^{(J)}} \right) \left(\sum_{J=1}^{2^{n-k}} \bar{\lambda}_{j\alpha}^{(J)} \bra{\psi_j^{(J)}} \right), \quad j=\chi +1 ,\dots,2^k, \, \alpha=1,\dots,2^{n-k}, \quad \left\lvert \lambda_{j\alpha}^{(J)} \right\rvert^2 = \frac{1}{2^{n-k}}
\end{aligned}
\end{align}
where $\left\{ \ket{\psi_k^{(J)}} \right\}_{k=1}^{2^k}$ is an orthonormal basis of the $J$ irrep, the phases of $\lambda_{j\alpha}^{(J)}$ are chosen such that the projectors to be orthogonal and $\bar{\lambda}$ denotes the complex conjugate. Let us compute the following quantity
\begin{align}
    \langle \mathds{1}_{J^\prime} , \mathcal{D}_\chi (\mathds{1}_J ) \rangle &= \Tr \left[ \mathds{1}_{J^\prime} \left( \sum_{i=1}^{\chi} \sum_{J_1=1}^{2^{n-k}} \Pi_i^{(J_1)} \, \mathds{1}_J \, \Pi_i^{(J_1)} \right.\right.\nonumber \\
    &\left.\left.\hspace{65pt}+ \sum_{j=\chi +1}^{2^k} \sum_{\alpha=1}^{2^{n-k}} \sum_{J_2,J_3,J_4,J_5=1}^{2^{n-k}} \lambda_{j\alpha}^{(J_2)} \bar{\lambda}_{j\alpha}^{(J_3)}  \lambda_{j\alpha}^{(J_4)} \bar{\lambda}_{j\alpha}^{(J_5)}  \ket{\psi_j^{(J_2)}} \bra{\psi_j^{(J_3)}} \mathds{1}_J \ket{\psi_j^{(J_4)}}  \bra{\psi_j^{(J_5)}} \right) \right] \nonumber \\
    &=  \Tr \left[ \sum_{i=1}^{\chi} \sum_{J_1=1}^{2^{n-k}} \delta_{J_1 J^\prime} \, \delta_{J_1J} \, \Pi_i^{(J_1)} \right. \nonumber \\
    &\hspace{65pt} \left. +  \sum_{j=\chi +1}^{2^k} \sum_{\alpha=1}^{2^{n-k}} \sum_{J_2,J_3,J_4,J_5=1}^{2^{n-k}} \delta_{J_2 J^\prime} \, \delta_{J_3J} \, \delta_{J_3J_4} \, \lambda_{j\alpha}^{(J_2)} \bar{\lambda}_{j\alpha}^{(J_3)}  \lambda_{j\alpha}^{(J_4)} \bar{\lambda}_{j\alpha}^{(J_5)}  \ket{\psi_j^{(J_2)}} \bra{\psi_j^{(J_5)}} \right] \nonumber \\
    &= \sum_{i=1}^\chi \delta_{J^\prime J}  +  \sum_{j=\chi +1}^{2^k} \sum_{\alpha=1}^{2^{n-k}} \sum_{J_5=1}^{2^{n-k}} \lambda_{j\alpha}^{(J^\prime)} \bar{\lambda}_{j\alpha}^{(J)}  \lambda_{j\alpha}^{(J)} \bar{\lambda}_{j\alpha}^{(J_5)} \, \delta_{J^\prime J_5} \nonumber \\
    &= \chi \, \delta_{J^\prime J} +  \sum_{j=\chi +1}^{2^k} \sum_{\alpha=1}^{2^{n-k}} \left\lvert \lambda_{j\alpha}^{(J^\prime)}\right\rvert^2 \, \left\lvert \lambda_{j\alpha}^{(J)}\right\rvert^2 = \chi \, \delta_{J^\prime J}+ \left(2^k - \chi \right) \, 2^{n-k} \frac{1}{(2^{n-k})^2} \nonumber \\
    &= \chi \, \delta_{J^\prime J}+ \frac{2^k - \chi}{2^{n-k}} \label{eqa43}
\end{align}
Now, using \cref{eqa43} the $\mathcal{A}$-OTOC is
\begin{align}
    G_{\mathcal{A}_{st}}(\mathcal{D}_\chi ) &= \frac{1}{2^n} \left( \sum_{J=1}^{2^{n-k}} \langle \mathcal{D}_\chi (\mathds{1}_J) , \mathcal{D}_\chi (\mathds{1}_J) \rangle - \sum_{J,J^\prime =1}^{2^{n-k}} \left\lvert \left\langle \frac{\mathds{1}_{J^\prime}}{\sqrt{2^k}} , \mathcal{D}_\chi (\mathds{1}_J ) \right\rangle \right\rvert^2\right) \nonumber \\
    &= \frac{1}{2^n} \left( \sum_{J=1}^{2^{n-k}} \langle \mathds{1}_J , \mathcal{D}_\chi (\mathds{1}_J) \rangle - \frac{1}{2^{k}} \sum_{J,J^\prime =1}^{2^{n-k}}  \left\lvert \chi \, \delta_{J^\prime J}+ \frac{2^k - \chi}{2^{n-k}} \right\rvert^2 \right) \nonumber \\
    &= \frac{1}{2^n} \left( \sum_{J=1}^{2^{n-k}} \left( \chi + \frac{2^k - \chi}{2^{n-k}} \right) - \frac{1}{2^k} \sum_{J,J^\prime =1}^{2^{n-k}} \chi^2 \, \delta_{J^\prime J} + \left(\frac{2^k - \chi}{2^{n-k}} \right)^2 + 2 \, \chi \left(\frac{2^k - \chi}{2^{n-k}} \right) \, \delta_{J^\prime J} \right) \nonumber \\
    &= \frac{1}{2^n} \left( \chi \, 2^{n-k} + 2^k - \chi - \frac{1}{2^k} \left( \chi^2 \, 2^{n-k} + \left( 2^k - \chi \right)^2 + 2 \, \chi \left(2^k - \chi \right) \right)\right) \nonumber \\
    &= \left( 1- \frac{2^k}{2^n} \right) \frac{\chi}{2^k} \left( 1-\frac{\chi}{2^k} \right) \label{eqa44}
\end{align}

\end{document}